\documentclass[preprint2,flushrt]{aastex}

\def\mbar{\ifmmode\overline{m}\else$\overline{m}$\fi}
\def\Mbar{\ifmmode\overline{M}\else$\overline{M}$\fi}
\def\mibar{\ifmmode\overline{m}_I\else$\overline{m}_I$\fi}
\def\MIbar{\ifmmode\overline{M}_I\else$\overline{M}_I$\fi}
\def\Nbar{\ifmmode\overline{N}\else$\overline{N}$\fi}
\def\solar{\ifmmode_{\mathord\odot}\else$_{\mathord\odot}$\fi}
\def\hst{\emph{HST}}
\def\etal{et al{.}}
\def\ho{\ifmmode H_0\else$H_0$\fi}
\def\sna{SN~Ia}
\def\batm{{\sc BATM}}

\def\dmod{\ifmmode(m{-}M)_0\else$(m{-}M)_0$\fi}
\def\mM{\ifmmode(m{-}M)_0\else$(m{-}M)_0$\fi}
\def\vi{\ifmmode(V{-}I)\else$(V{-}I)$\fi}
\def\viz{\ifmmode(V{-}I)_0\else$(V{-}I)_0$\fi}
\def\EBV{\ifmmode E_{B-V}\else$E_{B-V}$\fi}

 \shorttitle{Cosmology from High-z Supernovae}
 \shortauthors{Tonry et al.}

\begin{document}
\received{27 Feb 2003}
\title{Cosmological Results from High-z Supernovae\altaffilmark{1,2}}

\author{John L. Tonry,\altaffilmark{3}
Brian P. Schmidt,\altaffilmark{4}
Brian Barris,\altaffilmark{3}
Pablo Candia,\altaffilmark{5}
Peter Challis,\altaffilmark{6}
Alejandro Clocchiatti,\altaffilmark{7} 
Alison L. Coil,\altaffilmark{8}
Alexei V. Filippenko,\altaffilmark{8}
Peter Garnavich,\altaffilmark{9}
Craig Hogan,\altaffilmark{10}
Stephen T. Holland,\altaffilmark{9}
Saurabh Jha,\altaffilmark{6,8}
Robert P. Kirshner,\altaffilmark{6}
Kevin Krisciunas,\altaffilmark{5,11}
Bruno Leibundgut,\altaffilmark{12}
Weidong Li,\altaffilmark{8}
Thomas Matheson,\altaffilmark{6}
Mark M. Phillips,\altaffilmark{11}
Adam G. Riess,\altaffilmark{13}
Robert Schommer,\altaffilmark{5,15}
R. Chris Smith,\altaffilmark{5}
Jesper Sollerman,\altaffilmark{14}
Jason Spyromilio,\altaffilmark{12}
Christopher W. Stubbs,\altaffilmark{10}
and Nicholas B. Suntzeff\altaffilmark{5}}

\altaffiltext{1}{Based in part on observations with the NASA/ESA \emph{Hubble
Space Telescope,} obtained at the Space Telescope Science Institute,
which is operated by the Association of Universities for Research in
Astronomy (AURA), Inc., under NASA contract NAS 5-26555. This
research is primarily associated with proposal GO-8177, but also
uses and reports results from proposals GO-7505, 7588, 8641, and 9118.}
\altaffiltext{2}{
CFHT: Based in part on observations taken with the
Canada-France-Hawaii Telescope, 
operated by the National Research Council of Canada, le Centre National
de la Recherche Scientifique de France, and the University of Hawaii.
CTIO: Based in part on observations taken at the Cerro Tololo Inter-American
Observatory.
Keck: Some of the data presented herein were obtained at the W. M. Keck
Observatory, which is operated as a scientific partnership among
the California Institute of Technology, the University of California,
and the National Aeronautics and Space Administration.  The Observatory
was made possible by the generous financial support of the W. M. Keck
Foundation.
UH: Based in part on observations with the University of Hawaii 2.2-m telescope
at Mauna Kea Observatory, Institute for Astronomy, University of
Hawaii.
UKIRT: Based in part on observations with the United Kingdom Infrared Telescope
(UKIRT) operated by the Joint Astronomy Centre on behalf of the
U.K. Particle Physics and Astronomy Research Council.
VLT: Based in part on observations obtained at the European Southern
Observatory, Paranal, Chile, under programs ESO 64.O-0391 and ESO 64.O-0404.
WIYN: Based in part on observations taken at the WIYN Observatory, a
joint facility 
of the University of Wisconsin-Madison, Indiana University, Yale
University, and the National Optical Astronomy Observatories.}
\altaffiltext{3}{Institute for Astronomy, University of Hawaii,
2680 Woodlawn Drive, Honolulu, HI 96822; {jt@ifa.hawaii.edu},
{barris@ifa.hawaii.edu}} 
\altaffiltext{4}{The Research School of Astronomy and Astrophysics,
  The Australian National University, Mount Stromlo and Siding Spring
  Observatories, via Cotter Rd, Weston Creek PO 2611, Australia;
  {brian@mso.anu.edu.au}}
\altaffiltext{5}{Cerro Tololo Inter-American Observatory, Casilla
  603, La Serena, Chile; {nsuntzeff@noao.edu}, {kkrisciunas@noao.edu},
  {pcandia@ctiosz.ctio.noao.edu}} 
\altaffiltext{6}{Harvard-Smithsonian Center for Astrophysics, 60
  Garden Street, Cambridge, MA 02138; {kirshner@cfa.harvard.edu},
  {pchallis@cfa.harvard.edu}, {tmatheson@cfa.harvard.edu}} 
\altaffiltext{7}{Pontificia Universidad Cat\'{o}lica de Chile,
  Departamento de Astronom\'{i}a y Astrof\'{i}sica,
  Casilla 306, Santiago 22, Chile; {aclocchi@astro.puc.cl}}
\altaffiltext{8}{University of California, Berkeley, Department of
  Astronomy, 601 Campbell Hall, Berkeley, CA 94720-3411;
  {alex@astro.berkeley.edu}, {weidong@astro.berkeley.edu},
  {acoil@astro.berkeley.edu}, {sjha@astro.berkeley.edu}} 
\altaffiltext{9}{University of Notre Dame, Department of Physics, 225
  Nieuwland Science Hall, Notre Dame, IN 46556-5670;
  {pgarnavi@miranda.phys.nd.edu}, {sholland@nd.edu}} 
\altaffiltext{10}{University of Washington, Department of Astronomy,
  Box 351580, Seattle, WA 98195-1580; {hogan@astro.washington.edu},
  {stubbs@astro.washington.edu}} 
\altaffiltext{11}{Las Campanas Observatory, Casilla 601, La Serena,
  Chile; {mmp@lco.cl}} 
\altaffiltext{12}{European Southern Observatory, Karl-Schwarzschild-Strasse
  2, Garching, D-85748, Germany; {bleibund@eso.org}, {jspyromi@eso.org}}
\altaffiltext{13}{Space Telescope Science Institute, 3700 San Martin Drive,
Baltimore, MD 21218; {ariess@stsci.edu}}
\altaffiltext{14}{Stockholm Observatory, SCFAB, SE-106 91 Stockholm,
  Sweden; {jesper@astro.su.se}}
\altaffiltext{15}{Deceased 12 December 2001}
\begin{abstract}

The High-$z$ Supernova Search Team has discovered and observed 8 new
supernovae in the redshift interval $z=0.3$--1.2. These
independent observations, analyzed by similar but distinct methods,
confirm the result of Riess et al. (1998a) and Perlmutter et al. (1999)
that supernova luminosity distances imply an accelerating universe.
More importantly, they extend the redshift range of consistently
observed \sna\ to $z\approx 1$, where the signature of cosmological
effects has the opposite sign of some plausible systematic effects.
Consequently, these measurements not only provide another quantitative
confirmation of the importance of dark energy, but also constitute a
powerful qualitative test for the cosmological origin of cosmic
acceleration. We find a rate for \sna\ of 
$(1.4 \pm 0.5) \times 10^{-4} \; h^3 \; \hbox{Mpc}^{-3} \;
\hbox{yr}^{-1}$ at a mean redshift of 0.5.
We present distances and host extinctions for 230 \sna.
These place the following constraints on cosmological
quantities: if the equation of state parameter of the dark energy is
$w=-1$, then 
$H_0\,t_0 = 0.96\pm0.04$, and 
$\Omega_\Lambda-1.4\Omega_M=0.35\pm0.14$. 
Including the constraint of a flat Universe, we find
$\Omega_M=0.28\pm0.05$, independent of any large-scale structure
measurements. Adopting a prior based on the 2dF redshift survey constraint on
$\Omega_M$ and assuming a flat universe, we find that the equation of state
parameter of the dark energy lies in the range $-1.48<w<-0.72$ at 95\%
confidence.  If we further assume that $w>-1$, we obtain $w<-0.73$ at
95\% confidence.  These constraints are similar in precision and in
value to recent results reported using the WMAP satellite, also in
combination with the 2dF redshift survey.
\end{abstract}

\keywords{galaxies: distances and redshifts ---
cosmology: distance scale --- supernovae: general}

\section{Introduction}
\subsection{\sna\ and the Accelerating Universe}

Discovering Type Ia supernovae (\sna) with the intent of measuring the
history of cosmic expansion began in the 1980s with pioneering efforts
by N{\o}rgaard-Nielsen et al. (1989).  Their techniques, extended by the
Supernova Cosmology Project (Perlmutter et al. 1995) and by the
High-$z$ Supernova Search Team (HZT; Schmidt et al. 1998), started to
produce interesting results once large-format CCDs were introduced on
fast telescopes.  Efforts to improve the use of \sna\ as standard
candles by Phillips (1993), Hamuy \etal\ (1995), and Riess, Press \&
Kirshner (1996) meant that data from a modest number of these objects
at $z\approx 0.5$ should produce a significant measurement of cosmic
deceleration. Early results by Perlmutter \etal\ (1997) favored a
large deceleration which they attributed to $\Omega_M$ near 1.  But
subsequent analysis of an augmented sample (Perlmutter \etal\ 1998)
and independent work by the HZT (Garnavich \etal\ 1998a)
showed that the deceleration was small, and far from
consistent with $\Omega_M = 1$.

Both groups expanded their samples and both reached the surprising
conclusion that cosmic expansion is accelerating (Riess et al. 1998a;
Perlmutter et al. 1999). Cosmic acceleration requires the presence of
a large, hitherto undetected component of the Universe with negative
pressure: the signature of a cosmological constant or other form of
``dark energy.'' If this inference is correct, it points to
a major gap in current understanding of the fundamental physics of
gravity (e.g., Carroll 2001, Padmanabhan 2002).  The consequences of
these astronomical 
observations for theoretical physics are important and have led to a
large body of work related to the cosmological constant and its
variants.  For astronomers, this wide interest creates the obligation
to test and repeat each step of this investigation that concludes that
an unexplained energy is the principal component of the Universe.

Formally, the statistical confidence in cosmic acceleration is high ---
the inference of dark energy is not likely to result simply from
random statistical fluctuations.  But systematic effects that change
with cosmic epoch could masquerade as acceleration (Drell, Loredo, \&
Wasserman 2000; Rowan-Robinson 2002).  In this paper we not only provide
a statistically independent sample of well-measured supernovae, but
through the design of the search and execution of the follow-up, we
expand the redshift range of supernova measurements to the region
$z\approx 1$.  This increase in redshift range is important because
plausible systematic effects that depend on cosmic epoch, such as the
age of the stellar population, the ambient chemical abundances, and
the path length through a hypothesized intergalactic absorption (Rana
1979, 1980; Aguirre 1999a,b) would all increase with redshift.  But,
for plausible values of $\Omega_{\Lambda}$ and $\Omega_M$, 0.7 and 0.3
for example, the matter-dominated deceleration era would lie just
beyond $z = 1$. As a result, the {\it sign} of the observed effect on
luminosity distance would change: at $z\approx 0.5$ supernovae are {\it
dimmer} relative to an empty universe because of recent cosmic
acceleration, but at $z\approx 1$ the integrated cosmological effect
diminishes, while systematic effects are expected to be larger
(Schmidt et al. 1998).

Confidence that the Universe is dominated by dark energy has been
boosted by recent observations of the power spectrum of fluctuations
in the cosmic microwave background (de Bernardis et al. 2002; Spergel
\etal\ 2003).  Since the CMB observations strongly favor
$\Omega_{total} = 1 $ to high precision ($\pm 0.02$), and direct
measurements of $\Omega_M$ from galaxy clusters seem to lie around 0.3
(Peacock et al. 2001), mere subtraction shows there is a need for
significant dark energy that is not clustered with galaxies.  However,
this argument should not be used as an excuse to avoid scrutiny of
each step in the supernova analysis.  Supernovae provide the {\em
only} qualitative signature of the acceleration itself, through the
relation of luminosity distance with redshift, and most of that effect
is produced in the recent past, from $z=1$ to the present, and
not at redshift 1100 where the imprint on the CMB is formed.  This is
why more supernova data, better supernova data, and supernova data
over a wider redshift range are needed to confirm cosmic
acceleration. This paper is a step in that direction.
 
The most likely contaminants of the cosmological signal from \sna\ are
luminosity evolution, gray intergalactic dust, gravitational lensing,
or selection biases (see Riess 2000, Filippenko \& Riess 2001, and
Leibundgut 2001 for reviews). 
These have the potential to cause an apparent dimming of high-redshift
\sna\ that could mimic the effects of dark energy. But each of these
effects would also leave clues that we can detect.  By searching for
and limiting the additional observable effects we can find out whether
these potential problems are important.

If luminosity evolution somehow made high-redshift \sna\ intrinsically
dimmer than their local counterparts, supernova spectra, colors, rise
times and light-curve shapes should show some concomitant effects that
result from the different velocities, temperatures, and abundances of
the ejecta.  Comparison of spectra between low-redshift and high-redshift \sna\
(Coil et al. 2000) yields no significant difference, but the precision
is low and the predictions from theory (H\"{o}flich, Wheeler, \&
Thielemann 1998) are not easily translated into a limit on possible
variations in luminosity. Some troubling differences in the intrinsic
colors of the high and low-redshift samples have been pointed out by
Falco et al. (1999) and Leibundgut (2001).  Larger, well-observed
samples, including the one reported here, will show whether this
effect is real.  

Gray dust that absorbs without producing as much reddening as Galactic
dust could dim high-redshift \sna\ without leaving a measurable
imprint on the observed colors.  Riess et al. (1998a) argued that dust
of this sort would need to dim distant supernovae by 0.25 mag at
$z\approx 0.5$ in a matter-only universe. To produce the dimming we
attribute to acceleration, dust would also increase the variance of the
apparent \sna\ luminosities more than is observed.  However, such
dust, if smoothly dispersed between the galaxies, could appear
degenerate with cosmic acceleration (Rana 1979, 1980; Aguirre
1999a,b).  Aguirre's electromagnetic calculations of the scattering
properties for intergalactic dust show that it can produce less
reddening than galactic dust, but it cannot be perfectly gray.
Near-infrared observations of one \sna\ at $\approx 0.5$ (Riess et al. 2000) do
not show the presence of this form of dust. Further observations
over a wide wavelength range have been obtained by our team to
construct a more stringent limit; these will be reported in a future
paper (Jha et al. 2003a).  Recent work by Paerels et al. (2002), which
failed to detect X-ray scattering off gray dust around a
$z=4.3$ quasar, seems to indicate that gray, smoothly distributed,
intergalactic dust has a density which is too low by a factor of 10 to
account for the 0.25 mag dimming seen in the \sna\ Hubble diagram.

Selection biases could alter the cosmological inferences derived from
\sna\ if the properties of distant supernovae are systematically
different from those of the supernovae selected nearby. Simple
luminosity bias is minor because the
scatter in supernova luminosities, after correction for the
light-curve shape, is so small (Schmidt et al. 1998; Perlmutter et
al. 1999; Riess et al. 1998a).  However, the nearby sample of \sna\
currently spans a larger range of extinctions and intrinsic
luminosities than has been probed for \sna\ at $z\approx 0.5$.  Most
searches to date have only selected the tip of the iceberg: most
supernovae at high redshift lie below the sensitivity limit.  We assume
that we are drawing from the same population as nearby, but it would
be prudent to test this rigorously.  To make the SN evidence for dark
energy robust against sample selection effects, we need more sensitive
searches for \sna\ at $z\approx 0.5$ that could detect intrinsically
dim or extinguished \sna\ so we can verify that the distant supernovae
have a similar range of extinctions and intrinsic luminosities as the
nearby sample. This was one goal of the 1999 search reported here.
       
Extending the data set to higher redshift is a more ambitious and
difficult way to test for the cosmological origin of the observed
dimming effect.  Evolution or dust would most naturally lead to increased
dimming at higher redshift, while cosmic deceleration in the early
matter-dominated era ($z \ge 1.2$) would imprint the opposite
sign on luminosity distances.  A search
at higher redshift demand that we search to fainter flux limits in
bands that are shifted to the red to detect rest-frame $B$ and $V$.
In this paper, we describe a sensitive search in the $R$ and $I$ bands
carried out at the Canada-France-Hawaii Telescope (CFHT) and at the Cerro
Tololo Inter-American Observatory (CTIO) in 1999. For the highest
redshifts, we detect flux emitted in the ultraviolet at the source.
Members of the HZT have also embarked on an extensive study of
the $U$-band properties of nearby supernovae that will help with the
interpretation of these high-redshift objects (Jha 2003c).

Performing these tests requires searching for \sna\ with a deeper
magnitude limit in redder bands than previous searches.  This approach
can sample the full range of extinctions and luminosities at $z\approx
0.5$ and test for a turn-down in the Hubble diagram at $z\ge1$.  Thus
far, direct measurement of deceleration at early epochs rests in
observations of one \sna, SN 1997ff, at the remarkably high redshift
of $\sim1.7$.  These observations match best with a dark-energy source
for the observed behavior of \sna\ (Riess et al. 2001).  But this
single object represents just one data point isolated from the body of
\sna\ observations --- and there is evidence that this object could be
significantly magnified by gravitational lensing (Ben\'\i tez \etal\
2002).  A continuous sample from $z\approx 0.5$ through $z\approx 1$
out to $z\ge1.5$ is required to make this crucial test convincing.
The present paper represents a step in that direction by extending the
sample to $z\approx 1$. Future discoveries of very high-$z$ supernovae
using the Advanced Camera for Surveys on the {\it Hubble Space
Telescope (HST)}, successfully installed in 2002, will help bridge the
gap from the high-redshift end.

Section 2 of this paper describes the search, shows spectra of the
supernovae, and provides our photometric results. Section 3 gives an
account of the analysis including K-corrections, fits to the light
curves, and luminosity distances.  In \S 4 we discuss the inferred
cosmological parameters, and in \S 5 we discuss how these results
can be used to test for systematic errors in assessing cosmic
acceleration from \sna\ observations.  Some novel features of the
analysis are described more thoroughly in the Appendix.
 
\section{Observations}
\subsection{Search}
The \sna\ reported here were discovered at the
CFHT using the CFH-12K camera and at the CTIO 4-m Blanco
telescope using the CTIO Mosaic camera.  The CFH-12K camera provides
0\farcs206 pixels and a field of view of 0.33~deg$^2$, and the CTIO
mosaic has 0\farcs270 pixels with a field of view of 0.38~deg$^2$.

We obtained templates in 1999 October and obtained subsequent images
in November to find new objects, plausibly supernovae, with a rise
time in the observer frame of $\sim 1$ month.  On the nights of
1999~October~3 and 1999~October~7 (UT dates are used throughout this
paper), we obtained CFH-12K images of 15 
fields (5~deg$^2$) in the $I$ and $R$ bands with median seeing of
0\farcs65 and 0\farcs72, respectively.  We integrated for a total of
60 minutes on 6 of the $I$-band fields and 3 of the $R$-band fields,
and 30 minutes on the rest.  Each integration was of 10 minutes
duration, with the frames dithered by small offsets to help in
removing cosmic rays and CCD defects.  The photometric zero-points
were approximately $I=35.0$ and $R=35.6$ mag for the 60 minute exposures
--- in other words, a source of that magnitude would produce 1~e$^-$.  This
permits detection of stars with $m=24$ mag at a signal-to-noise ratio
(S/N) of 12 and 16, 
respectively, or detection of a supernova with $m=24$ mag at a S/N of 9
and 11 in a difference search.  A first-epoch search scheduled in
October at CTIO was clouded out.

Despite a generous time allocation, we ran into difficulties at CFHT a
month later because of weather. The first night (1999~November~2)
had clouds and bad seeing ($>1$\arcsec).  We accumulated 60
minutes in the $I$ band for each of 8 fields, but only one of these fields
contributed to the search, yielding one \sna: SN~1999ff.  The second
night (1999~November~3) had a mean extinction of about 0.5 mag
from clouds, but the seeing was very good (median 0\farcs59). We
obtained reasonably good data in the $I$ band for 8 fields, with
integrations of 60 minutes for five of these, and 30 minutes for three.
Three nights later (1999~November~6) we searched at CTIO, using 
$R$-band observations from CFHT as the first-epoch observations.  Although
the CTIO data enabled us to confirm the \sna\ found in the $I$ band at
CFHT, and produced several candidates that required spectroscopic
follow-up, we found no new supernovae from these observations.
Spectra of several CTIO candidates revealed the disappointing fact
that our attempt to compare November CTIO observations with October
CFHT templates produced subtly false candidates where a mismatch
between the CFHT and CTIO filters caused the difference between the
two to give the illusion of a new objects appearing between October and
November.  In fact, these were M stars or emission-line galaxies whose
H$\alpha$ emission line fell at the red end of the $R$-band filter.  Only the
comparison of CFHT data with CFHT templates provided genuine
detections of supernovae.

Our reduction and search procedure, detailed by Schmidt \etal\ (1998),
consists of bias subtraction, flatfielding, subtraction of $I$-band
fringes, masking bad columns, and finally combining dithered images to reject
cosmic rays and moving objects. In each field, we identified stars,
performed an astrometric solution, and then remapped the image onto a
common coordinate system. (This mapping is flux conserving but uses a
Jacobian to restore the photometric accuracy that is destroyed by
flatfielding.)  For each pair of images of a field, from the first
(October) and second (November) epoch, we then used the algorithms
described by Alard \& Lupton (1998) to convolve the image with the
better seeing into agreement with the worse image.  Next we matched the flux
levels and subtracted.  

These difference images were then searched
automatically for residuals resembling the point-spread function
(PSF), with possible \sna\ flagged for 
further inspection by humans.  Simultaneously, we searched all image pairs by
inspecting the difference images by eye.  We found little difference
in the detection efficiency between the semi-automatic approach and
inspection by experienced observers.  For all the \sna\
candidates, we inspected each of the individual exposures to ensure
that the putative \sna\ was not a cosmic ray, moving object, or CCD
flaw.  All 37 final candidates were tabulated along with estimated $I$
and $R$ magnitudes (from the CTIO observations), including the
properties of the possible host galaxy and distance from the center of
the host.  This candidate list became our observing list for
subsequent spectroscopic investigation.

Different exposure times in different fields, variable extinction due
to clouds, changes in seeing, and sensitivity variations among the
chips in the CFH-12K mosaic make it hard to quantify a single
detection threshold for genuine events in this search.  We believe
that, overall, our search is 95\% effective for objects brighter than
$I=23.5$ mag, and substantially better than 50\% at $I=24$ mag. In some
favorable cases we found objects as faint as $I=24.5$ mag.

\begin{figure}[t]
\epsscale{1.00}
\plotone{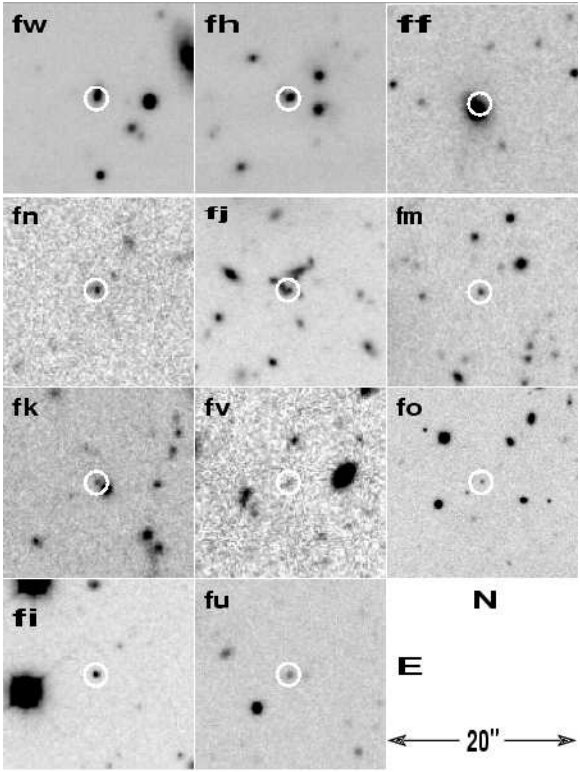}
\caption{Host galaxies for the eleven supernovae. Each
image is 20$''$ on a side and taken from
an average of several I-band frames. A 2$''$
radius circle marks the position of the supernova.
North is at the top and east to the left in each
image.}
\label{hoststamps}
\end{figure}

We began observing our 37 candidate objects at the Keck-II telescope
on 1999~November~8.  Our aim was to determine the nature of each
object and to measure its redshift.  Of these 37, two turned out to be
active galactic nuclei at redshifts of 1.47 and 1.67; one was an M
star and two were galaxies with H$\alpha$ emission admitted because of
mismatched $R$ filters; five had disappeared, suggesting either that
the discovery was well past maximum or the detection was spurious;
four were judged to be Type II supernovae from spectra or blue color;
two were too bright to be of interest for our purposes, $I\approx 21$
mag in bright host galaxies, possibly nearby SN II, for which we did
not spend time to get spectra; three were \sna\ with $0.3<z<0.7$ which
we chose not to follow; seven were faint candidates for which we did
not have time to get spectra; and 11 were the candidate \sna\ which we
chose to follow.  These are shown in Figure~\ref{hoststamps}. The
selection of objects was based on color information (we chose objects
with colors consistent with \sna\ with any amount of reddening, in the
range $0.1 < z < 1.5$); the amount of variation of the SN, with
preference to objects that were not seen in the previous epoch, but
that had varied significantly; the host-galaxy brightness at the
supernova position, avoiding supernova candidates on extremely bright
backgrounds; and host brightness/size --- avoiding supernovae in
low redshift galaxies ($z<0.1$). These selection
effects are all undesirable, but with limited telescope time, they were
necessary compromises to achieve the goals of the search.

Table~\ref{obstab} lists the J2000 coordinates of the \sna, the
Galactic extinction from Schlegel, Finkbeiner, \& Davis (1998), the
number of photometric observations in various bandpasses accumulated
from various telescopes, 
the modified Julian date of maximum light,
and our nickname for the supernova, assigned before
there was a designation for each event in the IAU Circulars (see Tonry et
al. 1999).  The last three objects on the list are real objects which
we pursued, but based on our spectra and subsequent light curves, the
evidence is too weak to consider them \sna.  SN~1999fi (Boris) was
very close to SN~1999fj, so it did not cost extra observing time to
get photometry, but the other two (SN~1999fo and SN~1999fu) fooled us
into making 22 photometric observations and expending many hours of
Keck time to get spectra that were, in the end, inconclusive.
Although these two are real events, with a definite flux increase from
October to November and a decrease in flux after November, we have not
shown they are \sna, and they are not included in the analysis.

\begin{deluxetable}{lllcrrl}
\tablewidth{0pt}   
\tablecaption{Fall 1999 Observations\label{obstab}} 
\tablehead{ 
\colhead{SN name} & 
\colhead{RA~(J2000)} & 
\colhead{Dec~(J2000)} & 
\colhead{$E(B{-}V)$} & 
\colhead{$N_{obs}$} & 
\colhead{MJD$_{max}$} & 
\colhead{Nickname}
}
\startdata
SN~1999fw & 23:31:53.03&+00:09:32.3 & 0.039 & 25 & 51482 & Nell  \\
SN~1999fh & 02:27:58.33&+00:39:36.8 & 0.031 & 15 & 51488 & Fearless Leader  \\
SN~1999ff & 02:33:54.39&+00:32:55.6 & 0.026 & 25 & 51494 & Bashful  \\
SN~1999fn & 04:14:03.88&+04:17:55.0 & 0.326 & 46 & 51498 & Scooby Doo  \\
SN~1999fj & 02:28:23.72&+00:39:09.6 & 0.030 & 20 & 51475 & Natasha  \\
SN~1999fm & 02:30:35.62&+01:09:43.3 & 0.023 & 21 & 51488 & Fred  \\
SN~1999fk & 02:28:53.88&+01:16:24.2 & 0.030 & 22 & 51492 & Velma  \\
SN~1999fv & 23:30:35.80&+00:16:40.0 & 0.041 &  8 & 51457 & Dudley Doright \\
SN~1999fo & 04:14:45.75&+06:38:34.3 & 0.291 & 11 &$\ldots$& Gertie  \\
SN~1999fi & 02:28:11.67&+00:43:39.3 & 0.031 & 12 &$\ldots$& Boris  \\
SN~1999fu & 23:29:48.21&+00:08:27.0 & 0.048 & 11 &$\ldots$& Alvin  \\
\enddata
\end{deluxetable}

\subsection{Spectral Observations and Reductions} 

We obtained spectra of our \sna\ candidates with LRIS (Oke et
al. 1995) on the Keck-II telescope during five nights between
1999 November 8 and 14.  The seeing varied from 0\farcs8 to
1\farcs1 during the first three nights and was $\sim$0\farcs5 on the
last two nights.  We used a 1\farcs0 slit for all of our observations,
except for SN~1999fv where we used a 0\farcs7 slit on one night.  We
used a 150 line mm$^{-1}$ grating for the first half of the run, then
switched to a 400 line mm$^{-1}$ grating for the second half in order
to better remove night-sky lines.  The resulting spectral resolution is
$\sim$20~\AA\ for the 150 line mm$^{-1}$ grating and $\sim$8~\AA\ for
the 400 line mm$^{-1}$ grating.  The pixel size for the 150 and 400 line
mm$^{-1}$ gratings is $\sim$5\ \AA\ pix$^{-1}$ and
$\sim$2\ \AA\ pix$^{-1}$, respectively.  The total exposure
times for each \sna\ are listed in Table \ref{spectab}.  We moved the
object along the slit between integrations to reduce the effects of
fringing.  The slit was oriented either near the parallactic angle, or to
include the nucleus of the host galaxy or a nearby bright star to
provide quantifiable astrometry along the slit and to define the trace
of a source along the spectral direction.  

When using the 150 line
mm$^{-1}$ grating, internal flatfield exposures and standard stars
were observed both with and without an order-blocking filter to remove
contamination from second-order light.  For the high-$z$ \sna\
observations no order-blocking filter was needed, as there is very
little light from the object or from the sky below an observed
wavelength of 4000~\AA.  We used BD+17\arcdeg4708 as a standard star
for the first two nights and Feige 34 (Massey et al. 1988) for the
last three nights.  Standard CCD processing and optimal spectral
extraction were done with IRAF\footnote[16]{IRAF is distributed by the 
National Optical Astronomy Observatories, which are operated by the
Association of Universities for Research in Astronomy, Inc., under
cooperative agreement with the National Science Foundation}.  
We used our own IDL
routines to calibrate the wavelengths and fluxes of the spectra and to
correct for telluric absorption bands.  For \sna\ that were observed
on more than one night, the data were binned to the larger pixel size
if different gratings were used, scaled to a common flux level, and
combined with weights depending on the exposure time to form a single
spectrum.

\subsection{\sna\ Spectra and Redshifts}

\begin{figure}[t]
\plotone{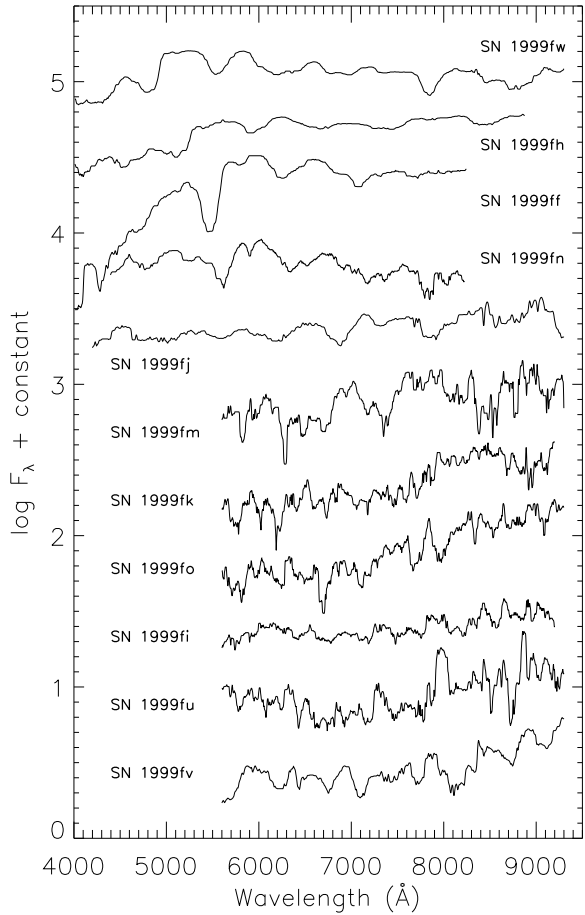}
\caption{High-$z$ supernova spectra shown at their observed
wavelengths, smoothed by a 32-pixel running median and a Gaussian of 
4-pixel sigma.
\label{restspec}}
\end{figure}

The \sna\ spectra are shown in Figure~\ref{restspec} at their
observed wavelengths.  For each \sna\ we list in Table \ref{spectab} the
redshift, the supernova type, the number of nights observed with Keck, the
total exposure time, and an explanation of how the redshift was
derived.  For \sna\ with narrow emission or absorption lines from the
host-galaxy light, we determined the redshift by using the centroid of
the line(s) with a Gaussian fit.  For the three \sna\ spectra without
narrow lines from the host galaxy, we use the broad \sna\ features to
determine the redshift by cross-correlating the high-$z$ spectrum with
low-$z$ \sna\ spectra obtained near maximum light.  We also include
galaxy and M-star spectra among our templates for cross-correlation
with spectra without narrow emission lines.  Errors on the redshifts
are $\pm$ 0.001 ($1\sigma$) when based on a narrow line and $\pm$ 0.01
when based on broad \sna\ features.  The redshift of SN~1999fv has a
large uncertainty (1.17--1.22) due to the low S/N
of the spectrum (see Coil et al. 2000 for further discussion).

\begin{figure}[t]
\plotone{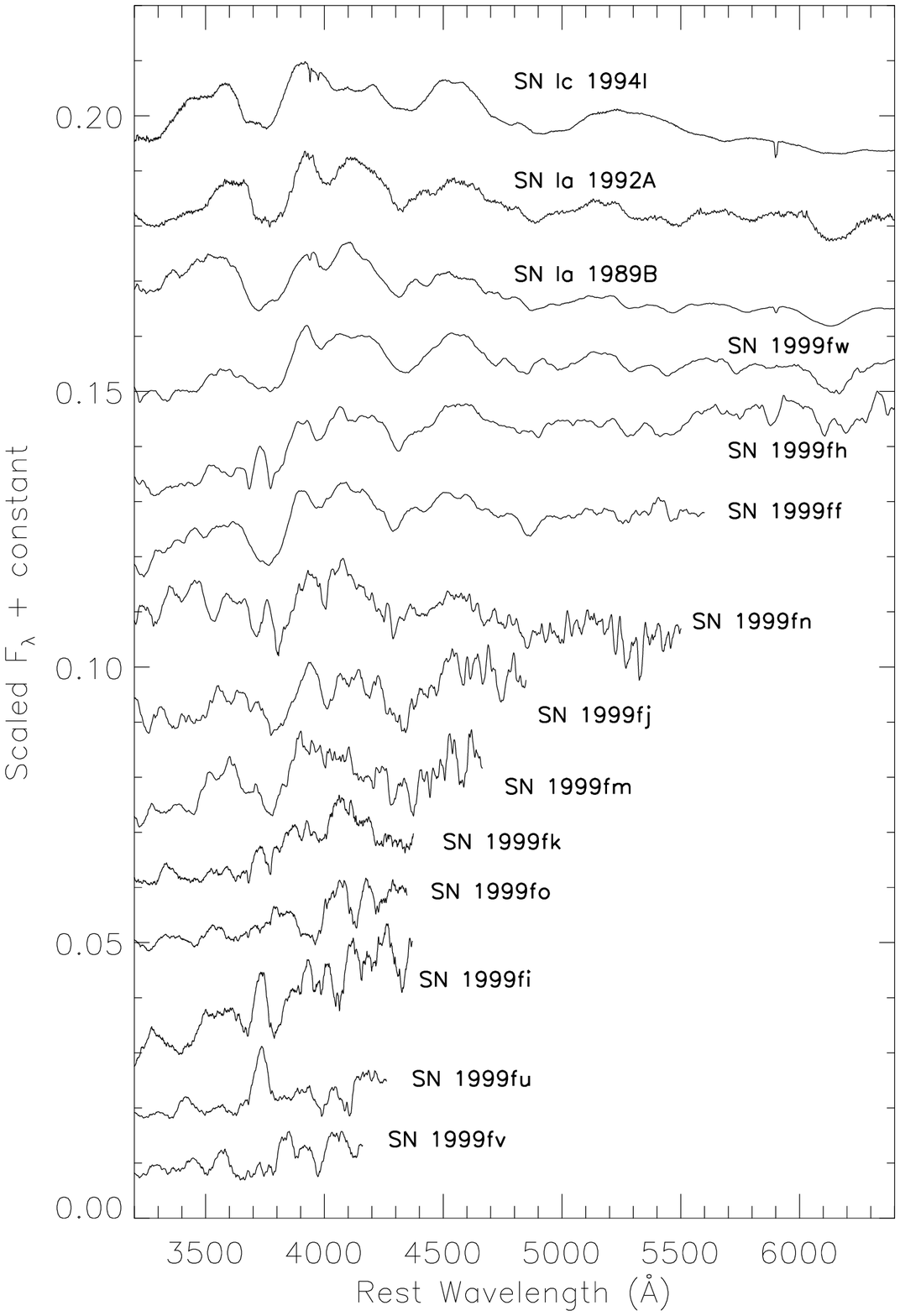}
\caption{High-$z$ SN spectra, shifted to rest wavelengths, smoothed by a 
Savitsky-Golay filter with a width of 100~\AA\ and compared 
to two low-$z$ \sna\  (SN~1992A and SN~1989B) and a low-$z$ 
SNIc (SN~1994I).  The high-$z$ SN spectra are shown sorted by redshift,
with the lowest-$z$ SN at the top. Based on these spectra, we cannot conclude
that SN 1999fi, 1999fo, and 1999fu are \sna.
\label{smoothspec}}
\end{figure}

Figure~\ref{smoothspec} presents the high-$z$ spectra smoothed with a
Savitsky-Golay filter of width 100~\AA, sorted by redshift, with the
lowest-$z$ \sna\ at the top.  This polynomial smoothing filter
preserves line features better than boxcar smoothing, which damps out
peaks and valleys in a spectrum. The width of the smoothing filter is
apparent in the [O~II] $\lambda$3727 emission 
line seen in many of the spectra.  We also
plot two low-$z$ \sna\ (SN~1989B, Wells et al. 1994; SN~1992A,
Kirshner et al. 1993) and a low-$z$ SNIc [SN~1994I dereddened by
$E(B{-}V)=0.45$ mag; Filippenko et al. 1995] for comparison, all
with ages before or at maximum light.  Features which distinguish a
\sna\ (Filippenko 1997) are deep \ion{Ca}{2} H\&K absorption near 3750~\AA, the
\ion{Si}{2} $\lambda$4130 dip blueshifted to 4000~\AA, and \ion{Fe}{2}
$\lambda$4555 and/or \ion{Mg}{2} $\lambda$4481, the combination of
which create a distinct double-bump feature centered at 4000~\AA.  For
the lower-$z$ spectra we have enough wavelength coverage to see
6150~\AA, where the \ion{Si}{2} $\lambda$6355 absorption feature is
prominent in \sna, but beyond $z\approx$ 0.4 this feature becomes
difficult to detect.  Early SNIc spectra look similar to \sna\
blueward of 5000~\AA, but lack the prominent \ion{Si}{2} dip at 4000~\AA\
which leads to the double-bump feature seen only in \sna.

SN 1999fw, 1999fh, 1999ff, 1999fn, and 1999fj are all clearly \sna.
SN~1999fm is a Type I SN, but we cannot distinguish from the
spectrum alone whether it is a Type Ia or Ic.  SN~1999fk shows a
definite rise around 4000~\AA\ and has some hints of a double peak at
this location, but is not definitively a \sna.  SN~1999fv shows
some bumps centered at 4000~\AA, which, as discussed in Coil et
al. (2000), are plausibly \sna\ features at a redshift of
$\sim 1.2$. All eight of these supernovae are used in this paper.

SN 1999fi, 1999fu, and 1999fo do not show any SN features in their
spectra.  For each of these systems, we have a redshift; for SN
1999fi and 1999fu this comes from a single line we assume to be
[O~II] $\lambda$3727.  For SN~1999fo the spectrum appears to have Ca~II H\&K
absorption at $z = 1.07$.  As stated above, these three objects are not
used in the analysis.

\begin{deluxetable}{lcccrl}
\tablewidth{0pt}   
\tablecaption{Fall 1999 Spectroscopic Data \label{spectab}} 
\tablehead{ 
\colhead{SN name} & 
\colhead{$z$} & 
\colhead{Type} & 
\colhead{$\#$ nights} & 
\colhead{Exp. (sec)} &
\colhead{$z$ determination}  
}
\startdata
SN~1999fw & 0.278 & Ia & 1 & 1200 & broad SN features \\
SN~1999fh & 0.369  & Ia & 1 & 1800 & [O~II] $\lambda$3727 in SN spectrum \\
SN~1999ff & 0.455 & Ia & 1 & 2200 & Balmer lines in galaxy spectrum \\
SN~1999fn & 0.477 & Ia & 2 & 5900 & [O~II] $\lambda$3727 in SN spectrum \\
SN~1999fj & 0.816 & Ia & 1 & 3600 & [O~II] $\lambda$3727 in SN spectrum \\
SN~1999fm & 0.95  & Ia\rlap{?} & 2 & 6600 & broad SN features \\
SN~1999fk & 1.057 & Ia\rlap{?} &2 & 7000  & [O~II] $\lambda$3727 in SN spectrum \\
SN~1999fv & 1.17-1.22 & Ia & 2 & 10000 & broad SN features \\
SN~1999fo & 1.07 & ?  & 3 & 14200 & Ca H\&K dip in spectrum\\
SN~1999fi & 1.10 & ? & 1 & 5000 &  [O~II] $\lambda$3727 in spectrum \\
SN~1999fu & 1.135  & ? & 2 & 7200 & [O~II] $\lambda$3727 in spectrum \\  
\enddata
\end{deluxetable}

\subsection{Properties of the Host Galaxies}

Host galaxies and the environment around each supernova are shown in
Figure~\ref{hoststamps}. The images are the average of the $I$-band
images with the best seeing. No host was detected for SN~1999fm.
SN~1999fj appears in a small group of galaxies and its host is assumed
to be the elongated galaxy just east of the supernova.  Host
magnitudes were measured from images taken after the supernova had
faded. For the brightest hosts, the flux was summed out to an isophote
of 26 mag/square-arcsec, but for the faint hosts the flux was summed
in a 4$''$ diameter aperture. The results, without correction for
extinction, are given in Table~\ref{hosttab} .

Offsets between the supernovae and their hosts are also given in Table
\ref{hosttab}, as is the projected separation in kpc assuming a
flat $\Lambda$-dominated cosmology with $\Omega_M=0.3$. The distribution of
projected separations for these new supernovae is narrower than that
of earlier searches (Farrah et al. 2002), meaning our supernovae were
discovered closer to the hosts. This may be due to the greater depth
of this search and better image quality as it attempted to detect
objects at $z>1$.

\begin{deluxetable}{lccccc}
\tablewidth{0pt}   
\tablecaption{Fall 1999 Host Galaxies$^a$\label{hosttab}} 
\tablehead{ 
\colhead{SN name} & 
\colhead{Host $R$} & 
\colhead{Host $I$} & 
\colhead{Host $Z$} & 
\colhead{Offset (\arcsec)} &
\colhead{Proj. Sep. (kpc) }  
}
\startdata
SN~1999fw &    21.1&    20.4& $\ldots$&     0.51& 2.3\\   
SN~1999fh &    21.2&    20.8&     20.5&     0.03& 0.2\\   
SN~1999ff &    19.9&    19.2&     19.1&     2.01& 12.6\\   
SN~1999fn &    24.5&    23.5& $\ldots$&     0.31& 2.0\\   
SN~1999fj &    23.3&    22.3&     21.9&     0.89& 7.2\\   
SN~1999fm &$\ldots$& $>24.5$& $\ldots$& $\ldots$& $\ldots$\\   
SN~1999fk &$\ldots$&    21.9&     23.7&     1.36& 11.9\\   
SN~1999fv & $>26.0$&    25.8& $\ldots$&     0.30& 2.7\\   
SN~1999fo &$\ldots$&    23.1&     23.2&     0.19& 1.7\\   
SN~1999fi &    24.8&    23.1&     22.9&     0.16& 1.4\\   
SN~1999fu &    24.2&    23.6&     23.5&     0.35& 3.1\\   
\enddata
\tablenotetext{a} {The offset is the angular separation between the SN and
host in arcsec, and the projected separation is in kpc. SN~1999fm had no 
detectable host. Ellipsis indicate no image available in that band.}
\end{deluxetable}

\subsection{Photometric Calibrations and Reductions}

The photometric data gathered of the supernovae discovered at CFHT and
CTIO came from a wide variety of instruments and telescopes including
Keck LRIS, Keck ESI, Keck NIRC, VLT FORS, VLT ISAAC, UKIRT UFTI, UH
2.2-m, VATT, WIYN, and \hst. We accumulated a total of 216
observations of our 11 candidate \sna\ over the span of approximately
one year, in the $V$, $R$, $I$, $Z$, and $J$ bandpasses, totaling
$\sim 200$ hours of exposure time.  The seeing of the combined,
remapped images had a median of 0\farcs83, with quartiles at 0\farcs72
and 1\farcs04.

Calibration of these data turned out to be especially challenging, due
to a variety of difficulties. The photometric calibration of stellar
sequences near our \sna\ was intended to come from observations with
the UH 2.2-m and the CTIO 1.5-m telescopes.  To this end, we observed
Landolt (1992) standards in $V$, $R$, $I$, and $Z$ on a variety of
photometric nights, and also observed our \sna\ fields to establish
the magnitudes of local photometric reference stars.  We took a
sequence of short exposures of a constant flatfield to establish the
shutter timing error, and checked the linearity of the CCD by
exposures of different duration.  For standard-star reductions we
summed the flux (with background subtraction) of each star through a
14\arcsec\ aperture, and calculated an atmospheric extinction term for
each night. A typical scatter in this fit to airmass and color was
0.02~mag, and arises from the usual causes: sky errors, imperfect
flatfielding, changes in atmospheric transparency, CCD non-linearity,
shutter timing errors, and PSF or scattered-light variations. In the
case of the UH 2.2-m data, there were some observations which deviated
significantly from the root-mean-square ({\it rms}) error of the
majority of data. This was eventually tracked down to an error in dome
control, causing observations to be vignetted by the dome shutter.
The UH 2.2-m data are therefore not completely reliable photometric
calibrations.  The CTIO 1.5-m data had no such problems, and were
taken on absolutely photometric nights. However, they do not cover all
fields, and the overlap in brightness between data with high
photometric accuracy from the 1.5-m and the unsaturated data from the
larger telescopes was not large.

To bridge these calibration difficulties, we combined two additional
calibration sources, the Sloan Digital Sky Survey (SDSS) (Stoughton et
al. 2002) and the 2001 HZT campaign (Barris et al. 2002). The 2001 HZT
campaign took great pains to obtain very high accuracy photometry at
the CTIO 1.5-m telescope, which was reduced in a manner identical to
the 1999 CTIO data, as described above. These observations of more
than 200 stars were used to calibrate a catalog of $R$, $I$, and $Z$
photometry of about 4000 stars near $\alpha$ = 02$^h$ 28$^m$, $\delta$
= +00\arcdeg 35\arcmin, derived from CFHT+12K mosaic images.  The $R$
and $I$ bandpasses of these data were transformed to the standard
Kron-Cousins system (Kron \& Smith 1951; Cousins 1976), using the CTIO
derived values, and the $Z$ band was transformed to the CTIO system,
as described below.

This 2001 HZT catalog and all of our supernovae at RA = $23^h$ and
$02^h$ are encompassed by the SDSS catalog. Using the matched stars at
$02^h$ between our 2001 HZT photometric catalog and the SDSS (using
PSF-derived magnitudes of the Sloan data --- not the default available
on the website), we derived the following transformations from the SDSS
$g'$, $r'$, $i'$, and $z'$ magnitudes and our Vega-based magnitudes:

$$  V-g' = +0.172 - 0.745 (g'-r'),$$
$$  R-r' = -0.122 - 0.316 (r'-i'),$$
$$  I-i' = -0.423 - 0.378 (i'-z'),$$
$$  Z-z' = -0.511 - 0.037 (r'-i'),$$
$$  Z-z' = -0.508 - 0.267 (r'-i') + 0.424 (i'-z').$$

\noindent
The two-color fit to $Z-z'$ is slightly better, but obscures the fact
that $Z$ and $z'$ are nearly identical, apart from a zero-point
offset. 

For all colors, the scatter in the transformation for individual stars
(which are not dominated by shot noise like the faint SDSS stars, or by 
saturation like the CFHT stars of roughly the same magnitude) is
approximately 0.02 mag {\it rms}. These transformations enable us to
calibrate the $R$, $I$, and $Z$ 1999 CFHT survey fields via the SDSS
early-release data.

Since we did not observe in the $V$ band during our Fall 2001
campaign, we needed extra steps to derive the calibration of the $V$
observations for our 1999 data.  We used Landolt (1992) stars in our
fields and derived a satisfactory relation for $0 < (R{-}I) < 1$ mag:
$$ (V{-}R) = -0.024 + 1.104 (R{-}I). $$ For redder stars, there are
significant differences in the $V$-band colors for giants and dwarfs,
which we sought to avoid.  This relation allowed us to produce ``$V$''
magnitudes for the Fall 2001 catalog, using the $R$ and $I$
magnitudes.  We then compared the stars that were also available from
SDSS, and derived a $g'$ to $V$ transformation. While individual stars
can have substantial errors via these transformations, in the mean of
hundreds of stars that we used these transformations are better than
0.01 mag.

\begin{figure}[t]
\plotone{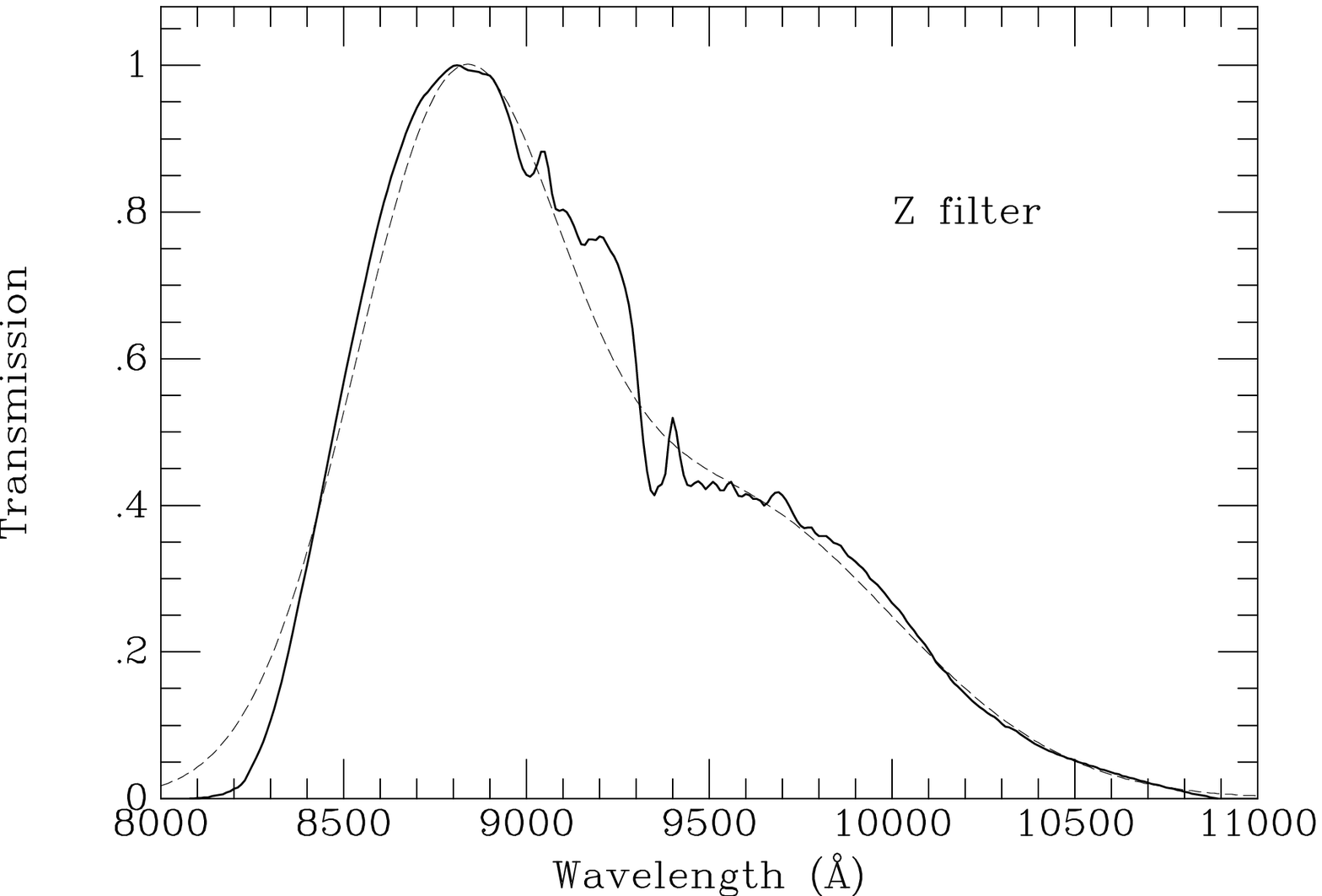}
\caption{$Z$ bandpass used for these observations. The dashed line
is an approximation of the actual transmission, made up of two Gaussians
as described in the text. \label{zfilt}}
\end{figure}

The $Z$ band was both important and problematic for us.  At $z>1$, the
rest-frame $B$ band starts to shift beyond the observer's $I$ band,
and the systematics of rest-frame $U$-band observations were not yet
well established at the time of these observations (but see Jha 2003c),
so we needed a color for our \sna\ based on bandpasses redder than $I$.
There is a variety of $Z$ bands; ours is a Vega-normalized bandpass
defined by the CTIO natural system (RG850, two aluminum reflections,
atmosphere, and a SITe CCD red cutoff), and it is quite well described as a
sum of two Gaussians, one centered at 8800~\AA\ with a full-width at
half-maximum (FWHM) of
280~\AA\ and a height of 0.86, and the other at 9520~\AA\ with a FWHM
of 480~\AA\ and a height of 0.41.  The $Z$ bandpass and this
approximation are illustrated in Figure~\ref{zfilt}. Our $Z$-band
observations were calibrated by observing a series of Landolt stars,
whose magnitudes were derived by integrating their spectrophotometry
(N. Suntzeff, in preparation) with this bandpass.  This system is defined
to have $(V-Z)=0$ mag for Vega.

The SDSS observations covered all our fields except for those
containing SN~1999fn and SN~1999fo. For all \sna, except these two, we
extracted SDSS stars which fell within the supernova fields, and used
the above relations to produce Johnson/Kron-Cousins magnitudes for
each star. For SN~1999fn and SN~1999fo, we used our UH 2.2-m
photometric observations, verified with a few overlapping stars from
the 1999 CTIO 1.5-m observations, to set the $R$ and $I$ magnitudes of
a stellar sequence around the SN. The $V,Z$ magnitudes were set using color
relations derived above.  As stated above, these transformations,
while good in the mean, may not be appropriate for unusual stars (or
quasars), and consequently, our derived magnitudes are based relative
to the sequence of stars, rather than weighting any individual star
too highly. The adopted standard-star sequences in the vicinity of
each supernova are given in Table~\ref{stdstar}. This calibration
procedure is far from ideal, but cross-checks suggest it is as good as
the Sloan zero-point uncertainty, 0.04 mag.

We calibrated the $J$-band data from VLT and UKIRT using observations
of ten standard stars from Persson et al. (1998).  The Keck data were
calibrated by transferring the photometric zero-points from the UKIRT
observations to a galaxy in the field of SN~1999fn.

\begin{deluxetable}{lllrrrrrrrr}
\tablewidth{0pt}   
\tablecaption{Local Standard Stars\label{stdstar}} 
\tablehead{ 
\colhead{SN name} & 
\colhead{RA} & 
\colhead{Dec} & 
\colhead{$V$} & \colhead{$\pm$} &
\colhead{$R$} & \colhead{$\pm$} &
\colhead{$I$} & \colhead{$\pm$} &
\colhead{$Z$} & \colhead{$\pm$}
}
\startdata
SN1999fu & 23:29:46.8 & +00:11:11 & 21.43&0.04 & 20.25&0.06 & 18.57&0.03 & 17.92&0.04\\
         & 23:29:46.5 & +00:04:45 & 20.70&0.03 & 19.69&0.04 & 18.42&0.03 & 17.93&0.04\\
         & 23:29:45.4 & +00:09:08 & 20.74&0.03 & 19.77&0.04 & 18.52&0.03 & 18.05&0.05\\
         & 23:29:41.7 & +00:10:18 & 19.40&0.02 & 18.87&0.03 & 18.36&0.03 & 18.20&0.05\\
         & 23:29:41.3 & +00:11:08 & 18.58&0.02 & 18.46&0.02 & 18.37&0.03 & 18.37&0.06\\
         & 23:29:37.8 & +00:05:06 & 20.73&0.03 & 19.80&0.04 & 18.79&0.04 & 18.40&0.06\\
         & 23:29:37.5 & +00:10:26 & 20.99&0.04 & 19.94&0.05 & 18.42&0.03 & 17.82&0.04\\
         & 23:29:35.3 & +00:10:06 & 20.94&0.03 & 19.87&0.05 & 18.45&0.03 & 17.89&0.04\\
SN1999ff & 02:34:00.2 & +00:34:20 & 20.03&0.02 & 18.99&0.03 & 17.62&0.02 & 17.08&0.03\\
         & 02:33:54.2 & +00:29:53 & 19.37&0.02 & 18.87&0.03 & 18.38&0.03 & 18.21&0.05\\
         & 02:33:51.7 & +00:29:48 & 18.71&0.02 & 17.88&0.02 & 16.96&0.02 & 16.62&0.02\\
         & 02:33:57.1 & +00:31:18 & 19.90&0.02 & 19.10&0.03 & 18.31&0.03 & 18.02&0.04\\
         & 02:33:53.3 & +00:28:18 & 19.71&0.02 & 18.79&0.03 & 17.74&0.02 & 17.33&0.03\\
         & 02:33:46.5 & +00:27:08 & 20.85&0.03 & 19.70&0.04 & 18.02&0.03 & 17.36&0.03\\
         & 02:33:36.6 & +00:28:39 & 20.02&0.03 & 19.29&0.03 & 18.56&0.03 & 18.30&0.06\\
         & 02:33:55.0 & +00:28:02 & 20.01&0.02 & 19.10&0.03 & 18.02&0.03 & 17.61&0.03\\
SN1999fi & 02:28:24.4 & +00:42:08 & 19.96&0.02 & 19.12&0.03 & 18.26&0.03 & 17.94&0.04\\
         & 02:28:17.3 & +00:39:06 & 21.20&0.04 & 20.02&0.05 & 18.36&0.03 & 17.70&0.04\\
         & 02:28:20.5 & +00:46:42 & 20.49&0.03 & 19.58&0.04 & 18.50&0.03 & 18.09&0.05\\
         & 02:28:10.9 & +00:37:53 & 21.38&0.05 & 20.33&0.07 & 18.90&0.04 & 18.33&0.06\\
         & 02:28:13.2 & +00:38:32 & 20.03&0.03 & 19.53&0.04 & 19.07&0.05 & 18.91&0.10\\
         & 02:28:01.1 & +00:42:07 & 21.64&0.06 & 20.58&0.08 & 19.14&0.05 & 18.57&0.07\\
         & 02:28:10.6 & +00:46:24 & 21.27&0.05 & 20.36&0.07 & 19.31&0.06 & 18.91&0.09\\
         & 02:28:20.3 & +00:39:40 & 21.72&0.06 & 20.74&0.10 & 19.46&0.06 & 18.96&0.10\\
SN1999fv & 23:30:39.0 & +00:19:24 & 19.77&0.03 & 19.36&0.03 & 18.89&0.04 & 18.73&0.08\\
         & 23:30:20.7 & +00:17:28 & 19.20&0.02 & 18.85&0.03 & 18.39&0.03 & 18.24&0.05\\
         & 23:30:45.7 & +00:17:16 & 20.62&0.03 & 19.69&0.04 & 18.53&0.03 & 18.07&0.05\\
         & 23:30:29.0 & +00:12:18 & 19.44&0.02 & 18.96&0.03 & 18.55&0.03 & 18.42&0.06\\
         & 23:30:33.9 & +00:16:19 & 21.86&0.07 & 20.82&0.10 & 19.31&0.06 & 18.70&0.08\\
         & 23:30:39.8 & +00:14:41 & 21.36&0.05 & 20.45&0.07 & 19.26&0.05 & 18.80&0.09\\
         & 23:30:24.3 & +00:18:58 & 22.01&0.07 & 20.86&0.11 & 19.09&0.05 & 18.38&0.06\\
         & 23:30:46.6 & +00:20:25 & 21.80&0.06 & 20.60&0.09 & 18.96&0.04 & 18.31&0.06\\
SN1999fh & 02:28:04.7 & +00:40:21 & 19.32&0.02 & 18.79&0.03 & 18.27&0.03 & 18.09&0.05\\
         & 02:28:01.6 & +00:44:19 & 22.64&0.09 & 21.11&0.14 & 19.23&0.05 & 18.47&0.06\\
         & 02:28:01.1 & +00:42:07 & 21.64&0.06 & 20.58&0.08 & 19.14&0.05 & 18.57&0.07\\
         & 02:27:58.6 & +00:35:55 & 21.92&0.07 & 20.82&0.10 & 19.21&0.05 & 18.57&0.07\\
         & 02:28:02.6 & +00:40:23 & 19.90&0.03 & 19.35&0.03 & 18.79&0.04 & 18.60&0.07\\
         & 02:28:05.7 & +00:41:12 & 22.54&0.11 & 21.38&0.17 & 19.65&0.07 & 18.96&0.10\\
         & 02:28:06.7 & +00:41:30 & 21.87&0.07 & 20.79&0.10 & 19.48&0.07 & 18.97&0.10\\
         & 02:27:58.6 & +00:35:46 & 21.62&0.06 & 20.76&0.10 & 19.78&0.08 & 19.41&0.15\\
SN1999fm & 02:30:31.2 & +01:07:49 & 19.15&0.02 & 18.27&0.02 & 17.29&0.02 & 16.92&0.03\\
         & 02:30:39.4 & +01:07:56 & 20.15&0.03 & 19.18&0.03 & 18.03&0.03 & 17.58&0.03\\
         & 02:30:41.5 & +01:08:03 & 21.02&0.03 & 19.90&0.05 & 18.44&0.03 & 17.86&0.04\\
         & 02:30:26.3 & +01:08:59 & 17.56&0.02 & 17.05&0.02 & 16.57&0.02 & 16.41&0.02\\
         & 02:30:25.6 & +01:09:21 & 19.62&0.02 & 18.48&0.02 & 16.80&0.02 & 16.14&0.02\\
         & 02:30:46.3 & +01:09:31 & 20.66&0.03 & 19.58&0.04 & 18.16&0.03 & 17.61&0.03\\
         & 02:30:24.2 & +01:10:57 & 20.93&0.03 & 19.89&0.05 & 18.47&0.03 & 17.91&0.04\\
         & 02:30:37.7 & +01:12:00 & 19.01&0.02 & 18.03&0.02 & 16.88&0.02 & 16.44&0.02\\
SN1999fo & 04:14:46.4 & +06:38:46 & 18.87&0.03 &\ldots&\ldots&17.40&0.03 & 17.28&0.03\\
         & 04:14:40.4 & +06:39:46 & 19.17&0.03 &\ldots&\ldots&17.74&0.03 & 17.61&0.03\\
         & 04:14:41.4 & +06:38:40 & 21.09&0.03 &\ldots&\ldots&18.26&0.03 & 17.91&0.03\\
         & 04:14:49.6 & +06:37:35 & 19.45&0.02 &\ldots&\ldots&18.30&0.03 & 18.25&0.03\\
         & 04:14:46.5 & +06:38:29 & 20.13&0.02 &\ldots&\ldots&18.81&0.03 & 18.74&0.03\\
         & 04:14:42.8 & +06:37:52 & 22.04&0.07 &\ldots&\ldots&19.14&0.03 & 18.77&0.03\\
         & 04:14:43.6 & +06:37:23 & 21.61&0.05 &\ldots&\ldots&19.08&0.03 & 18.81&0.03\\
         & 04:14:45.6 & +06:37:47 & 21.60&0.05 &\ldots&\ldots&19.12&0.03 & 18.86&0.03\\
SN1999fj & 02:28:13.9 & +00:39:05 & 19.80&0.02 & 18.77&0.03 & 17.46&0.02 & 16.94&0.03\\
         & 02:28:28.1 & +00:38:25 & 20.81&0.03 & 19.71&0.04 & 18.05&0.03 & 17.40&0.03\\
         & 02:28:17.3 & +00:39:06 & 21.20&0.04 & 20.02&0.05 & 18.36&0.03 & 17.70&0.04\\
         & 02:28:26.5 & +00:40:53 & 20.86&0.03 & 19.78&0.04 & 18.33&0.03 & 17.76&0.04\\
         & 02:28:19.0 & +00:36:59 & 21.83&0.06 & 20.69&0.09 & 19.05&0.05 & 18.41&0.06\\
         & 02:28:31.7 & +00:37:01 & 20.79&0.03 & 19.89&0.05 & 18.89&0.04 & 18.51&0.07\\
         & 02:28:15.7 & +00:40:27 & 23.91&0.30 & 22.51&0.48 & 19.86&0.09 & 18.78&0.08\\
         & 02:28:20.3 & +00:39:40 & 21.72&0.06 & 20.74&0.10 & 19.46&0.06 & 18.96&0.10\\
SN1999fw & 23:31:36.2 & +00:15:10 & 21.19&0.04 & 20.00&0.05 & 18.24&0.03 & 17.54&0.03\\
         & 23:31:37.3 & +00:10:14 & 18.92&0.02 & 18.50&0.02 & 18.09&0.03 & 17.95&0.04\\
         & 23:31:56.4 & +00:15:47 & 21.19&0.04 & 20.24&0.06 & 19.18&0.05 & 18.77&0.08\\
         & 23:32:01.7 & +00:13:41 & 21.61&0.05 & 20.55&0.08 & 18.95&0.04 & 18.31&0.06\\
         & 23:31:58.0 & +00:14:12 & 20.69&0.03 & 19.67&0.04 & 18.27&0.03 & 17.72&0.04\\
         & 23:31:58.3 & +00:13:09 & 20.57&0.03 & 19.51&0.04 & 18.21&0.03 & 17.71&0.04\\
         & 23:31:52.6 & +00:09:32 & 20.22&0.03 & 19.56&0.04 & 18.89&0.04 & 18.66&0.08\\
         & 23:31:59.9 & +00:11:08 & 20.79&0.03 & 19.66&0.04 & 18.07&0.03 & 17.45&0.03\\
SN1999fn & 04:14:09.2 & +04:19:06 & 18.35&0.03 & 17.79&0.03 & 17.19&0.03 & 17.19&0.03\\
         & 04:13:59.9 & +04:18:46 & 19.47&0.03 & 18.24&0.03 & 16.74&0.03 & 16.44&0.03\\
         & 04:14:07.8 & +04:19:15 & 19.40&0.03 & 18.24&0.03 & 17.14&0.03 & 16.91&0.03\\
         & 04:13:59.7 & +04:19:08 & 19.90&0.05 & 18.68&0.03 & 17.42&0.03 & 17.11&0.03\\
         & 04:14:07.9 & +04:17:58 & 19.92&0.02 & 18.75&0.03 & 17.60&0.03 & 17.37&0.03\\
         & 04:14:05.6 & +04:17:40 & 20.67&0.03 & 19.41&0.03 & 18.08&0.03 & 17.79&0.03\\
         & 04:13:58.6 & +04:16:37 & 20.81&0.04 & 19.63&0.03 & 18.20&0.03 & 17.88&0.03\\
         & 04:14:02.0 & +04:17:41 & 21.22&0.05 & 19.94&0.03 & 18.51&0.03 & 18.20&0.03\\
SN1999fk & 02:28:48.2 & +01:16:17 & 19.33&0.02 & 18.43&0.02 & 17.43&0.02 & 17.05&0.03\\
         & 02:28:45.6 & +01:15:39 & 19.96&0.02 & 18.90&0.03 & 17.47&0.02 & 16.91&0.03\\
         & 02:29:01.1 & +01:13:42 & 21.89&0.06 & 20.57&0.08 & 18.59&0.03 & 17.80&0.04\\
         & 02:28:54.6 & +01:14:43 & 20.79&0.03 & 19.86&0.05 & 18.81&0.04 & 18.41&0.06\\
         & 02:29:05.3 & +01:13:46 & 21.50&0.05 & 20.46&0.08 & 18.95&0.04 & 18.35&0.06\\
         & 02:28:42.9 & +01:15:31 & 20.21&0.03 & 19.92&0.05 & 19.45&0.06 & 19.29&0.13\\
         & 02:29:02.1 & +01:13:14 & 22.63&0.12 & 21.49&0.19 & 19.92&0.10 & 19.30&0.13\\
         & 02:28:42.3 & +01:15:16 & 20.69&0.04 & 20.32&0.07 & 19.95&0.10 & 19.84&0.22\\
\enddata
\end{deluxetable}

All our supernova images were compared with the above stellar
sequences to derive zero-points.  The star fluxes were obtained using
the Vista ``psf'' routine, which sums up the light within a radius of
20 pixels around a star, subtracts a fitted sky value, and rejects
contamination from neighboring objects.  Comparison of the fluxes and
magnitudes, with due regard for saturation and magnitude errors, gave
us a flux-magnitude zero-point and error for most images.  For \hst\
images, we used the precepts of Dolphin (2000) to convert fluxes
to magnitudes. 

\subsection{Photometry of \sna}

Our reductions of deep supernova observations obtained after the
candidates were identified are very similar to the reductions carried
out during the search.  After bias subtraction, flatfielding,
$I$-band and $Z$-band fringe correction, masking bad
columns, and combination of dithered images while rejecting cosmic rays
and moving objects, we identified stars relative to a fiducial image
and performed an astrometric solution.  We remapped a $1024 \times 1024$ pixel
subarray of the source image onto a tangent plane coordinate system
with 0\farcs2 pixels.  The \hst\ observations had a small enough field
of view and large enough distortion that some amount of star selection
by hand was necessary to obtain a satisfactory astrometric match.

We had expected to use \hst\  to acquire very high-quality light curves
for a subset of our objects, but the gyro failure in Fall 1999 made
\hst\  unavailable until after this group of \sna\ had disappeared.  We did
succeed in using \hst\  to get template observations of the host galaxies
after the supernovae had faded for most of our targets.  Late-time
templates are very helpful to subtract the host galaxy accurately and
to set the \sna\ flux zero-point.

Most modern supernova light curves are derived by subtraction of a
template image taken before the supernova explosion or long after, so
that it carries no supernova flux.  While this sets a good flux zero-point 
for the subtraction, it also inflicts a common, systematic flux
error on the light curve which can be substantial if the template
observation has poor S/N or very different seeing.

To avoid these correlated errors, we subtracted all $N(N-1)/2$ pairs of
observations from one another.  We carried this out using the Alard 
\& Lupton (1998) code
mentioned previously.  This procedure produces negative fluxes for
some subtractions, but it has the advantage that {\em every}
observation serves as a template for the entire light curve.  The
details of this procedure are in the Appendix.  We have found that
this procedure reduces the errors by a factor of about $\sqrt 2$
(Novicki \& Tonry 2000).

There are two unresolved questions after the $N(N-1)/2$ procedure has
been followed.  We do not know a flux zero-point, since it has been
removed when we difference images, and we need a flux-to-magnitude
conversion which incorporates the independent magnitude zero-points
established for each image.

The flux zero-point is generally established by late-time observations
when there is no remaining supernova flux.  This is an improvement
over the usual ``template'' method, because the flux zero-point which
comes out of the $N(N-1)/2$ procedure for the late-time point comes from
the average comparison with all the other observations, and the
relative fluxes of all the other observations are not dependent solely
on the final observation.  In some cases where we had several
late-time points without supernova flux, we adjusted the flux zero-point 
to be an average of those points, or to include a first point
that was obtained prior to the supernova explosion.

Infrared template images were available only for SN~1999fm and
SN~1999fk, so we used standard PSF photometry with the DaoPhot
II/Allstar package (Stetson 1987; Stetson \& Harris 1988) to measure
the $J$-band magnitudes of the SN corrected to the same aperture used
for the standard stars.  SN~1999fm was not visible in the $J$ band.
SN~1999fk is near the edge of its host so contamination from the
galaxy is negligible.  SN~1999fn and SN~1999fv are embedded in their
hosts, so contamination may be significant.  SN~1999ff is located in a
bright ($J\approx17.5$ mag) elliptical galaxy for which it was essential
to subtract the host.  We used the {\sc iraf/stsdas} analysis isophote
tasks to construct a model of the host, subtracted our model, and then
performed PSF photometry on the supernova.  We were unable to define a
PSF for the Keck observations of SN~1999fn due to the lack of
suitable stars in the Keck/NIRC field of view, so we performed
aperture photometry in an aperture with a radius of one arcsecond.

Tables \ref{dist99fw}--\ref{dist99fv} give the derived supernova light
curves, and Figure~\ref{mlcslc} shows the data points and typical
fitted light curves.


\begin{deluxetable}{lrcrcl}
\tablewidth{0pt}   
\tablecaption{Observations of SN~1999fw\label{dist99fw}}    
\tablehead{ \colhead{MJD} & \colhead{$m$} & \colhead{$\pm$} & \colhead{$K$} &
\colhead{$\pm$} & \colhead{$obs$}} \startdata
         &     $V$  &            &  $K_{V\rightarrow B}$  &  &  \\
51496.25 &    22.18 &     0.15   & $-0.34$ &    0.04 & 991114\_uh\\
51514.20 &    23.84 &     0.86   & $-0.24$ &    0.03 & 991202\_uh   \\
51529.21 & $>$24.50 &   $\ldots$ & $-0.21$ &    0.04 & 991217\_uh   \\
51875.25 & $>$24.50 &   $\ldots$ & $-0.26$ &    0.05 & 001127\_esi  \\
         &     $R$  &            &  $K_{R\rightarrow V}$  &  &  \\
51455.27 & $>$24.50 &  $\ldots$  & $-0.20$ &   0.15  & 991004\_12k  \\
51488.15 &    21.30 &     0.01   & $-0.42$ &   0.04  & 991106\_8k   \\
51496.22 &    21.43 &     0.02   & $-0.30$ &   0.06  & 991114\_uh   \\
51514.26 &    22.40 &     0.13   & $-0.15$ &   0.06  & 991202\_uh   \\
51517.21 &    22.66 &     0.09   & $-0.14$ &   0.07  & 991205\_lris \\
51529.28 &    23.54 &     0.21   & $-0.09$ &   0.07  & 991217\_uh   \\
51875.26 & $>$24.50 &  $\ldots$  & $-0.43$ &   0.09  & 001127\_esi  \\
         &     $I$  &            &  $K_{I\rightarrow R}$  &  &  \\
51454.28 & $>$24.50 &  $\ldots$  & $-0.53$ &   0.05  & 991003\_12k  \\
51484.34 &    21.16 &     0.03   & $-0.54$ &   0.01  & 991102\_12k  \\
51485.33 &    21.14 &     0.04   & $-0.54$ &   0.01  & 991103\_12k  \\
51496.23 &    21.34 &     0.12   & $-0.52$ &   0.03  & 991114\_uh   \\
51512.31 &    21.98 &     0.07   & $-0.56$ &   0.04  & 991130\_uh   \\
51517.21 &    22.03 &     0.03   & $-0.57$ &   0.04  & 991205\_lris \\
51875.28 & $>$24.50 &  $\ldots$  & $-0.55$ &   0.10  & 001127\_esi  \\
         &     $Z$  &            &  $K_{Z\rightarrow I}$  &  &  \\
51496.28 & $>$24.00 &  $\ldots$  & $-0.11$& $0.07$& 991114\_uh   \\
51517.20 &    21.92 &     0.48   & $-0.08$& $0.15$& 991205\_lris \\
51527.22 & $>$24.50 &  $\ldots$  & $ 0.00$& $0.16$& 991216\_ufti \\
\enddata \end{deluxetable}

\begin{deluxetable}{lrcrcl}
\tablewidth{0pt}   
\tablecaption{Observations of SN~1999fh\label{dist99fh}}   
\tablehead{ \colhead{MJD} & \colhead{$m$} & \colhead{$\pm$} & \colhead{$K$} &
\colhead{$\pm$} & \colhead{$obs$}} \startdata
         &     $R$  &            &  $K_{R\rightarrow B}$  &  &  \\
51455.47 & $>$24.50 &  $\ldots$  & $-0.85$ &   0.12 & 991004\_12k  \\
51488.25 &    22.74 &     0.02   & $-0.96$ &   0.07 & 991105\_8k   \\
51496.51 &    23.02 &     0.10   & $-1.02$ &   0.11 & 991114\_uh   \\
51528.23 & $>$24.50 &  $\ldots$  & $-1.44$ &   0.10 & 991216\_uh   \\
51570.29 & $>$24.50 &  $\ldots$  & $-1.32$ &   0.07 & 000127\_lris \\
51875.30 & $>$24.50 &  $\ldots$  & $-1.28$ &   0.20 & 001127\_esi  \\
         &     $I$  &            &  $K_{I\rightarrow V}$  &  &  \\
51454.51 & $>$24.50 &  $\ldots$  & $-0.88$ &   0.07 & 991003\_12k  \\
51485.36 &    22.40 &     0.07   & $-0.92$ &   0.02 & 991103\_12k  \\
51496.46 &    22.41 &     0.07   & $-0.91$ &   0.05 & 991114\_uh   \\
51570.27 & $>$24.50 &  $\ldots$  & $-0.94$ &   0.07 & 000127\_lris \\
51875.29 & $>$24.50 &  $\ldots$  & $-0.95$ &   0.10 & 001127\_esi  \\
\enddata \end{deluxetable}

\begin{deluxetable}{lrcrcl}
\tablewidth{0pt}   
\tablecaption{Observations of SN~1999ff\label{dist99ff}}   
\tablehead{ \colhead{MJD} & \colhead{$m$} & \colhead{$\pm$} & \colhead{$K$} &
\colhead{$\pm$} & \colhead{$obs$}} \startdata
         &     $R$  &            &  $K_{R\rightarrow B}$  &  &  \\
51455.41 & $>$24.50 &  $\ldots$  & $-0.68$ &  0.05   & 991004\_12k  \\
51496.26 &    22.66 &     0.04   & $-0.70$ &  0.02   & 991114\_8k   \\
51496.35 &    22.72 &     0.08   & $-0.70$ &  0.02   & 991114\_uh   \\
51497.48 &    22.58 &     0.03   & $-0.70$ &  0.02   & 991115\_uh   \\
51517.34 &    24.16 &     0.22   & $-0.77$ &  0.03   & 991205\_lris \\
51529.35 &    24.39 &     0.37   & $-0.82$ &  0.03   & 991217\_uh   \\
51530.28 &    24.83 &     0.52   & $-0.82$ &  0.03   & 991218\_uh   \\
51876.26 & $>$24.50 &  $\ldots$  & $-0.78$ &  0.10   & 001128\_esi  \\
         &     $I$  &            &  $K_{I\rightarrow V}$  &  &  \\
51454.49 & $>$24.50 &  $\ldots$  & $-0.82$ &  0.05   & 991003\_12k  \\
51484.46 &    22.86 &     0.07   & $-0.84$ &  0.00   & 991102\_12k  \\
51496.28 &    22.49 &     0.09   & $-0.84$ &  0.01   & 991114\_8k   \\
51496.36 &    22.37 &     0.12   & $-0.84$ &  0.01   & 991114\_uh   \\
51497.47 &    22.27 &     0.12   & $-0.84$ &  0.01   & 991115\_uh   \\
51517.32 &    23.26 &     0.09   & $-0.84$ &  0.01   & 991205\_lris \\
51528.30 &    23.59 &     0.20   & $-0.85$ &  0.01   & 991216\_uh   \\
51876.27 & $>$24.50 &  $\ldots$  & $-0.86$ &  0.05   & 001128\_esi  \\
         &     $Z$  &            &  $K_{Z\rightarrow R}$  &  &  \\
51496.37 &    22.85 &     0.18   & $-0.79$ &  0.03   & 991114\_uh   \\
51517.37 &    22.65 &     0.14   & $-0.85$ &  0.07   & 991205\_lris \\
51527.30 &    23.51 &     1.19   & $-0.91$ &  0.06   & 991216\_ufti \\
51873.22 &    23.67 &     1.56   & $-0.94$ &  0.10   & 001125\_wiyn \\
         &     $J$  &            &  $K_{J\rightarrow I}$  &  &  \\
51501.29 &    22.61 &     0.10   &$-0.76 $ &  0.03   & 991119\_nirc \\
51526.31 &    23.03 &     0.23   &$-0.79 $ &  0.06   & 991214\_nirc \\
\enddata \end{deluxetable}

\begin{deluxetable}{lrcrcl}
\tablewidth{0pt}   
\tablecaption{Observations of SN~1999fn\label{dist99fn}}   
\tablehead{ \colhead{MJD} & \colhead{$m$} & \colhead{$\pm$} & \colhead{$K$} &
\colhead{$\pm$} & \colhead{$obs$}} \startdata
         &     $R$  &            &  $K_{R\rightarrow B}$  &  &  \\
51459.59 & $>$24.50 &  $\ldots$  & $-0.75$ &   0.05  & 991008\_12k  \\
51494.30 &    22.93 &     0.04   & $-0.69$ &   0.01  & 991112\_8k   \\
51496.58 &    22.78 &     0.06   & $-0.69$ &   0.01  & 991114\_uh   \\
51497.50 &    22.71 &     0.04   & $-0.69$ &   0.01  & 991115\_uh   \\
51512.29 &    23.17 &     0.08   & $-0.72$ &   0.02  & 991130\_8k   \\
51517.53 &    23.32 &     0.12   & $-0.73$ &   0.02  & 991205\_lris \\
51518.27 &    23.49 &     0.08   & $-0.73$ &   0.02  & 991206\_8k   \\
51529.44 &    24.28 &     0.06   & $-0.75$ &   0.02  & 991217\_uh   \\
51530.32 &    24.19 &     0.03   & $-0.75$ &   0.02  & 991218\_uh   \\
51545.30 &    24.92 &     0.27   & $-0.77$ &   0.02  & 000102\_8k   \\
51570.36 &    25.63 &     0.34   & $-0.76$ &   0.02  & 000127\_lris \\
51581.10 & $>$24.50 &  $\ldots$  & $-0.74$ &   0.01  & 000207\_fors \\
51603.02 & $>$24.50 &  $\ldots$  & $-0.73$ &   0.02  & 000228\_fors \\
51637.33 & $>$24.50 &  $\ldots$  & $-0.73$ &   0.02  & 000403\_hst  \\
         &     $I$  &            &  $K_{I\rightarrow V}$  &  &  \\
51459.62 & $>$24.50 &  $\ldots$  & $-0.85$ &    0.02 & 991008\_12k  \\
51485.54 &    23.26 &     0.09   & $-0.86$ &    0.01 & 991103\_12k  \\
51494.32 &    22.52 &     0.12   & $-0.87$ &    0.01 & 991112\_8k   \\
51496.59 &    22.31 &     0.10   & $-0.87$ &    0.01 & 991114\_uh   \\
51497.51 &    22.37 &     0.05   & $-0.87$ &    0.01 & 991115\_uh   \\
51501.34 &    22.59 &     0.12   & $-0.87$ &    0.01 & 991119\_vatt \\
51512.32 &    22.55 &     0.15   & $-0.85$ &    0.01 & 991130\_8k   \\
51512.49 &    22.79 &     0.07   & $-0.85$ &    0.01 & 991130\_uh   \\
51517.47 &    23.03 &     0.08   & $-0.85$ &    0.01 & 991205\_lris \\
51518.28 &    22.97 &     0.12   & $-0.85$ &    0.01 & 991206\_8k   \\
51528.41 &    23.34 &     0.09   & $-0.86$ &    0.01 & 991216\_uh   \\
51570.34 & $>$24.50 &  $\ldots$  & $-0.88$ &    0.01 & 000127\_lris \\
51579.09 & $>$24.50 &  $\ldots$  & $-0.87$ &    0.01 & 000205\_fors \\
51637.40 & $>$24.50 &  $\ldots$  & $-0.86$ &    0.02 & 000403\_hst  \\
         &     $Z$  &            &  $K_{Z\rightarrow R}$  &  &  \\
51496.61 &    22.30 &     0.56   & $-0.89$ &    0.03 & 991114\_uh   \\
51497.52 &    22.02 &     0.26   & $-0.88$ &    0.03 & 991115\_uh   \\
51512.54 &    22.47 &     0.29   & $-0.83$ &    0.08 & 991130\_uh   \\
51517.49 &    22.41 &     0.05   & $-0.86$ &    0.09 & 991205\_lris \\
51518.33 &    22.75 &     0.36   & $-0.87$ &    0.09 & 991206\_8k   \\
51528.42 &    23.38 &     0.27   & $-0.97$ &    0.08 & 991217\_ufti \\
51530.37 &    23.31 &     0.14   & $-0.98$ &    0.08 & 991218\_uh   \\
51570.24 &    23.32 &     0.62   & $-1.08$ &    0.13 & 000127\_lris \\
51637.47 & $>$24.50 &  $\ldots$  & $-0.98$ &    0.10 & 000403\_hst  \\
51873.35 & $>$24.50 &  $\ldots$  & $-0.99$ &    0.20 & 001125\_wiyn \\
         &     $J$  &            &  $K_{J\rightarrow I}$  &  &  \\
51498.52 &    22.30 &     0.20   & $-0.77$ & 0.02    & 991116\_ufti \\
51501.36 &    22.22 &     0.05   & $-0.76$ & 0.03    & 991119\_nirc \\
51526.45 &    22.86 &     0.21   & $-0.79$ & 0.06    & 991214\_nirc \\
51530.30 &    22.53 &     0.39   & $-0.79$ & 0.07    & 991218\_ufti \\
\enddata \end{deluxetable}

\begin{deluxetable}{lrcrcl}
\tablewidth{0pt}   
\tablecaption{Observations of SN~1999fj\label{dist99fj}}   
\tablehead{ \colhead{MJD} & \colhead{$m$} & \colhead{$\pm$} & \colhead{$K$} &
\colhead{$\pm$} & \colhead{$obs$}} \startdata
         &     $I$  &            &  $K_{I\rightarrow B}$  &  &  \\
51454.51 &    24.90 &     0.13   & $-1.02$ &    0.10 & 991003\_12k \\
51485.36 &    23.14 &     0.03   & $-1.20$ &    0.02 & 991103\_12k \\
51493.22 &    23.28 &     0.05   & $-1.18$ &    0.03 & 991111\_fors\\
51496.41 &    23.70 &     0.13   & $-1.17$ &    0.03 & 991114\_uh  \\
51497.12 &    23.53 &     0.06   & $-1.17$ &    0.03 & 991115\_fors\\
51497.42 &    23.60 &     0.15   & $-1.17$ &    0.03 & 991115\_uh  \\
51512.40 &    24.45 &     0.23   & $-1.18$ &    0.02 & 991130\_uh  \\
51518.06 &    24.93 &     0.40   & $-1.20$ &    0.02 & 991206\_8k  \\
51527.34 &    25.11 &     0.50   & $-1.20$ &    0.02 & 991215\_lris\\
51538.31 & $>$24.50 &  $\ldots$  & $-1.20$ &    0.03 & 991226\_lris\\
51808.58 & $>$24.50 &  $\ldots$  & $-1.23$ &    0.03 & 000921\_hst \\
         &     $Z$  &            &  $K_{Z\rightarrow V}$  &  &  \\
51493.24 &    23.37 &     0.18   & $-1.07$ &    0.06 & 991111\_fors\\
51497.14 &    23.36 &     0.15   & $-1.01$ &    0.07 & 991115\_fors\\
51498.42 &    23.91 &     0.29   & $-0.99$ &    0.08 & 991116\_ufti\\
51526.07 &    24.33 &     0.32   & $-0.85$ &    0.10 & 991214\_fors\\
51808.64 & $>$24.50 &  $\ldots$  & $-0.98$ &    0.25 & 000921\_hst \\
\enddata \end{deluxetable}

\begin{deluxetable}{lrcrcl}
\tablewidth{0pt}   
\tablecaption{Observations of SN~1999fm\label{dist99fm}}   
\tablehead{ \colhead{MJD} & \colhead{$m$} & \colhead{$\pm$} & \colhead{$K$} &
\colhead{$\pm$} & \colhead{$obs$}} \startdata
         &     $I$  &            &  $K_{I\rightarrow B}$  &  &  \\
51459.43 &    24.44 &     0.02   & $-1.35$ &    0.13 & 991008\_12k  \\
51485.47 &    23.05 &     0.11   & $-1.16$ &    0.06 & 991103\_12k  \\
51494.17 &    23.20 &     0.06   & $-1.16$ &    0.09 & 991112\_fors \\
51496.40 &    23.22 &     0.05   & $-1.15$ &    0.10 & 991114\_lris \\
51497.17 &    23.25 &     0.07   & $-1.15$ &    0.10 & 991115\_fors \\
51512.34 &    23.83 &     0.15   & $-1.07$ &    0.10 & 991130\_uh   \\
51517.26 &    24.32 &     0.18   & $-1.02$ &    0.09 & 991205\_lris \\
51528.19 &    25.00 &     0.46   & $-0.93$ &    0.07 & 991215\_lris \\
51527.29 &    24.54 &     0.06   & $-0.93$ &    0.07 & 991215\_fors \\
51538.22 &    25.19 &     0.24   & $-0.89$ &    0.07 & 991226\_lris \\
51936.22 & $>$24.50 &  $\ldots$  & $-0.93$ &    0.07 & 010127\_lris \\
         &     $Z$  &            &  $K_{Z\rightarrow V}$  &  &  \\
51494.19 &    23.13 &     0.24   & $-1.16$ &    0.13 & 991112\_fors \\
51496.37 &    23.14 &     0.11   & $-1.11$ &    0.14 & 991114\_lris \\
51497.19 &    23.30 &     0.18   & $-1.09$ &    0.15 & 991115\_fors \\
51528.02 &    24.68 &     0.26   & $-0.59$ &    0.17 & 991216\_fors \\
51529.24 &    24.14 &     0.32   & $-0.58$ &    0.18 & 991218\_ufti \\
51936.24 & $>$24.50 &  $\ldots$  & $-0.75$ &    0.30 & 010127\_lris \\
         &     $J$  &            &  $K_{J\rightarrow R}$  &  &  \\
51500.33 &    22.48 &     0.18   & $-1.40$ &    0.06 & 991118\_nirc \\
51526.28 &    23.27 &     0.18   & $-1.49$ &    0.09 & 991214\_nirc \\
\enddata \end{deluxetable}

\begin{deluxetable}{lrcrcl}
\tablewidth{0pt}   
\tablecaption{Observations of SN~1999fk\label{dist99fk}}   
\tablehead{ \colhead{MJD} & \colhead{$m$} & \colhead{$\pm$} & \colhead{$K$} &
\colhead{$\pm$} & \colhead{$obs$}} \startdata
         &     $I$  &            &  $K_{I\rightarrow U}$  &  &  \\
51459.43 & $>$24.50 &  $\ldots$  & $-0.62$ &    0.14 & 991008\_12k  \\
51485.47 &    23.96 &     0.13   & $-0.63$ &    0.09 & 991103\_12k  \\
51493.19 &    23.79 &     0.12   & $-0.70$ &    0.07 & 991111\_fors \\
51496.41 &    23.45 &     0.11   & $-0.73$ &    0.08 & 991114\_lris \\
51498.37 &    23.77 &     0.23   & $-0.76$ &    0.08 & 991116\_uh   \\
51512.45 &    23.69 &     0.17   & $-0.89$ &    0.07 & 991130\_uh   \\
51517.40 &    24.02 &     0.52   & $-0.92$ &    0.07 & 991205\_lris \\
51527.18 &    25.04 &     0.67   & $-0.97$ &    0.06 & 991214\_fors \\
51527.39 &    24.58 &     0.29   & $-0.97$ &    0.06 & 991215\_lris \\
51804.76 & $>$24.50 &  $\ldots$  & $-1.06$ &    0.10 & 000917\_hst  \\
51964.21 & $>$24.50 &  $\ldots$  & $-1.06$ &    0.15 & 010224\_hst  \\
         &     $Z$  &            &  $K_{Z\rightarrow B}$  &  &  \\
51496.42 &    23.07 &     0.19   & $-1.42$ &    0.02 & 991114\_lris \\
51499.38 &    23.88 &     0.67   & $-1.42$ &    0.01 & 991117\_ufti \\
51501.20 & $>$24.50 &  $\ldots$  & $-1.42$ &    0.00 & 991119\_isaac\\
51527.02 &    24.21 &     0.55   & $-1.34$ &    0.03 & 991215\_fors \\
51804.82 & $>$24.50 &  $\ldots$  & $-1.31$ &    0.03 & 000917\_hst  \\
51964.35 & $>$24.50 &  $\ldots$  & $-1.31$ &    0.10 & 010224\_hst  \\
         &     $J$  &            &  $K_{J\rightarrow V}$  &  &  \\
51501.16 &    23.26 &     0.09   & $-1.49$ &    0.04 & 991118\_isaac\\
\enddata \end{deluxetable}

\begin{deluxetable}{lrcrcl}
\tablewidth{0pt}   
\tablecaption{Observations of SN~1999fv\label{dist99fv}}   
\tablehead{ \colhead{MJD} & \colhead{$m$} & \colhead{$\pm$} & \colhead{$K$} &
\colhead{$\pm$} & \colhead{$obs$}} \startdata
         &     $I$  &            &  $K_{I\rightarrow U}$  &  &  \\
51454.34 & $>$24.50 &  $\ldots$  & $-0.61$ &    0.13 & 991003\_12k   \\
51485.20 &    24.17 &     0.09   & $-0.59$ &    0.02 & 991103\_12k   \\
         &     $J$  &            &  $K_{J\rightarrow V}$  &  &  \\
51502.60 &    23.61 &     0.20   & $-1.59$ &    0.05 & 991120\_isaac \\
51503.53 &    23.06 &     0.15   & $-1.58$ &    0.05 & 991121\_isaac \\
\enddata \end{deluxetable}

\section{\sna\ Analysis}

\subsection{K-Corrections}
K-corrections were calculated using the formulas described by Kim,
Goobar, \& Perlmutter (1996) and Schmidt et al. (1998). The apparent
brightness of a supernova observed in filter $j$, $m_j(t)$, but with a $z=0$
absolute magnitude light curve in filter $i$, $M_i(t)$, is given by

$$m_j(t) = M_i((1+z)t^\prime)+25 + 5\log\left( {D_L} \over {\rm Mpc} \right) + K_{ij}((1+z)t^\prime),$$

\noindent
where $K_{ij}$ is given by 

\begin{equation}K_{ij} = 2.5\log\left[(1+z) {\int F_\lambda(\lambda) S_i(\lambda)d\lambda \over
\int F_\lambda(\lambda/(1+z))S_j(\lambda)d\lambda}\right] + {\cal Z}_j - {\cal Z}_i, \label{eq:newKcorr}
\end{equation}
for filter energy sensitivity functions $S_i$ and $S_j$, with zero-points 
${\cal Z}_i$ and ${\cal Z}_j$, and supernova spectrum $F_\lambda$.
The luminosity distance, $D_L$, depends on the cosmological
parameters, and is discussed extensively by Schmidt \etal\ (1998).

For multi-filter light curve shape (MLCS) fitting, 
K-corrections were calculated in the same manner as
for Riess \etal\ (1998a), as part of the fitting process, using a
prescription similar to that of Nugent \etal\ (2002).  For
K-corrections presented in Tables \ref{dist99fw}--\ref{dist99fv}, we
used a set of 135 \sna\ spectra ranging from 14~d before maximum
light to 92~d after maximum light. Extending the work of Nugent
\etal\ (2002), before applying the K-correction formulae, the \sna\
spectra of each epoch were matched to the de-reddened $U,B,V,R,I$
color curves of a well-observed set of SN templates, by warping the
spectral energy response with a spline containing knots at the
effective wavelengths of each filter. In our case we used the
compilation of Jha (2002) for SN~1991T, 1994D, 1996X, 1997bp, 1997bq,
1997br, 1998ab, 1998aq, 1998bu, 1998es, 1998V, 1999aa, 1999ac, 1999by,
1999dq, 1999gp. The spectra were corrected for the estimated host
extinction in the rest frame, as well as for Galactic extinction in
the observer's frame using the Schlegel et al. (1998) values and the
Savage \& Mathis (1979) reddening law.  The modified
spectra were then used to calculate the K-corrections, providing a
series of K-correction estimates as a function of SN age, and of SN
template type. These values were fitted with a 3-knot spline, and
uncertainties estimated from the scatter of allowable templates as
dictated by the light-curve fits (in this case, the modified dm15
method of Germany \etal\ 2001).

Our K-corrections were used to
fit the light curve and update the time of maximum light, as well as
the allowable supernova template type and host extinction, and the whole 
process was
iterated until convergence (only two iterations were
required). Tables \ref{dist99fw}--\ref{dist99fv} list the derived
K-corrections for each observation along with their uncertainties.

\subsection{Light-Curve Fits using MLCS}

\begin{figure}[t]
\plotone{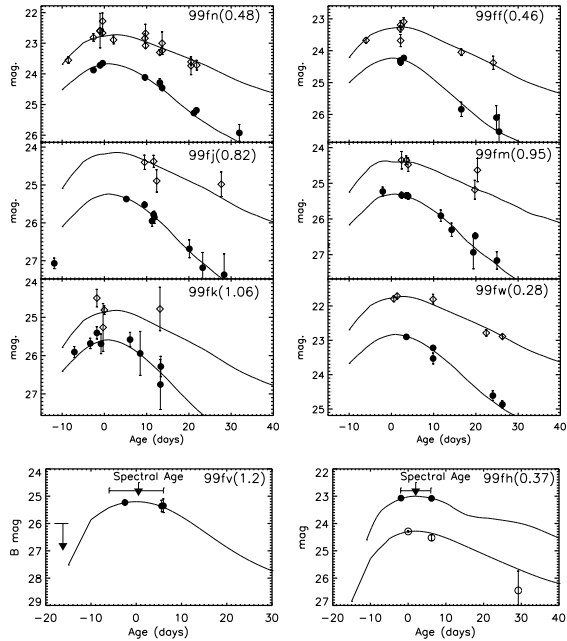}
\caption{MLCS fits to our eight
\sna. The open diamonds represent K-corrected $V$-band fits, while the
filled-in circles represent K-corrected $B$-band fits, offset by +1 mag,
except for SN~1999fh, which is $V+1$ and $R$. SN 1999fo, 1999fi, and
1999fu have no light curves because we could not establish from the
spectra that these were indeed \sna, so they were not followed.
Redshifts are indicated in parentheses.
\label{mlcslc}}
\end{figure}

We fit the photometry of the eight \sna\ shown in Figure~\ref{mlcslc}
using either the MLCS algorithm
as described by Riess et al. (1998a) or Riess et al. (2001), depending
on the quality of the data.

For \sna\ whose light curve constrains the time of maximum well and
whose colors provide sufficient leverage on potential reddening, we
fit the photometry using the MLCS algorithm as described by Riess et
al. (1998a).  This method employs the observed correlation between
light-curve shape, luminosity, and colors of \sna\ in the Hubble
flow, leading to significant improvements in the precision of distance
estimates derived from \sna\ (Hamuy et al. 1995, 1996b; Riess, Press,
\& Kirshner 1995, 1996a).

The empirical model for a \sna\ light and color curve is described by
four parameters: a date of maximum ($t$), a luminosity difference (at maximum, 
$\Delta$), an apparent distance modulus ($\mu_B$), and an extinction
($A_B$). Due to the redshifts of the supernova host galaxies we first correct
the supernova light curves for Galactic extinction (Schlegel et al. 1998),
and then determined the host-galaxy extinction.

Although the best extinction estimates and most precise distances are
found from \sna\ light curves which have extended temporal and
spectral coverage, ideally to $\sim$100~d after maximum
and 4 or 5 bandpasses (Jha 2002), we only use data within 40~d of
maximum light and in rest-frame $B$ and $V$ to match the data which
can be sampled at high redshift.  This should only compromise our
accuracy slightly, however.

For SN~1999fv and SN~1999fh, failure of the \hst\ gyros led to data
with less than optimal coverage in time.  For these two objects we
used the snapshot distance method (Riess et al. 1998b) which employs
the observed \sna\ spectrum to constrain the age.  This approach is
significantly less precise than the distance from a well-sampled light
curve. 

\subsection{Light Curve Fits using $\Delta m_{15}$($B$) and dm15}

We also use the $\Delta$m$_{15}$ method of Phillips et al. (1999) to
estimate a distance to each of our \sna.  Originally, the method
characterized the light curve by a parameter tied to the $B$-band
photometry: the decline in magnitudes for a SN Ia in the first 15 days
after $B$-band maximum.  With the accumulation of a larger number of
well-sampled light curves, it became possible to expand the method so
that $\Delta$m$_{15}$($B$) is determined from the $B$, $V$, and
$I$-band light curves.  The Phillips et al. method then compares the
light curve of a new object with a small number (six) of template
light curves.
\footnote[17]{The six templates are based on the light curves of seven
objects: SN 1991T, 1991bg, 1992A, 1992al, 1992bc, and 1992bo+1993H.
The method now compares the light curve of a new object with 6
template light curves (Hamuy et al. 1996b).}

In contrast to the MLCS method of Riess et al. (1996, 1998a), the
$\Delta$m$_{15}$ method does not produce a {\em continuum} of ``synthetic''
templates.  There is a small number of {\em discrete} templates.  The
reddening of a \sna\ is determined from the $B-V$ and $V-I$
colors at maximum, and from the $B-V$ colors between 30 and 90 days after
$V$-band maximum.  It is assumed that unreddened \sna\ of all
decline rates show uniform late-time color curves, as found by Lira
(1995).  Jha (2002, \S 3) found that the ``Lira law'' holds well for
the 44 \sna\ studied by him.  Only exceptional cases like SN~2000cx
seem to deviate greatly from the relation (Li et al. 2001; Candia et
al. 2003).

Distances using the modified dm15 method are described in detail by
Germany (2001). In summary, well-observed \sna\ (approximately 15
objects) are fitted with a spline in $BVRI$ and assigned a value of
$\Delta$m$_{15}$($B$) based on the method of Phillips (1993) and
Phillips et al. (1999). These templates are extended to explosion
times using the rise-time measurements of Riess et al. (1999b).  \sna\
at $cz > 4000$ km s$^{-1}$ with extinction values from Phillips et al. (1999)
are fitted with these templates, with the best-fitting template used to
assign these objects a dm15 value and maximum-light values $B_{max}$,
$V_{max}$, $R_{max}$, and $I_{max}$.  The Hubble velocity then
provides an absolute magnitude in each band, from which $M_{B,V,R,I}$
{\it vs} dm15 relationships are derived. The values found from these
relationships are applied to the set of well-observed templates to
create a set of absolute magnitude light curves in $BVRI$. To measure
the distance to a new supernova, each template is fitted to
the data, marginalizing over time of maximum and $A_V$, to find the
distance modulus of the object. The distance is then estimated as the
$\chi^2$-weighted average of each template distance, with the distance
uncertainty derived from the union of all templates and distances that
have acceptable $\chi^2$ fits.

\subsection{Light Curve Fits using \batm}

The Bayesian Adapted Template Match (\batm) Method, described by Tonry
(2003), uses 
a set of approximately 20 nearby supernova light curves which are well
observed in multiple colors and which span the range of
absolute luminosity to predict what they would
look like at an arbitrary redshift with an arbitrary host extinction
through the actual filters used for observation.  Since these \sna\ do
not in general have a significant amount of spectrophotometry, the \batm\
method uses approximately 100 spectral energy distributions (SEDs) of
other \sna\ at a variety of \sna\ ages, and warps the SEDs into
agreement with the photometry.  Given each of the template predictions and a
light curve of a new \sna, the \batm\ method then calculates the
likelihood distribution as a function of explosion time $t_0$, host
extinction $A_V$, and distance $d$, and marginalizes over $t_0$ to get
a likelihood as a function of $A_V$ and $d$.  This is multiplied by a
prior for $A_V$ which is a Gaussian of $\sigma=0.2$ mag for negative
$A_V$, a Gaussian of $\sigma=1.0$ for positive $A_V$, and a delta
function at $A_V=0$ of equal weight.  Unfortunately, for these
observations (and probably most extant high redshift \sna\ distances)
there is strong covariance between $d$ and $A_V$ and the choice of
prior makes a non-negligible difference in the final estimate for $d$.
The \batm\ method then combines the probability distributions for all of
the template \sna\ and marginalizes over $A_V$ to get a final estimate
for the distance.

The \batm\ method is still undergoing improvement and is not currently
``tuned'' to minimize scatter of the \sna\ in the Hubble flow.  For
example, one might expect that certain bandpasses or time spans during
a supernova's light curve might be better predictors of distance (the
MLCS method is designed to optimize those choices).  The \batm\ method
simply fits all the data available in the observed bandpasses
(avoiding the use of K-corrections) without regard for how a set of
\sna\ in the nearby Hubble flow behave. As a result, the scatter of
the \batm\ method for these Hubble-flow \sna\ is somewhat larger than
that of MLCS (0.22 mag versus 0.18 mag), but it has a different, and perhaps
smaller, set of systematic errors.

\subsection{Fall 1999 Distances}

We list the distance estimates and uncertainties based on our light
curves from MLCS, $\Delta m_{15}$($B$), dm15, and \batm\ in
Table~\ref{disttab}, along with the empty universe distance modulus
corresponding to each redshift.  The uncertainties are formal error
estimates derived from each method, and we do not attempt here to
evaluate whether these estimates are accurate.  Below, in
Table~\ref{bigtable}, we do pay attention to the self consistency of
the various error estimates.  This assumes a nominal value of $H_0 =
65$ km s$^{-1}$ Mpc$^{-1}$, though the cosmological inferences do not
generally depend on the value of $H_0$.  The estimates are generally
consistent with each other, and we list a ``final'' distance which is
the median of the estimates of each of the four methods.  The uncertainty of
this ``final'' distance is taken to be the median of the errors of the
contributing methods, because we regard the process of taking a median
of different analysis methods not so much sharpening our precision as
rejecting non-Gaussian results.

\begin{deluxetable}{lcccccc}
\tablewidth{0pt}   
\tablecaption{Distance Moduli ($h=0.65$)\label{disttab}} 
\tablehead{ 
\colhead{SN name} & 
\colhead{MLCS} & 
\colhead{$\Delta m_{15}$} & 
\colhead{dm15} & 
\colhead{\batm} & 
\colhead{Final} &
\colhead{$\Omega=0$}
}
\startdata
SN~1999fw & 40.97 0.31 & 40.89 0.15 & 40.80 0.24 & 40.99 0.42 & 40.93 0.27 & 40.82 \\ 
SN~1999fh &  $\ldots$  & $\ldots$   & 41.77 0.26 & 41.46 0.47 & 41.61 0.36 & 41.52 \\ 
SN~1999ff & 42.38 0.29 & 42.60 0.22 & 42.35 0.43 & 42.61 0.32 & 42.49 0.30 & 42.05 \\ 
SN~1999fn & 42.69 0.26 & 42.36 0.16 & 41.98 0.40 & 42.19 0.23 & 42.27 0.24 & 42.18 \\ 
SN~1999fj & 44.03 0.22 & 43.59 0.37 & 43.76 0.35 & 43.55 0.43 & 43.67 0.36 & 43.62 \\ 
SN~1999fm & 44.18 0.22 & 44.06 0.23 & 43.84 0.37 & 43.77 0.35 & 43.95 0.29 & 44.05 \\ 
SN~1999fk &\llap{$<$}44.23 0.30 & $\ldots$   & 44.26 0.39 & 44.17 0.36 & 44.23 0.37 & 44.36 \\ 
SN~1999fv &\llap{$<$}44.59 0.39 & $\ldots$   &  $\ldots$  &\llap{$<$}44.19 0.45 &\llap{$<$}44.39 0.42 & 44.74 \\ 
\enddata
\end{deluxetable}

\subsection{Rates}

From observations of just three \sna\ at $z\approx 0.4$ Pain \etal\ (1996)
calculated a rate of \sna\ which amounts to
34~\sna~yr$^{-1}$~deg$^{-2}$, with a 1$\sigma$ uncertainty of
a factor of $\sim 2$, for objects found in the range of
$21.3<R<22.3$ mag.  A more recent estimate by Pain \etal\ (2002) is 
$(1.53\pm 0.3) \times10^{-4} \; h^3 \; \hbox{Mpc}^{-3} \; \hbox{yr}^{-1}$ 
at a mean redshift of 0.55.  Cappellaro et al. (1999) report a nearby rate
of \sna\ of $0.36\pm0.11\,h^2$ SNu (where 1 SNu 
$\equiv 10^{10}\,L_{B\solar}$ per century) or 
$(0.79 \pm 0.24) \times10^{-4} \; h^3\; \hbox{Mpc}^{-3} \; \hbox{yr}^{-1}$, 
using a local luminosity density of 
$\rho = 2.2 \times 10^{8} \; h \; L_{B\solar} \;\hbox{Mpc}^{-3}$.
[There is significant uncertainty in the luminosity density, however: Marzke
\etal\ (1998) obtained $1.7 \times 10^{8}$, and Folkes \etal\ (1999)
found $2.7 \times 10^{8}$.]  Pain \etal\ (2002) report a very
modest increase in the rate of \sna\ with redshift, perhaps tracking
the star formation rate which Wilson \etal\ (2002) estimate as being
proportional to $(1+z)^{1.7}$ in a flat universe with $\Omega_M=0.3$.

Although our Fall 1999 observing was focused on finding candidates for
detailed study, it still contains useful information on supernova
rates. Two facts help establish rates from this search.  First, our
observations were very deep in the $I$ band (since we were primarily
searching for $z>1$ \sna) and second, because we also sought to find
good $z\approx 0.5$ \sna\ for \hst\ observations, we obtained spectra
for every plausible candidate \sna\ with $z<0.7$.  While we did not
obtain light curves for all these candidates, we do know which were
Type Ia, and we know their redshifts.  This provides the raw material
for a computation of the supernova rate.

At the outset we need to make some reasonable assumption about the
underlying luminosity distribution of \sna, calculate what fraction
of these we would see in our search, and then derive the supernova
rate.  We assume a luminosity function for \sna\ at maximum which
consists of three Gaussians: 64\% ``normal'' \sna\ with mean
absolute magnitude $M_B=-19.5$ mag 
($h \equiv H_0/(100~{\rm km}~{\rm s}^{-1}~{\rm Mpc}^{-1}) = 0.65$)
and dispersion
$\sigma=0.45$~mag, 20\% ``SN 1991T'' \sna\ with $M_B=-19.8$ mag and
dispersion $\sigma=0.30$~mag, and 16\% ``SN 1991bg'' \sna\ with
$M_B=-18.0$ mag and dispersion $\sigma=0.50$~mag (Li et al. 2001a).

Table~\ref{peakmag} gives values that a ``normal'' \sna\ might achieve
at maximum, derived from the colors of SN~1995D at maximum and the
spectral energy distribution of SN~1994S, with parentheses indicating
magnitudes that are less accurate because they involve an
extrapolation beyond the data for either of these two objects.

\begin{deluxetable}{rcccccc}
\tablewidth{0pt}   
\tablecaption{Peak Magnitudes of Normal \sna\ \label{peakmag}} 
\tablehead{ 
\colhead{$z$} & 
\colhead{$B$} & 
\colhead{$V$} & 
\colhead{$R$} & 
\colhead{$I$} & 
\colhead{$Z$} & 
\colhead{$J$}
}
\startdata
 0.05 & 17.37 & 17.41 & 17.36 & 17.82 & 17.72 &(17.70)\\
 0.10 & 18.94 & 18.88 & 18.80 & 19.20 & 19.24 &(19.30)\\
 0.20 &(20.70)& 20.29 & 20.27 & 20.36 & 20.75 &(20.96)\\
 0.30 &(21.94)& 21.08 & 21.07 & 20.97 & 21.28 &(21.63)\\
 0.40 &(23.07)& 21.89 & 21.62 & 21.55 & 21.67 & 21.97 \\
 0.50 &(24.06)&(22.59)& 22.10 & 22.01 & 21.96 & 22.35 \\
 0.60 &(24.90)&(23.26)& 22.59 & 22.41 & 22.37 & 22.79 \\
 0.70 &(25.53)&(23.97)& 23.06 & 22.70 & 22.70 & 23.06 \\
 0.80 &(26.00)&(24.75)&(23.54)& 22.91 & 23.01 & 23.11 \\
 0.90 &(26.39)&(25.46)&(24.06)& 23.12 & 23.17 & 23.20 \\
 1.00 &(26.75)&(26.03)&(24.61)& 23.48 & 23.39 & 23.25 \\
 1.10 &(27.08)&(26.45)&(25.14)& 23.78 & 23.50 & 23.38 \\
 1.20 &(27.39)&(26.78)&(25.66)&(24.19)& 23.72 & 23.59 \\
 1.30 &(27.67)&(27.07)&(26.13)&(24.59)& 24.11 & 23.72 \\
 1.40 &(27.95)&(27.35)&(26.55)&(24.91)&(24.39)& 23.94 \\
 1.50 &(28.20)&(27.60)&(26.93)&(25.36)&(24.70)& 24.08 \\
 1.60 &(28.45)&(27.85)&(27.24)&(25.85)&(25.04)& 24.18 \\
 1.70 &(28.68)&(28.08)&(27.51)&(26.30)&(25.32)& 24.32 \\
 1.80 &(28.90)&(28.30)&(27.75)&(26.77)&(25.72)& 24.38 \\
\enddata
\end{deluxetable}

We work in a space of redshift $z$ and $m_{max}$, where $m$ is
the bandpass of interest ($I$ band here).  We use the magnitudes of
SN~1995D at maximum and the SED of SN~1994S to convert a point in this
space to a value for the $B$-band (rest frame) maximum and therefore a
density from our luminosity function.  We multiply in a volume factor
$dV/dz/d\Omega$ and a factor of $(1{+}z)^{-1}$ because we are interested in
the observer's rate of observing \sna.  All of this is done for a flat
universe with ($\Omega_M, \Omega_\Lambda$) = (0.3 , 0.7).  We then
convolve the resulting distribution with a dust extinction
distribution which approximates the detailed curves of Hatano, Branch,
\& Deaton (1998): 25\% of hosts are ``bulge systems'' with the
probability of a given host $A_B$ given by $f(A_B) \propto
0.02\,\delta(A_B) + 10^{-1.25-A_B}$, 75\% of hosts are ``disk
systems'' with probability $f(A_B) \propto 0.02\,\delta(A_B) +
e^{-3.77-A_B^2}$.  (The redshifted SEDs tell
us how the observed $I$-band extinction depends on a given host galaxy
extinction $A_B$.)  This gives us a distribution for the observed rate
of \sna\ as a function of peak apparent $I$-band magnitude.

However, two complications arise.  The ``observation time'' which must
be multiplied into this distribution is a convolution of the search time and
the intrinsic time that a given \sna\ is above the detection
threshold.  Also, our supernova detection threshold is not based only
on peak $I$-band magnitude.  We have two search epochs, separated by
31 days, and we have a sensitivity of $I\approx24$ mag to the flux {\em
difference} between the first and second epochs.

Using the formalism of the \batm\ method, we place SN~1995D ($\Delta m_{15}
= 0.99$, MLCS $\Delta = -0.42$) and SN~1999by ($\Delta m_{15} = 1.93$,
MLCS $\Delta = +1.48$) in the $z$--$m_{max}$ space.  Any point in the
space between them is assigned a light curve which is the flux
interpolation between the redshifted light curves of SN~1995D and SN~1999by.
Points which lie brighter or fainter than these two \sna\ are assigned
copies of one or the other light curves which are simply scaled
brighter or fainter to match the desired peak brightness.  Given a
light curve, we calculate the visibility time for each point in the
$(z,m_I)$ diagram, using a 31~d lag time and a sensitivity of
$I<24$ mag.  (Our actual probability of detecting a supernova rolls off
gradually at fainter magnitudes, but we had S/N~$\approx10$ at $I<24$ mag
and we ignored candidates which we thought would peak at
$I>24$ mag, hence the approximation of a step at $I=24$ mag.)  

The resulting difference light curve is identical to the actual light
curve when the explosion occurs after the first epoch, but for
explosions that erupted earlier than the first epoch, taking the
difference between our two observation epochs decreases the difference
and can be negative.  Low-luminosity \sna\ with rapid rise and
declines would be visible for a shorter time than higher-luminosity
\sna\ with their slowly declining light curves, so they are less
likely to be found.  For the purpose of calculating rates, the total
time that a supernova could be detected is a convolution of the time that
the difference flux from the supernova explosion exceeds the
sensitivity limit and the search time (a delta function in this case).

\begin{figure}[t]
\plotone{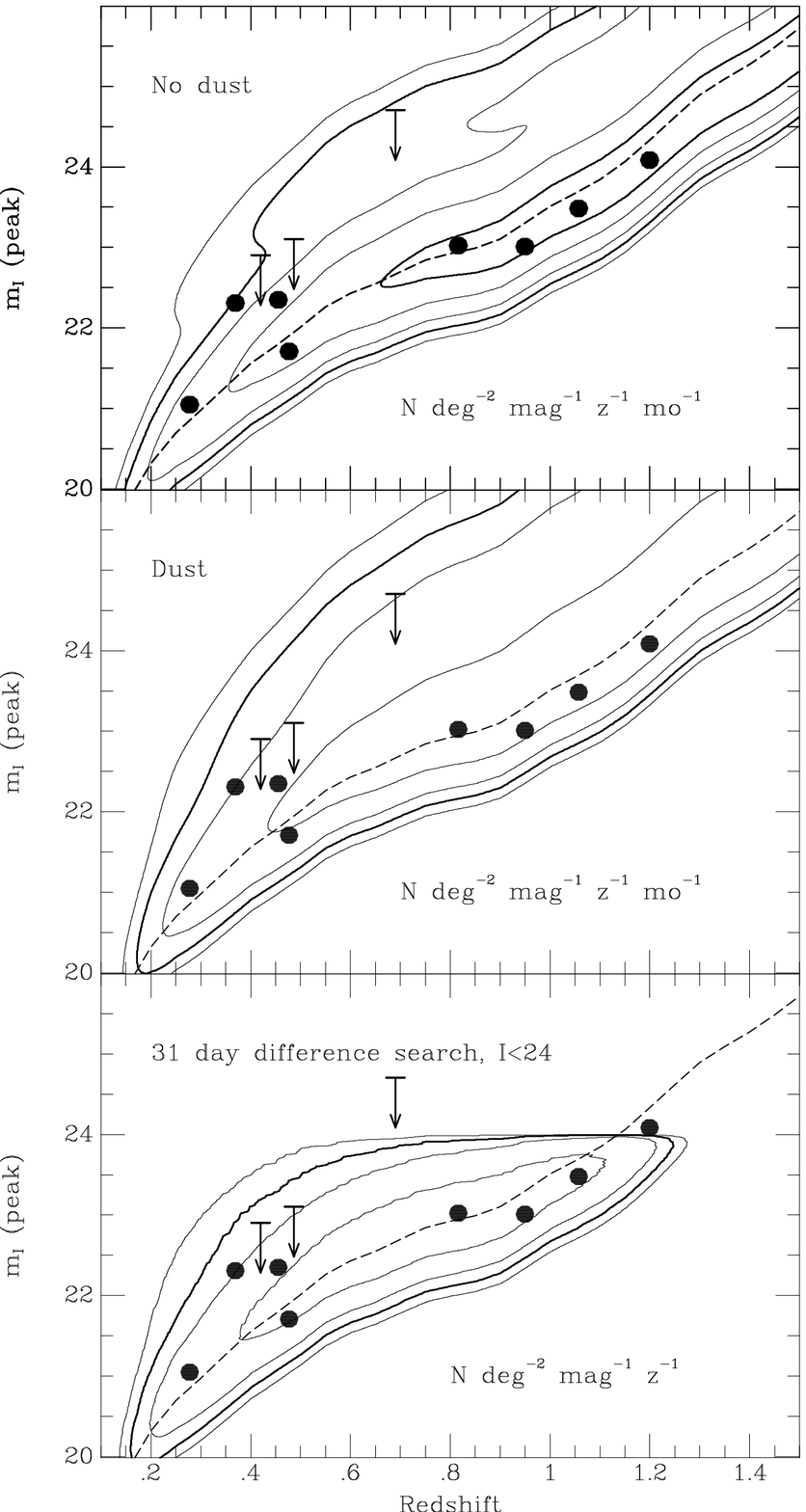}
\caption{Distribution of \sna\ occurrence as a function of peak $I$-band
magnitude and redshift.  Contours of predicted rates are drawn
with heavy curves at 1 and 10, and light curves at intervening
intervals of 2 and 5.  The ridge line of an unreddened ``normal''
\sna\ is drawn as a dashed line.  The upper panel shows the
distribution based on the \sna\ luminosity function if there were no
reddening; the middle panel shows the effect of convolving with the
model extinction distribution; and the bottom panel shows the expected
distribution for our search parameters.
\label{iceberg}}
\end{figure}

Figure \ref{iceberg} shows
contours of predicted rates (number per observer's month per square
degree per magnitude per unit redshift).  We plot our eight
\sna\ as well as three other \sna\ which we did not follow.  Without
light curves for these three, we can only show the discovery
(difference) magnitude which is an upper limit to the peak $I$ band
magnitude that was eventually reached.

\begin{figure}[t]
\plotone{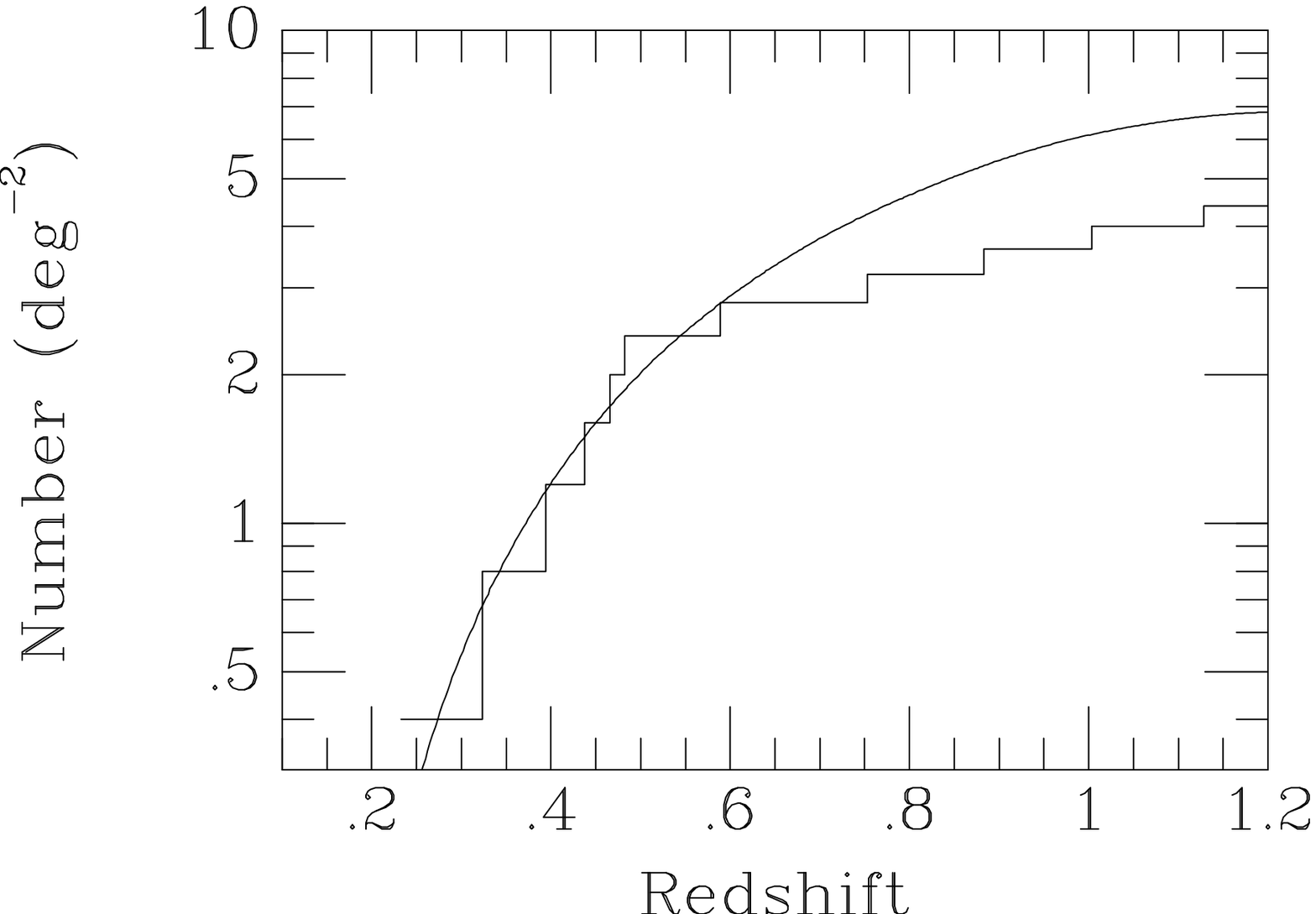}
\caption{Cumulative number of \sna\ discovery is shown as a function
of redshift.  The histogram is the observed \sna; the curve is
the prediction of what we should have seen given a flux
limit of $I<24$ mag and a 31~d difference search.  
We think that our search is nearly complete for
$z\le0.6$, but do {\it not} claim completeness for $z>0.6$, because of
the necessity to pick a subset for spectroscopy.
\label{rates}}
\end{figure}

The normalization of Figure~\ref{iceberg} is adjusted to bring the
cumulative distribution as a function of $z$ into agreement with our
observations.  Figure~\ref{rates} integrates the lower panel of 
Figure~\ref{iceberg} over magnitude and then assembles the cumulative
predicted number of \sna\ as a function of redshift.  We compare this
with the number observed to normalize the model.  Our sensitivity
limit was approximately $I<24$ mag, and the time between our first and
second search epochs was 31~d.  Our total area surveyed (including
the cloudy second epoch) was 3.0~deg$^{2}$, of which we lost about 15\%
due to chip edges, bad columns, stars, etc., so our effective
search area was about 2.5~deg$^{2}$.  Subsequent tests of our search
technique using synthetic \sna\ injected into the images indicate that
we miss very few objects down to the quoted sensitivity limit.

Uncertainties in the searched area, effective time sampling, and
modeling of supernova properties amount to perhaps 10--20\%, and a
similar uncertainty is the fraction of objects which we detected
photometrically but for which we did not obtain spectroscopy.  Since
our goal was to find suitable targets for \hst\ studies and to find
supernovae at $z \approx 0.5$, we were not complete in our
spectroscopic follow-up.  Subjective factors go into the decision of
which objects to follow.  For example, if we find a candidate at
$I=23.5$ mag, but it appears to be in a galaxy at $z\approx 0.3$, we
might very well choose not to get a spectrum, reasoning that ``the
supernova is well past maximum'' or ``it might be a type II'' or ``it
might be highly extinguished, or a SN 1991bg-type''.  Spectroscopic
resources are {\it so} valuable that our first priority, to obtain
targets, often outweighs our desire to control systematics.  The
factors in these decisions change with redshift of the supernova: we
believe we were complete in our search to $z<0.6$ (even though we did
not follow three \sna\ photometrically), but we were definitely not
complete for more distant and fainter \sna.  The lower panel of
Figure~\ref{iceberg} shows that the counts of \sna\ to a redshift of
$z<0.6$ are not very sensitive to our estimate of our completeness
limit.

Figure~\ref{rates} shows the comparison between this model cumulative
distribution and the observed rates for our Fall 1999 campaign.  The
Poisson statistics of the 8 \sna\ counted with $z<0.6$ are still the
largest source of uncertainty, and we estimate our overall
normalization is accurate to about 35\%.  The slope agreement between
the observations and the model in Figure~\ref{rates} reassures us that
our model is reasonably accurate.

These model predictions, which we calculate for comparison with the
observations, include complicated factors for visibility within a time
window and sensitivity limit, convolution with dust extinction, time
dilation factors of $(1+z)$, and volume factors of $dV/dz/d\Omega$.  The
fundamental normalization that goes into the calculation and that
brings the model into agreement with the observations shown in
Figure~\ref{rates} amounts to 40,000 \sna\ of all sub-types per Gpc$^3$ per
(rest frame) year for the Hubble constant used above ($H_0$ = 65
km s$^{-1}$ Mpc$^{-1}$) for the \sna\
luminosity function.  Put in more sensible units, this is a rate of
$r = (1.4 \pm 0.5) \times 10^{-4} \; h^3 \; \hbox{Mpc}^{-3} \;
\hbox{yr}^{-1}$ \sna\
at a mean redshift of $z=0.46$. So the observed rate of
\sna\ from a shell at redshift $z$ is
$$ dN = r\; d\Omega \;(dV/dz/d\Omega) \; (1+z)^{-1}\; \Delta t,$$
where $\Delta t$ has to be integrated over the luminosity function of
\sna\ and the search parameters.
These rates are in excellent agreement with those of Pain \etal\ (2002), and
are not inconsistent with the local rates from Cappellaro et al. (1999),
particularly given the uncertainty in the local luminosity density.

Although we were not complete in our counts of \sna\ at $z\approx1$,
we do not believe it is possible that the rate of \sna\ closely tracks
the star formation rate (SFR).  Application of the SFR from Wilson
\etal\ (2002) would suggest that the rate at $z\approx1.1$ should be
three times as great as occurs locally and nearly twice as great as
the rate at $z=0.46$.  Our model illustrated in Figure~\ref{rates}, if
altered from constant rate per comoving volume to one which tracks the
SFR of Wilson \etal\ (2002), would predict 16 \sna\ deg$^{-2}$ in our
search instead of 6.5 (we discovered 4).  It is our impression that
the constant rate per volume is closer to the truth --- \sna\ at
$m_I=24$ mag are simply not that common.  Obviously, \sna\ explosions
lag star formation by some unknown time distribution.  This suggests
the lag must span a substantial fraction of the time between the peak
of star formation and the time when a supernova observed at $z\sim1$
explodes, which is about 2 Gyr.  As pointed out by Pain \etal\ (2002),
we should obtain better constraints on any increase in \sna\ rate at
$z>1$ from ongoing search campaigns.

\subsection{A Homogeneous Set of \sna\ Distances}

As standardizable candles, \sna\ have played an extremely important
role in recent years.  However, observations of \sna\ are so difficult
and resource-intensive that the results have appeared in a large
number of papers, using a variety of distance zero-points, photometric
calibrations, etc.  We have therefore undertaken to draw all the
results available to us together into a single table with a common
calibration.

The most recent data apart from the observations presented here are
the results from the campaigns of the HZT in 1997 (Leibundgut et
al. 2003), 1998 (Suntzeff et al. 2003), 1999 (Clocchiatti et al. 2003),
and 2000 (Jha et al. 2003b).

In the 1997 data, there are four spectroscopically classified \sna\
(SN~1997as, 1997bb, 1997bh, and 1997bj) and one object in an
elliptical galaxy (SN~1997bd) at redshifts in the range 0.33 to
0.67. The redshifts are all from narrow emission lines associated with
the host galaxies, and on the prominent Ca~II H\&K absorption in the
elliptical galaxy.  The supernova spectrum in all cases yielded an age
concordant with the light curves. Light curves in the $R$ and $I$
filters have been established for all objects. The campaign in 1997
made use of ground-based telescopes only. The distance moduli, colors,
and absorptions are given in Table~\ref{bigtable}.

From our 1998 campaign, Suntzeff et al. (2003) report photometry and
spectroscopy of six \sna: 1998I, 1998J, 1998M, 1998ac, 1998ai, and
1998aj.  A seventh object (SN~1998ah) was only observed
spectroscopically.  These seven objects range in redshift from 0.43 to
0.887.  One of the significant results of the 1998 and 1999 (see below)
campaigns was the comparison of the photometry of the \sna\ and nearby
field stars from ground-based and \hst\ observations:  we found that
the zero-points of the photometry agree to within 0.02 mag, which is
reassuring.  Since the highest-weighted points in light curves of
high-$z$ supernovae are usually the \hst\ data, owing to their smaller
statistical uncertainties, we naturally wanted to compare photometry
obtained on the ground and in space. Because the evidence from supernova
observations for a non-zero cosmological constant hinges on a
systematic faintness of \sna\ at $z\approx 0.5$ by 0.22 mag, we are now
quite confident that the systematic faintness of supernovae at such redshifts
is {\em not} due to a zero-point error between space-based and
ground-based light curves.

In the Spring 1999 campaign, Clocchiatti et al. (2003) find five
spectroscopically classified \sna: SN~1999M, 1999N, 1999Q, 1999S, and
1999U.  All of them were followed up from ground-based observatories
and two of of them, SN~1999Q and SN~1999U, were observed with {\it HST}.  We
report in Table~\ref{bigtable} the distance moduli to the latter two
based on the {\it HST} photometry.  The redshift of SN 1999U was obtained
from emission lines in the parent galaxy, and that of SN 1999Q from
cross correlating its spectrum with nearby \sna\ close to
maximum light.

The 2000 campaign was intended primarily to investigate the
possibility that gray dust was causing the dimming we attribute to
cosmology.  Jha et al. (2003b) present photometry over a very wide
wavelength range.  There are six \sna\ at a redshift of $\sim 0.5$ whose
distances we include in Table~\ref{bigtable}: SN~2000dz, 2000ea,
2000ec, 2000ee, 2000eg, and 2000eh.

In addition to these observing campaigns of the HZT, we also drew from
the literature all \sna\ results we could find.  In cases where we had
access to the observational data and photometry we recomputed fits to
the distance and extinction using the MLCS98 method of Riess \etal\
(1998a), the MLCS2k2 improvement of Jha et al. (2003c), and the dm15
and \batm\ methods described above.  In other cases, such as the
Supernova Cosmology Project (SCP; Perlmutter et al.  1999), which has
not published their photometry, we took their published distances,
extinctions, and uncertainties at face value.  One might expect that
applying multiple fitting techniques as we have done helps decrease
errors in the derived distances.  The SCP-only results appear to have
at least a factor of 2 more scatter with respect to a cosmological fit
than our data.  This could be due to the quality of the underlying
photometry, or to a less varied approach to the analysis.

Assembling all of these results from the literature and performing new
fits when possible, we obtain long lists of \sna\ results for each
method which are internally self-consistent, but which have
unknown offsets with respect to one another because of differing and
inconsistent assumptions about Hubble constants and zero-points.  We
also have one more distance measure, derived from the redshift of the
\sna\ in the ``Hubble flow'' (defined as $0.01 < z < 0.1$) and
the luminosity distance for an empty universe:

\begin{equation}
D_L = \frac{cz}{H_0}\left(1 + \frac{z}{2}\right) \; .
\end{equation}

All of the methods overlap for many \sna, particularly the
Cal\'{a}n-Tololo sample (Hamuy et al. 1996a), with even the SCP
results having 17 or 18 overlaps with the rest, and the other methods
having 52--108 overlaps.  In order to determine the zero-point offsets
we compute the median difference between all pairs of methods, forming
an antisymmetric matrix, and then fit this matrix as differences
between components of a zero-point vector.  The residuals are quite
well behaved, with the worst agreement between SCP and MLCS98 at 0.07
mag, and the rest 0.02 mag or better.  We put all the methods on
the Hubble flow zero-point, i.e., distance in Mpc multiplied by $H_0$.

It is beyond the scope of this paper to analyze and optimize the
combination of the published data.  We do note that there actually
does appear to be a modest improvement in accuracy (judged by scatter
with respect to the cosmological line) by combining different fitting
methods.  Nevertheless, we are concerned about the susceptibility to
bias which might creep in if we optimize scatter with respect to the
parameters we seek to measure, so we choose to use medians,
a simple and robust combination scheme.
Our ``best distance'' for each \sna\ is just the median of all the
distances of all the different methods (excepting redshift-based
luminosity distance!).  The best distances for the Fall 1999 data
presented here differ slightly from those in Table~\ref{disttab}
because we did not have $\Delta m_{15}$ distances for more than the
Fall 1999 data and we thought that for the sake of consistency we
should treat the Fall 1999 data in the same way as the other \sna\
for which we have observations.  The scatter between the distances in
Table~\ref{disttab} and Table~\ref{bigtable} is 0.08 mag.

Obtaining accurate uncertainties is somewhat more complicated.  The
scatter of any method's results around any cosmological may not match
the error estimates.  We therefore examine this scatter for each
method, prune at 3$\sigma$, and multiply the error estimates for each
method by a factor to bring them into agreement with the scatter.
This appears to be a more realistic adjustment than adding a ``cosmic
scatter'' in quadrature (which we already do anyway when called for).
We then obtain a final error estimate as the median of all of the
contributing methods' errors diminished by the $1/4$ power of the
number of contributors (a factor empirically derived by examining
scatter around cosmological models).  As above, we believe the
covariance between the different methods applied to the same data is
such that we are getting only a minimal improvement in accuracy by
taking a median, but we hope that we are thereby ``Gaussianizing'' the
distribution.  The final list comprises 230 \sna.

We will use this list below for determining cosmological parameters,
and although we will not repeat the data we found in the literature,
we thought it important to publish the actual distances which are
being used for calculation of cosmological parameters as well as other
relevant information which may be of use to others.  In
Table~\ref{bigtable} we list the supernova name, its Galactic
coordinates, host, and literature reference.  We give the redshift as
$\log(cz)$ and the luminosity distance and error in the same units,
i.e., we do {\it not} want to use uncertain units of Mpc when we know
the Hubble flow zero-point so exquisitely well.  The median of the
estimates for host-galaxy extinction in the rest-frame $V$ band
follows, where we have used a standard Cardelli, Clayton, \& Mathis
(1989) extinction law to bring estimates in different bandpasses to
$V$.  Finally we identify the methods which contributed to the median:
MLCS98 = R, SCP/stretch = P, MLCS2k2 = J, dm15 = S, and \batm\ = T.

\begin{deluxetable}{lrrrllrrrrl}
\tablewidth{0pt}  
\tablecaption{SNIa Summary\label{bigtable}} 
\tablehead{ 
\colhead{SN} & 
\colhead{$l^{II}$} & 
\colhead{$b^{II}$} & 
\colhead{$z$} & 
\colhead{Host} & 
\colhead{Ref.$^a$} & 
\colhead{$\log(cz)$} & 
\colhead{$\langle \log (d H_0) \rangle$} & 
\colhead{$\pm$} & 
\colhead{$\langle A_V \rangle$} & 
\colhead{Contrib}
}
\startdata
sn72E &314.840&$ 30.080$&0.0023&N5253     &      16&2.839&2.399&0.033&0.10&RT   \\
sn80N &240.161&$-56.689$&0.0056&N1316     &       9&3.225&3.140&0.043&0.21&RT   \\
sn81B &292.970&$ 64.743$&0.0072&N4536     &       2&3.334&3.077&0.041&0.25&RT   \\
sn81D &240.161&$-56.689$&0.0056&N1316     &       9&3.225&3.044&0.055&0.31&JT   \\
sn86G &309.543&$ 19.401$&0.0027&N5128     &      26&2.908&2.440&0.035&0.49&RT   \\
sn88U &  8.737&$-81.227$&0.3100&Anon      &      24&4.968&5.096&0.072&0.10&RT   \\
sn89B &241.991&$ 64.403$&0.0036&N3627     &      37&3.033&2.844&0.030&0.74&RST  \\
sn90N &294.369&$ 75.987$&0.0044&N4639     &      21&3.120&3.204&0.035&0.30&RST  \\
sn90O & 37.654&$ 28.360$&0.0307&M+034403  &      10&3.964&3.977&0.025&0.09&RPJST\\
sn90T &341.503&$-31.526$&0.0400&P63925    &      10&4.079&4.101&0.042&0.25&RJT  \\
sn90Y &232.645&$-53.854$&0.0390&Anon      &      10&4.068&3.985&0.039&0.54&RJT  \\
sn90af&330.823&$-42.235$&0.0500&Anon      &      10&4.176&4.149&0.026&0.06&RPJST\\
sn91M & 30.392&$ 45.900$&0.0076&I1151     &       5&3.358&3.397&0.030&0.00&R    \\
sn91S &214.061&$ 57.426$&0.0560&U05691    &      10&4.225&4.283&0.042&0.11&RJT  \\
sn91T &292.609&$ 65.191$&0.0070&N4527     &      21&3.322&2.961&0.028&0.50&RST  \\
sn91U &311.824&$ 36.212$&0.0331&I4232     &      10&3.997&3.938&0.032&0.47&RJST \\
sn91ag&342.556&$-31.639$&0.0141&I4919     &      10&3.626&3.626&0.025&0.11&RJST \\
sn91bg&278.227&$ 74.464$&0.0042&N4374     &       4&3.100&3.064&0.035&0.02&RT   \\
sn92A &235.896&$-54.059$&0.0058&N1380     &      33&3.240&3.124&0.023&0.08&RST  \\
sn92G &184.623&$ 59.851$&0.0062&N3294     &       5&3.269&3.286&0.030&0.29&RT   \\
sn92J &263.540&$ 23.545$&0.0460&Anon      &      10&4.140&4.099&0.033&0.27&RJST \\
sn92K &306.275&$ 16.309$&0.0112&E269-57   &      10&3.526&3.432&0.036&0.56&RJST \\
sn92P &295.617&$ 73.110$&0.0265&I3690     &      10&3.900&3.939&0.027&0.13&RPJST\\
sn92ae&332.706&$-41.988$&0.0750&Anon      &      10&4.352&4.359&0.026&0.14&RPJST\\
sn92ag&312.490&$ 38.386$&0.0262&E508-67   &      10&3.895&3.864&0.031&0.49&RPJST\\
sn92al&347.337&$-38.490$&0.0141&E234-69   &      10&3.626&3.636&0.026&0.06&RPJT \\
sn92aq&  1.776&$-65.315$&0.1010&Anon      &      10&4.481&4.513&0.027&0.04&RPJST\\
sn92au&319.112&$-65.883$&0.0610&Anon      &      10&4.262&4.252&0.030&0.13&RJST \\
sn92bc&245.694&$-59.635$&0.0186&E300-09   &      10&3.746&3.777&0.021&0.03&RPJST\\
sn92bg&274.611&$-18.346$&0.0360&Anon      &      10&4.033&4.036&0.030&0.24&RPJST\\
sn92bh&267.849&$-37.328$&0.0450&Anon      &      10&4.130&4.191&0.026&0.33&RPJST\\
sn92bi& 63.264&$ 47.239$&0.4580&Anon      &      25&5.138&5.301&0.095&$\ldots$&P    \\
sn92bk&265.036&$-48.918$&0.0580&E156-08   &      10&4.240&4.222&0.025&0.11&RJST \\
sn92bl&344.129&$-63.925$&0.0430&E291-11   &      10&4.110&4.095&0.022&0.06&RPJST\\
sn92bo&261.995&$-80.348$&0.0178&E352-57   &      10&3.727&3.745&0.021&0.04&RPJST\\
sn92bp&208.832&$-51.096$&0.0790&Anon      &      10&4.374&4.336&0.026&0.05&RPJST\\
sn92br&288.014&$-59.428$&0.0880&Anon      &      10&4.421&4.441&0.027&0.07&RPJST\\
sn92bs&240.028&$-55.345$&0.0630&Anon      &      10&4.276&4.327&0.026&0.19&RPJST\\
sn93B &273.324&$ 20.460$&0.0710&Anon      &      10&4.328&4.345&0.026&0.32&RPJST\\
sn93H &318.223&$ 30.336$&0.0251&E445-66   &      10&3.876&3.845&0.026&0.32&RJST \\
sn93O &312.414&$ 28.926$&0.0520&Anon      &      10&4.193&4.231&0.026&0.08&RPJST\\
sn93ac&149.707&$ 17.212$&0.0490&P17787    &      30&4.167&4.206&0.043&0.49&RJST \\
sn93ae&144.629&$-63.220$&0.0180&U01071    &      30&3.732&3.684&0.024&0.17&RJST \\
sn93ag&268.435&$ 15.929$&0.0500&Anon      &      10&4.176&4.209&0.026&0.26&RPJST\\
sn93ah& 25.876&$-76.771$&0.0286&E471-27   &      10&3.933&3.933&0.032&0.18&RJST \\
sn94B &208.140&$ 26.669$&0.0900&Anon      &      29&4.431&4.507&0.030&0.00&R    \\
sn94C &174.632&$ 29.922$&0.0510&M+081523  &      29&4.184&4.139&0.036&0.00&R    \\
sn94D &290.151&$ 70.140$&0.0027&N4526     &      27&2.908&3.003&0.026&0.06&RT   \\
sn94F &258.613&$ 68.127$&0.3540&Anon      &      25&5.026&5.155&0.075&$\ldots$&P    \\
sn94G &162.893&$ 52.779$&0.4250&Anon      &      25&5.105&5.105&0.100&$\ldots$&P    \\
sn94H &173.057&$-53.517$&0.3740&Anon      &      25&5.050&5.023&0.060&$\ldots$&P    \\
sn94M &291.687&$ 63.033$&0.0244&N4493     &      30&3.864&3.835&0.026&0.40&RJST \\
sn94Q & 64.384&$ 39.680$&0.0290&P59076    &      30&3.939&3.955&0.026&0.26&RJST \\
sn94S &187.377&$ 85.142$&0.0161&N4495     &      30&3.684&3.693&0.024&0.04&RJST \\
sn94T &318.017&$ 59.838$&0.0360&P46640    &      30&4.033&4.014&0.025&0.17&RJST \\
sn94U &308.732&$ 54.772$&0.0056&N4948     &      29&3.225&3.149&0.030&0.00&R    \\
sn94ae&225.342&$ 59.665$&0.0054&N3370     &      30&3.209&3.295&0.028&0.29&RST  \\
sn94al&163.157&$-34.816$&0.4200&Anon      &      25&5.100&5.189&0.064&$\ldots$&P    \\
sn94am&173.104&$-53.563$&0.3720&Anon      &      25&5.047&5.131&0.057&$\ldots$&P    \\
sn94an& 69.409&$-49.082$&0.3780&Anon      &      25&5.054&5.195&0.081&$\ldots$&P    \\
sn95D &230.027&$ 39.659$&0.0077&N2962     &      30&3.363&3.373&0.024&0.08&RST  \\
sn95E &141.994&$ 30.262$&0.0116&N2441     &      30&3.541&3.559&0.034&2.18&RS   \\
sn95K &259.953&$ 43.327$&0.4780&Anon      &      28&5.156&5.284&0.038&0.02&RT   \\
sn95M &246.716&$ 28.833$&0.0530&Anon      &      29&4.201&4.243&0.044&0.10&RT   \\
sn95ac& 58.694&$-55.049$&0.0490&Anon      &      30&4.167&4.160&0.028&0.24&RJST \\
sn95ae& 76.817&$-56.240$&0.0680&Anon      &      29&4.309&4.315&0.086&0.00&R    \\
sn95ak&169.658&$-48.982$&0.0219&I1844     &      30&3.817&3.772&0.033&0.55&RJST \\
sn95al&192.179&$ 50.835$&0.0060&N3021     &      30&3.255&3.328&0.026&0.31&RST  \\
sn95ao&178.196&$-50.515$&0.3000&Anon      &      29&4.954&4.957&0.156&0.00&R    \\
sn95ap&179.359&$-46.149$&0.2300&Anon      &      29&4.839&4.893&0.120&0.00&R    \\
sn95aq&113.341&$-54.599$&0.4530&Anon      &      25&5.133&5.313&0.064&$\ldots$&P    \\
sn95ar&127.657&$-58.471$&0.4650&Anon      &      25&5.144&5.345&0.070&$\ldots$&P    \\
sn95as&127.757&$-58.339$&0.4980&Anon      &      25&5.174&5.421&0.064&$\ldots$&P    \\
sn95at&129.274&$-58.142$&0.6550&Anon      &      25&5.293&5.333&0.058&$\ldots$&P    \\
sn95aw&165.472&$-54.078$&0.4000&Anon      &      25&5.079&5.151&0.056&$\ldots$&P    \\
sn95ax&166.056&$-53.909$&0.6150&Anon      &      25&5.266&5.317&0.064&$\ldots$&P    \\
sn95ay&176.870&$-48.453$&0.4800&Anon      &      25&5.158&5.271&0.062&$\ldots$&P    \\
sn95az&202.114&$-31.504$&0.4500&Anon      &      25&5.130&5.181&0.061&$\ldots$&P    \\
sn95ba&215.987&$ 22.984$&0.3880&Anon      &      25&5.066&5.209&0.057&$\ldots$&P    \\
sn95bd&187.111&$-21.660$&0.0152&U03151    &      30&3.659&3.648&0.029&0.94&RJST \\
sn96C & 99.624&$ 65.036$&0.0276&M+082547  &      30&3.918&3.989&0.026&0.32&RJST \\
sn96E &253.120&$ 34.311$&0.4250&Anon      &      28&5.105&5.156&0.066&0.15&RT   \\
sn96H &290.752&$ 62.242$&0.6200&Anon      &      28&5.269&5.397&0.043&0.04&RT   \\
sn96I &276.852&$ 60.008$&0.5700&Anon      &      28&5.233&5.356&0.053&0.07&RT   \\
sn96J &253.221&$ 34.301$&0.3000&Anon      &      28&4.954&5.010&0.064&0.26&RT   \\
sn96K &224.381&$ 20.459$&0.3800&Anon      &      28&5.057&5.183&0.049&0.03&RT   \\
sn96R &259.070&$ 54.356$&0.1510&Anon      &      29&4.656&4.621&0.104&0.00&R    \\
sn96T &247.589&$ 37.000$&0.2410&Anon      &      29&4.859&4.941&0.112&0.00&R    \\
sn96U &259.357&$ 68.005$&0.4300&Anon      &      28&5.110&5.288&0.054&0.05&RT   \\
sn96V &257.577&$ 57.537$&0.0250&N3644     &      29&3.875&3.871&0.057&0.00&R    \\
sn96X &310.232&$ 35.649$&0.0078&N5061     &      30&3.369&3.266&0.024&0.08&RST  \\
sn96Z &253.609&$ 22.559$&0.0087&N2935     &      30&3.416&3.387&0.036&0.55&RST  \\
sn96ab& 43.159&$ 56.932$&0.1240&Anon      &      30&4.570&4.621&0.032&0.05&RJST \\
sn96af&319.640&$-43.222$&0.1000&Anon      &       7&4.477&4.428&0.040&0.00&S    \\
sn96ag&356.386&$-49.535$&0.1400&Anon      &       8&4.623&4.680&0.045&0.00&S    \\
sn96ai&101.583&$ 79.246$&0.0041&N5005     &      30&3.090&3.109&0.121&4.08&RS   \\
sn96am&355.950&$-49.361$&0.0650&A3809     &       8&4.290&4.356&0.043&0.32&S    \\
sn96ao&265.439&$-51.259$&0.0580&A3128     &       8&4.240&4.240&0.067&0.52&S    \\
sn96bk&111.255&$ 54.881$&0.0072&N5308     &      30&3.334&3.298&0.041&0.81&RST  \\
sn96bl&116.992&$-51.302$&0.0348&Anon      &      30&4.018&4.033&0.031&0.28&RJST \\
sn96bo&144.460&$-48.956$&0.0165&N0673     &      30&3.694&3.653&0.032&0.86&RST  \\
sn96bv&157.337&$ 17.972$&0.0167&U03432    &      30&3.700&3.663&0.036&0.73&RJST \\
sn96bx&263.413&$-46.747$&0.0580&Anon      &       8&4.240&4.256&0.130&0.36&S    \\
sn96cf&250.449&$ 50.009$&0.5700&Anon      &      25&5.233&5.333&0.060&$\ldots$&P    \\
sn96cg&220.767&$ 22.154$&0.4900&Anon      &      25&5.167&5.299&0.057&$\ldots$&P    \\
sn96ci&333.110&$ 62.084$&0.4950&Anon      &      25&5.171&5.245&0.056&$\ldots$&P    \\
sn96ck&301.409&$ 62.096$&0.6560&Anon      &      25&5.294&5.393&0.068&$\ldots$&P    \\
sn96cl&256.574&$ 48.668$&0.8270&Anon      &      25&5.394&5.609&0.108&$\ldots$&P    \\
sn96cm& 10.891&$ 46.743$&0.4500&Anon      &      25&5.130&5.313&0.061&$\ldots$&P    \\
sn96cn&334.314&$ 61.810$&0.4300&Anon      &      25&5.110&5.305&0.060&$\ldots$&P    \\
sn97E &140.201&$ 25.814$&0.0132&N2258     &      11&3.597&3.624&0.024&0.29&RJST \\
sn97F &204.472&$-28.455$&0.5800&Anon      &      25&5.240&5.371&0.061&$\ldots$&P    \\
sn97G &202.331&$-26.508$&0.7620&Anon      &      25&5.359&5.573&0.107&$\ldots$&P    \\
sn97H &202.374&$-26.214$&0.5260&Anon      &      25&5.198&5.309&0.057&$\ldots$&P    \\
sn97I &202.366&$-26.207$&0.1720&Anon      &      25&4.712&4.713&0.055&$\ldots$&P    \\
sn97J &209.921&$ 15.374$&0.6190&Anon      &      25&5.269&5.439&0.068&$\ldots$&P    \\
sn97K &216.354&$ 16.082$&0.5920&Anon      &      25&5.249&5.563&0.081&$\ldots$&P    \\
sn97L &220.026&$ 21.876$&0.5500&Anon      &      25&5.217&5.381&0.064&$\ldots$&P    \\
sn97N &220.656&$ 22.099$&0.1800&Anon      &      25&4.732&4.765&0.054&$\ldots$&P    \\
sn97O &220.066&$ 22.446$&0.3740&Anon      &      25&5.050&5.383&0.062&$\ldots$&P    \\
sn97P &256.583&$ 48.254$&0.4720&Anon      &      25&5.151&5.301&0.056&$\ldots$&P    \\
sn97Q &256.878&$ 48.379$&0.4300&Anon      &      25&5.110&5.193&0.055&$\ldots$&P    \\
sn97R &256.950&$ 48.501$&0.6570&Anon      &      25&5.294&5.445&0.061&$\ldots$&P    \\
sn97S &256.960&$ 48.704$&0.6120&Anon      &      25&5.264&5.417&0.058&$\ldots$&P    \\
sn97Y &124.772&$ 62.369$&0.0166&N4675     &      11&3.697&3.730&0.039&0.19&JT   \\
sn97ac&220.010&$ 22.485$&0.3200&Anon      &      25&4.982&5.051&0.055&$\ldots$&P    \\
sn97af&220.026&$ 22.419$&0.5790&Anon      &      25&5.239&5.375&0.060&$\ldots$&P    \\
sn97ai&249.958&$ 50.362$&0.4500&Anon      &      25&5.130&5.245&0.070&$\ldots$&P    \\
sn97aj&256.605&$ 48.217$&0.5810&Anon      &      25&5.241&5.297&0.060&$\ldots$&P    \\
sn97am&256.340&$ 49.057$&0.4160&Anon      &      25&5.096&5.193&0.057&$\ldots$&P    \\
sn97ap&333.646&$ 61.901$&0.8290&Anon      &      25&5.395&5.543&0.060&$\ldots$&P    \\
sn97as&224.726&$ 20.115$&0.5080&Anon      &      17&5.183&5.158&0.049&0.44&RST  \\
sn97aw&239.440&$ 47.883$&0.4400&Anon      &      17&5.120&5.355&0.050&0.36&RS   \\
sn97bb&290.721&$ 62.492$&0.5180&Anon      &      17&5.191&5.356&0.049&0.04&RST  \\
sn97bd&296.961&$ 62.650$&0.6710&Anon      &      17&5.304&5.200&0.065&0.99&ST   \\
sn97bh&330.050&$ 59.713$&0.4200&Anon      &      17&5.100&5.242&0.049&0.33&RST  \\
sn97bj&248.760&$ 48.844$&0.3340&Anon      &      17&5.001&5.070&0.051&0.20&ST   \\
sn97bp&301.157&$ 51.213$&0.0095&N4680     &      11&3.455&3.416&0.025&0.44&RJST \\
sn97bq&136.293&$ 39.485$&0.0096&N3147     &      11&3.459&3.484&0.023&0.42&JST  \\
sn97br&311.841&$ 40.328$&0.0080&E576-40   &   11:18&3.380&3.282&0.026&0.47&JT   \\
sn97by&312.690&$ 34.865$&0.0450&A1736     &       8&4.130&4.084&0.039&0.16&S    \\
sn97bz&259.751&$ 56.407$&0.0313&Anon      &       8&3.972&4.060&0.116&3.11&S    \\
sn97ce& 69.240&$ 36.620$&0.4400&Anon      &      28&5.120&5.228&0.041&0.03&RT   \\
sn97cj&125.800&$ 54.608$&0.5000&Anon      &      28&5.176&5.323&0.047&0.05&RT   \\
sn97ck& 57.602&$ 38.452$&0.9700&Anon      &      28&5.464&5.638&0.075&0.17&RT   \\
sn97cn&  9.137&$ 69.506$&0.0175&N5490     &   11:36&3.720&3.730&0.026&0.06&JT   \\
sn97cp&336.902&$-45.466$&0.1600&Anon      &       8&4.681&4.792&0.098&0.45&S    \\
sn97cu&265.042&$-51.213$&0.0620&A3128     &       8&4.269&4.240&0.045&0.00&S    \\
sn97cw&113.095&$-49.487$&0.0164&N0105     &      11&3.692&3.674&0.028&1.04&JST  \\
sn97dg&103.616&$-33.984$&0.0297&Anon      &      11&3.950&4.033&0.029&0.25&JST  \\
sn97do&171.001&$ 25.269$&0.0104&U03845    &      11&3.494&3.545&0.028&0.32&JST  \\
sn97dr&253.245&$-55.861$&0.0750&A3112     &       8&4.352&4.296&0.039&0.00&S    \\
sn97dt& 87.564&$-39.122$&0.0061&N7448     &      11&3.262&3.375&0.043&0.46&JT   \\
sn97fb&242.377&$-37.317$&0.0530&A3301     &       8&4.201&4.254&0.045&0.71&S    \\
sn97fc&242.691&$-37.596$&0.0540&A3301     &       8&4.209&4.230&0.039&0.10&S    \\
sn97fd&264.995&$-49.215$&0.1900&Anon      &       8&4.756&4.888&0.065&0.00&S    \\
sn97ff&125.906&$ 54.831$&1.7550&Anon      &      32&5.721&5.905&0.083&0.00&R    \\
sn98D & 63.777&$ 72.905$&0.0132&N5440     &      11&3.597&3.605&0.069&0.20&J    \\
sn98I &216.646&$ 18.718$&0.8870&Anon      &      34&5.425&5.575&0.053&0.25&JST  \\
sn98J &238.498&$ 32.106$&0.8330&Anon      &      34&5.397&5.615&0.052&0.26&JST  \\
sn98M &260.707&$ 60.423$&0.6300&Anon      &      34&5.276&5.396&0.045&0.28&RST  \\
sn98V & 43.942&$ 13.346$&0.0170&N6627     &      11&3.707&3.686&0.028&0.31&JST  \\
sn98ab&124.861&$ 75.194$&0.0278&N4704     &      11&3.921&3.876&0.025&0.59&JST  \\
sn98ac&239.012&$ 31.832$&0.4670&Anon      &      34&5.146&5.260&0.043&0.27&RST  \\
sn98aj&238.163&$ 31.312$&0.8600&Anon      &      34&5.411&5.698&0.070&0.20&ST   \\
sn98aq&138.837&$ 60.269$&0.0037&N3982     &       1&3.045&3.181&0.023&0.06&RST  \\
sn98bp& 43.643&$ 20.481$&0.0104&N6495     &      11&3.494&3.475&0.029&0.42&RJST \\
sn98br&  6.580&$ 50.757$&0.0810&A2029     &       8&4.385&4.442&0.063&0.39&S    \\
sn98bu&234.413&$ 57.020$&0.0043&N3368     &      12&3.110&2.912&0.028&1.03&RST  \\
sn98cm&332.822&$ 62.779$&0.0800&A1780     &       8&4.380&4.398&0.059&0.36&S    \\
sn98co& 41.520&$-44.941$&0.0171&N7131     &      11&3.710&3.729&0.046&0.28&JT   \\
sn98cs& 65.238&$ 43.339$&0.0327&U10432    &      11&3.991&3.956&0.037&0.03&RT   \\
sn98de&122.031&$-35.241$&0.0157&N0252     &   11:23&3.673&3.727&0.024&0.68&JST  \\
sn98dh& 82.828&$-50.644$&0.0076&N7541     &      11&3.358&3.403&0.034&0.50&ST   \\
sn98dk&102.856&$-62.161$&0.0120&U00139    &      11&3.556&3.571&0.028&0.80&JST  \\
sn98dm&145.975&$-67.406$&0.0054&M+010444  &      11&3.209&3.498&0.028&1.06&ST   \\
sn98do&135.234&$-61.849$&0.0920&Anon      &       8&4.441&4.462&0.069&0.00&S    \\
sn98dv&272.283&$-40.300$&0.1550&Anon      &       8&4.667&4.706&0.124&0.42&S    \\
sn98dw&143.355&$-77.658$&0.0490&A0151     &       8&4.167&4.208&0.051&0.00&S    \\
sn98dx& 77.674&$ 26.667$&0.0538&U11149    &      11&4.208&4.175&0.030&0.05&JT   \\
sn98dz&247.207&$-56.375$&0.0910&Anon      &       8&4.436&4.458&0.045&0.16&S    \\
sn98ea&272.198&$-39.821$&0.0570&A3266     &       8&4.233&4.314&0.043&0.58&S    \\
sn98ec&166.295&$ 20.711$&0.0200&U03576    &      11&3.778&3.801&0.039&0.57&JST  \\
sn98ef&125.882&$-30.566$&0.0170&U00646    &      11&3.707&3.650&0.028&0.08&JST  \\
sn98eg& 76.464&$-42.060$&0.0234&U12133    &      11&3.846&3.897&0.042&0.26&JT   \\
sn98es&143.189&$-55.177$&0.0095&N0632     &   11:20&3.455&3.438&0.031&0.33&JST  \\
sn98fb&264.638&$-51.045$&0.0600&A3128     &       8&4.255&4.262&0.059&0.00&S    \\
sn99Q &215.915&$ 17.925$&0.4600&Anon      &    31:3&5.140&5.336&0.042&0.07&RST  \\
sn99U &238.552&$ 30.694$&0.5000&Anon      &       3&5.176&5.321&0.070&0.02&ST   \\
sn99X &186.585&$ 39.591$&0.0257&C180022   &      11&3.887&3.888&0.037&0.33&JT   \\
sn99aa&202.725&$ 30.313$&0.0157&N2595     &11:20:15&3.673&3.684&0.032&0.01&RST  \\
sn99ac& 19.883&$ 39.943$&0.0099&N6063     &   11:20&3.472&3.478&0.023&0.42&JST  \\
sn99ao&243.830&$-20.023$&0.0550&A3392     &       8&4.217&4.242&0.039&0.26&S    \\
sn99by&166.914&$ 44.120$&0.0027&N2841     &    6:20&2.908&3.067&0.167&0.72&ST   \\
sn99cc& 59.667&$ 48.740$&0.0316&N6038     &      11&3.977&3.983&0.029&0.13&RJST \\
sn99cl&282.270&$ 76.509$&0.0082&N4501     &      15&3.391&3.009&0.090&1.68&RS   \\
sn99cp&334.851&$ 52.708$&0.0104&N5468     &      15&3.494&3.525&0.026&0.04&RJT  \\
sn99cw&101.768&$-67.906$&0.0112&M-010201  &      11&3.526&3.469&0.034&0.17&J    \\
sn99da& 89.732&$ 32.649$&0.0121&N6411     &   14:20&3.560&3.623&0.044&0.58&RJST \\
sn99dk&137.349&$-47.464$&0.0141&U01087    &      14&3.626&3.686&0.039&0.07&RT   \\
sn99dq&152.839&$-35.871$&0.0136&N0976     &   11:20&3.610&3.548&0.023&0.35&JST  \\
sn99ef&125.719&$-50.086$&0.0380&U00607    &      11&4.057&4.147&0.033&0.03&JT   \\
sn99ej&130.442&$-28.946$&0.0127&N0495     &      11&3.581&3.684&0.031&0.08&JT   \\
sn99ek&189.403&$ -8.234$&0.0176&U03329    &      11&3.722&3.687&0.044&0.62&JT   \\
sn99ff&168.784&$-52.949$&0.4550&Anon      &      35&5.135&5.289&0.048&0.16&RST  \\
sn99fh&166.731&$-53.793$&0.3690&Anon      &      35&5.044&5.135&0.061&0.70&ST   \\
sn99fj&166.879&$-53.734$&0.8150&Anon      &      35&5.388&5.564&0.051&0.09&RST  \\
sn99fk&166.405&$-53.167$&1.0560&Anon      &      35&5.500&5.659&0.054&0.03&RST  \\
sn99fm&167.075&$-52.994$&0.9490&Anon      &      35&5.454&5.580&0.053&0.04&RST  \\
sn99fn&188.216&$-31.901$&0.4770&Anon      &      35&5.155&5.249&0.039&0.17&RST  \\
sn99fv& 84.298&$-56.405$&1.1990&Anon      &      35&5.556&5.651&0.068&0.24&RST  \\
sn99fw& 84.665&$-56.682$&0.2780&Anon      &      35&4.921&5.007&0.047&0.34&RST  \\
sn99gd&198.835&$ 33.977$&0.0190&N2623     &      11&3.756&3.779&0.029&1.27&JT   \\
sn99gh&255.048&$ 23.736$&0.0088&N2986     &      11&3.421&3.356&0.023&0.15&JST  \\
sn99gp&143.251&$-19.504$&0.0260&U01993    &11:20:14&3.892&3.933&0.028&0.15&RJST \\
sn00B &166.353&$ 22.791$&0.0193&N2320     &      11&3.762&3.740&0.039&0.38&JST  \\
sn00bk&295.292&$ 55.233$&0.0266&N4520     &      14&3.902&3.911&0.030&0.26&RJT  \\
sn00ce&149.096&$ 32.004$&0.0165&U04195    &   11:14&3.694&3.705&0.030&1.02&RJT  \\
sn00cf& 99.883&$ 42.165$&0.0360&M+111925  &      11&4.033&4.112&0.032&0.13&RJST \\
sn00cn& 53.445&$ 23.318$&0.0233&U11064    &      11&3.844&3.847&0.023&0.27&JST  \\
sn00cx&136.506&$-52.482$&0.0070&N0524     &   11:19&3.322&3.320&0.023&0.02&RST  \\
sn00dk&126.834&$-30.344$&0.0164&N0382     &      11&3.692&3.677&0.023&0.01&JST  \\
sn00dz& 84.367&$-56.390$&0.5000&Anon      &      13&5.176&5.352&0.053&0.07&RST  \\
sn00ea&167.211&$-61.402$&0.4200&Anon      &      13&5.100&5.091&0.053&0.88&RST  \\
sn00ec&166.295&$-60.176$&0.4700&Anon      &      13&5.149&5.340&0.043&0.04&RST  \\
sn00ee&166.045&$-53.429$&0.4700&Anon      &      13&5.149&5.343&0.041&0.04&RST  \\
sn00eg&167.097&$-53.108$&0.5400&Anon      &      13&5.209&5.236&0.049&0.17&RST  \\
sn00eh&188.297&$-31.651$&0.4900&Anon      &      13&5.167&5.224&0.042&0.20&RST  \\
sn00fa&194.167&$ 15.479$&0.0218&U03770    &      11&3.815&3.837&0.026&0.34&JST  \\
sn01V &218.929&$ 77.733$&0.0162&N3987     &      22&3.686&3.662&0.023&0.20&JST  \\
\enddata
\tablenotetext{a}{
\scriptsize
References:
(1) Boffi et al. 2003 (in preparation);
(2) Buta \& Turner 1983, PASP, 95, 72, and Tsvetkov 1982, SvAL, 8, 115;
(3) Clocchiatti et al. 2003 (in preparation);
(4) Filippenko et al. 1992, AJ, 104, 1543, and Leibundgut et al. 1993, AJ, 105, 301, and Turatto et al. 1996, MNRAS, 283, 1;
(5) Ford et al. 1993, AJ, 106, 3;
(6) Garnavich et al. 2001, astro-ph/0105490;
(7) Germany 2001 (PhD thesis ANU);
(8) Germany et al. 2003, A\&A, in press;
(9) Hamuy et al. 1991, AJ, 102, 208;
(10) Hamuy et al. 1996, AJ, 112, 2408;
(11) Jha 2002 (PhD thesis Harvard);
(12) Jha et al. 1999, ApJS, 125, 73, and Suntzeff et al. 1999, AJ, 117, 1175;
(13) Jha et al. 2003, ApJ (in preparation);
(14) Krisciunas 2001, AJ, 122, 1616;
(15) Krisciunas et al. 2000, ApJ, 539, 658;
(16) Leibundgut et al. 1991, A\&AS, 89, 537;
(17) Leibundgut et al. 2003 (in preparation);
(18) Li et al. 1999, AJ, 117, 2709;
(19) Li et al. 2001, PASP, 113, 1178;
(20) Li et al. 2004, (in preparation);
(21) Lira et al. 1998, AJ, 115, 234;
(22) Mandel et al. 2001, AAS, 199, 4704;
(23) Modjaz et al. 2001, PASP, 113, 308;
(24) Norgaard-Nielsen et al. 1989, Nature, 339, 523;
(25) Perlmutter et al. 1999, ApJ, 517, 565;
(26) Phillips et al. 1997, PASP, 99, 592;
(27) Richmond et al. 1995, AJ, 109, 2121, and Patat et al. 1996, MNRAS, 278, 111, and Meikle et al. 1996, 281, 263;
(28) Riess et al. 1998 AJ, 116, 1009;
(29) Riess et al. 1998, ApJ, 504, 935;
(30) Riess et al. 1999, AJ, 117, 707;
(31) Riess et al. 2000, ApJ, 536, 62;
(32) Riess et al. 2001, ApJ, 560, 49;
(33) Suntzeff 2001 (private communication);
(34) Suntzeff et al. 2003 (in preparation);
(35) Tonry et al. 2003, ApJ (this paper);
(36) Turatto et al. 1998, AJ, 116, 2431;
(37) Wells et al. 1994, AJ, 108, 2233
}
\end{deluxetable}

\section{Cosmological Implications}


\subsection{Hubble Diagrams}

At a redshift of 0.5, the difference in luminosity distance
between a flat and open universe with $\Omega_M = 0.3$ is 0.22 mag.
With 230 \sna\ and 79 at redshifts greater than 0.3, we might expect
to be able to distinguish between these cosmologies at better than
6$\sigma$.  The real concern, however, is whether we are making a
systematic error or hitting a systematic noise floor for \sna, and
that we are fooling ourselves in thinking that we are making progress
by averaging together more and more \sna.  

The goal of this particular campaign was to strive to find and follow
\sna\ at distinctly higher redshift than the majority of results at
$z=0.5$, since the effect of a dark-energy dominated cosmology
looks very different from most plausible systematic effects at
$z\approx 1$.  If the universe really has no dark energy and the
dimness of \sna\ by 0.22 mag at $z=0.5$ is the result of a systematic
error proportional to the redshift, one might expect \sna\ to be dimmer
by 0.44 mag at $z=1$ instead of 0.26 mag which would be the case for a
flat universe.

\begin{figure}[t]
\plotone{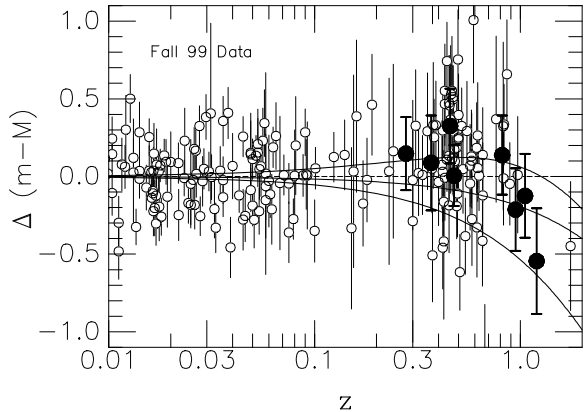}
\caption{The Fall 1999 and other data points are shown in a residual
Hubble diagram with respect to an empty universe.  From top to bottom
the curves show $(\Omega_M,\Omega_{\Lambda}) = (0.3,0.7)$, 
(0.3,0.0), and 
(1.0,0.0), respectively, and the Fall 1999 points are highlighted.
\label{hub99}}
\end{figure}

Figure~\ref{hub99} shows the results from this campaign plotted over
the full set of data from Table~\ref{bigtable}.  Evidently there is a
tendency for the \sna\ to be brighter at $z > 0.9$ (although the
uncertainties are still large), suggesting that we are truly seeing a
change in the deceleration of the universe and the effects of dark energy.

Given the heterogeneous nature of the data and the continual worry
about non-Gaussian distributions, it is difficult to be quantitative
about how well we can reject the possibility that we are seeing a
systematic error in \sna\ brightness in a universe with no dark
energy, and we discuss this in more detail below.

\begin{figure}[t]
\plotone{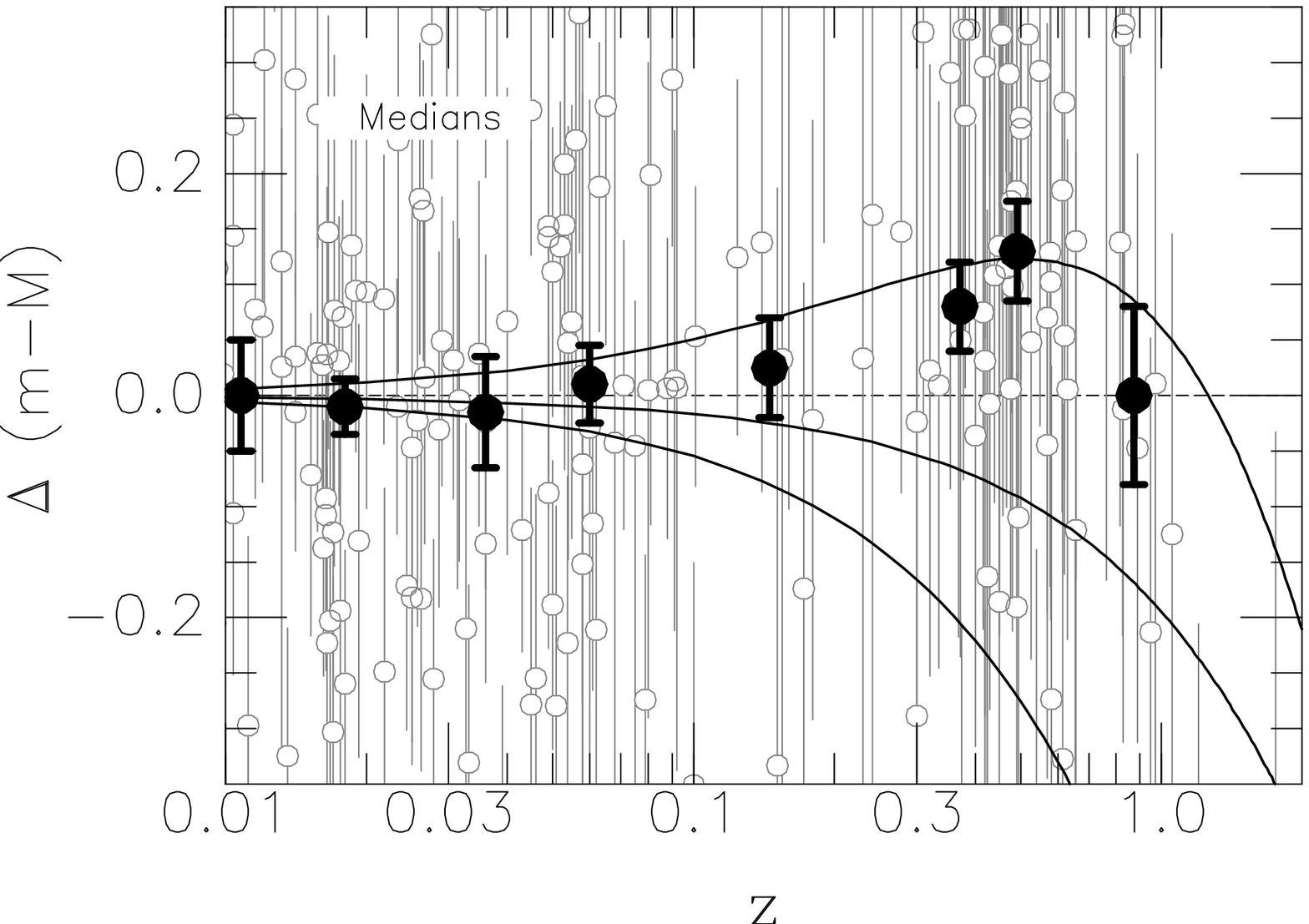}
\caption{The Fall 1999 and other data points are shown in a residual
Hubble diagram with respect to an empty universe.  In this plot the
highlighted points correspond to median values in eight redshift bins.
From top to bottom
the curves show $(\Omega_M,\Omega_\Lambda) = (0.3,0.7)$, 
(0.3,0.0), and (1.0,0.0), respectively.  
\label{hubmed}}
\end{figure}

The scatter of the points from Table~\ref{bigtable} obscures the trend
with redshift, so we take medians over redshift bins and illustrate
the result in Figure~\ref{hubmed}.  Our redshift bins are derived from
the data by choosing bins which have at least 12 \sna\ and a bin width
of more than 0.25 in $\log z$ but no more than 50 \sna.  The
uncertainties, estimated from the 68\% percentiles of the scatter of
the points in each bin, are all less than 0.05 mag except for 0.08 mag
in the highest $z$ bin.  We believe that this is approximately where
systematics will start to dominate statistical error.  That the points
mostly lie below the $(0.3,0.7)$ cosmology gets at the heart of what we
believe may the systematic uncertainty floor of the \sna\ distances in
Table~\ref{bigtable}; we discuss this below.  Nevertheless, we are
apparently starting to see \sna\ become brighter at $z\approx 0.9$,
rather than becoming ever dimmer because of a systematic effect.

\subsection{Cosmological Parameters}

We can fit the data from Table~\ref{bigtable} with a variety of
cosmological models.  We do not want to fit all 230 \sna\ because the
nearby ones are affected by peculiar motions which are beyond the
scope of this paper to characterize.  However, we do need to allow for
the velocity uncertainty in modeling, so when we compute
$\chi^2$ we add 500~km~s$^{-1}$ divided by the redshift in quadrature to
the distance error.  For all fits we report $\chi^2$ at the minimum as
a function of $H_0$, since this is equivalent to marginalizing over
$H_0$.

Fitting all 230 \sna\ yields $\chi^2 = 252$.  If we restrict our fits
to $z>0.01$, we obtain $\chi^2 = 206$ for 195 points.  We are also
very wary of ``extinguished'' \sna\ since we are not certain we
understand host extinction and reddening very well,
\footnote[19]{Phillips et al. (1999) give in their Table 2 the
Galactic and host reddening $E(B-V)$ for nearby supernovae.  If, as
suggested by the 44 \sna\ studied by Jha (2002, \S 3), most \sna\ obey
Lira's (1995) relation, then the colors at maximum and in the tail of
the $B-V$ color curve {\em can} give us a good handle on the {\em
reddening} at optical wavelengths.  However, while most galaxies may
have dust like that in our Galaxy, with $R_V \equiv A_V /E(B-V)
\approx 3.1$ (Cardelli et al. 1989), there are heavily reddened
objects such as SN~1999cl whose host galaxy has dust with $R_V < 2$
(Krisciunas et al. 2001).  Furthermore, high-redshift supernovae are
typically too faint to detect with sufficient precision 30 to 90
rest-frame days after maximum light.  Finally, while a combination of
optical and infrared data gives us a good handle on the extinction
$A_V$ (Krisciunas et al. 2000, 2001, 2003), \sna\ with $z \gtrsim 0.10$
are too faint in the rest-frame $H$-band and $K$-band even for 10-m
class telescopes to apply the $V$ minus infrared method of determining
extinction.}
and we are mindful of the covariance between $A_V$ and $d$.  If we
further restrict the sample to $A_V<0.5$ mag we get $\chi^2 = 169.5$
for 172 \sna, and we regard this as the most useful sample for
deriving cosmological parameters.  There are further cuts which could
be made: exclusion of the SCP data points (where there is no
information on host galaxy extinction) gives us $\chi^2 = 121.6$ for
130 \sna.  However the constraints on cosmological parameters do not
change significantly and we are concerned about acquiring systematic
problems along with spurious precision.

In the analyses that follow we use various subsamples of the 172 \sna\
with $z>0.01$ and $A_V<0.5$ mag. For each calculation we determine
$\chi^2$ as a function of $H_0$, $\Omega_M$, and $\Omega_\Lambda$.
Given that $\chi^2/N$ is close to unity, we convert this to a
probability as $\exp(-0.5\chi^2)$, normalize, marginalize over $H_0$,
and locate contours of constant probability density which enclose 68\%,
95\%, and 99.5\% of the probability.  In those calculations where
we additionally constrain cosmology by using the galaxy clustering
shape measurements, we use observations from the 2dF redshift survey (Percival
et al. 2001), $\Omega_M h = 0.20\pm0.03$, and 
$H_0 = 72 \pm8$~km~s$^{-1}$~Mpc$^{-1}$ (Freedman \etal\ 2001). In all
cases, contours are calculated for a cosmological constant equation of
state ($w=-1$).

\begin{figure}[t]
\plotone{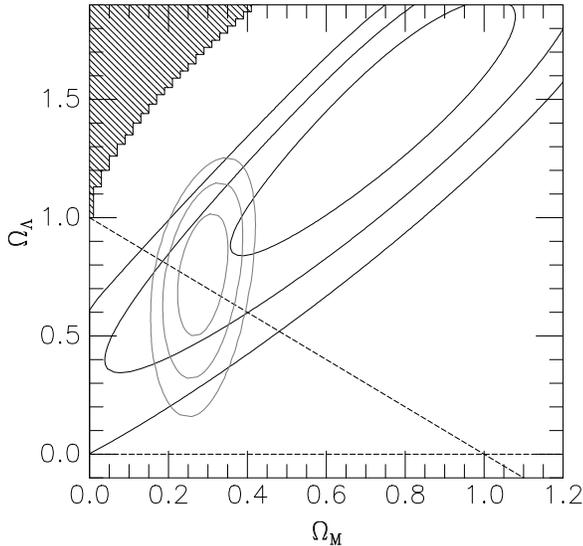}
\caption{Probability contours for $\Omega_\Lambda$ versus $\Omega_M$ 
are shown at 1$\sigma$, 2$\sigma$, and 3$\sigma$ with $w=-1$.  We 
also give 1$\sigma$, 2$\sigma$,
and 3$\sigma$ contours when we adopt a prior of $\Omega_M h =
0.20\pm0.03$ from the 2dF survey (Percival et al. 2001).  
These constraints use only the 26 new
\sna\ at $z>0.3$ (which are completely independent of any which have
been used before for cosmological constraints).
\label{omom_new}}
\end{figure}

Figure~\ref{omom_new} shows these contours for a sample of 94 \sna\
($\chi^2 = 103$) which include only the recent results from Leibundgut
\etal\ (2003), Suntzeff \etal\ (2003), Jha \etal\ (2003b), and the
data presented here.  The error ellipses are virtually identical to
those drawn by Riess \etal\ (1998) and Perlmutter \etal\ (1999),
although the data and analyses are completely different, confirming
the conclusion that the universe is accelerating.  This independent
sample gives a consistent result: the acceleration inferred
several years ago was not a statistical fluke.

\begin{figure}[t]
\plotone{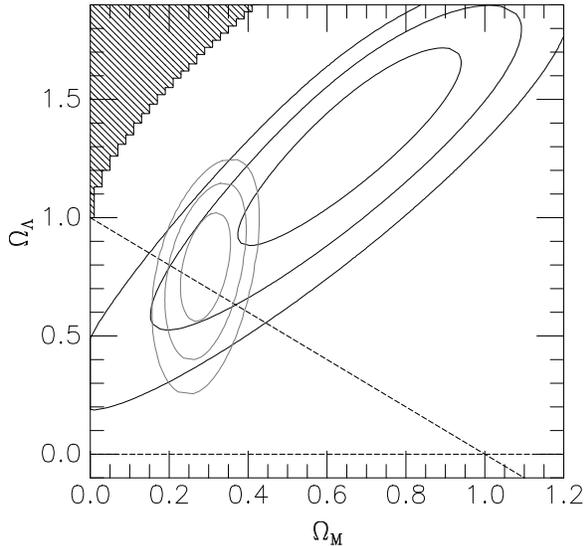}
\caption{Probability contours for $\Omega_\Lambda$ versus $\Omega_M$
are shown at 1$\sigma$, 2$\sigma$, and 3$\sigma$ with $w=-1$.  We 
also give 1$\sigma$, 2$\sigma$,
and 3$\sigma$ contours when we adopt a prior of $\Omega_M h =
0.20\pm0.03$ from the 2dF survey (Percival et al. 2001).  
These constraints use the 113 \sna\ published by the HZT.
\label{omom_hiz}}
\end{figure}

Figure~\ref{omom_hiz} shows these contours for the sample of 130 \sna\
($\chi^2=122$) which includes all the \sna\ which have been published
by the HZT.

\begin{figure}[t]
\plotone{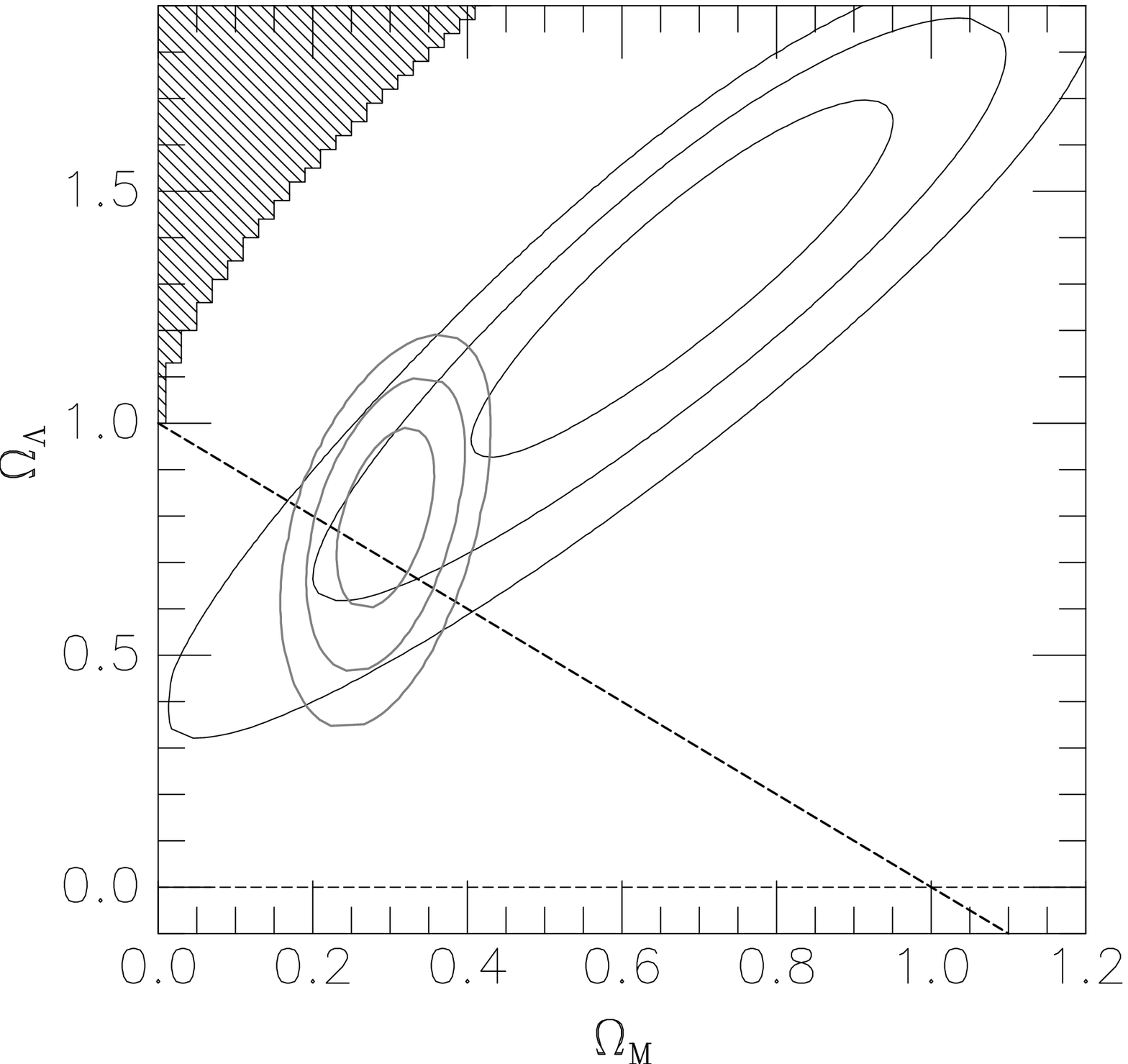}
\caption{Probability contours for $\Omega_\Lambda$ versus $\Omega_M$
are shown at 1$\sigma$, 2$\sigma$, and 3$\sigma$ with $w=-1$.  We 
also give 1$\sigma$, 2$\sigma$,
and 3$\sigma$ contours when we adopt a prior of $\Omega_M h = 
0.20\pm0.03$ from the 2dF survey (Percival et al. 2001).  
These constraints use the full   
sample of 172 \sna\ with $z>0.01$ and $A_V<0.5$ mag.
\label{omom_all}}
\end{figure}

Figure~\ref{omom_all} shows these
contours for the sample of 172 \sna\ ($\chi^2=170$) which includes all
the \sna\ for which we have either data or published results.

In all cases, the derived probability contours with a 2dF prior are
remarkably tightly constrained to the $\Omega_{tot} = 1$ line which is
being delineated so well by CMB observations.

As with previous results from \sna, the approximate 
combination $\Omega_\Lambda-\Omega_M$
is relatively well constrained. For these data
$$\Omega_\Lambda-1.4\Omega_M = 0.35\pm0.14.$$
However, the presence of these higher-redshift observations from Fall
1999 as well as SN~1997ff (Riess et al. 2001) has started to close off
the contours from running away to very large $\Omega$.  If we assume a
flat universe as well as $w=-1$, the \sna\ data require $\Omega_M =
0.28\pm0.05$, independent of large-scale structure estimates.

\begin{figure}[t]
\plotone{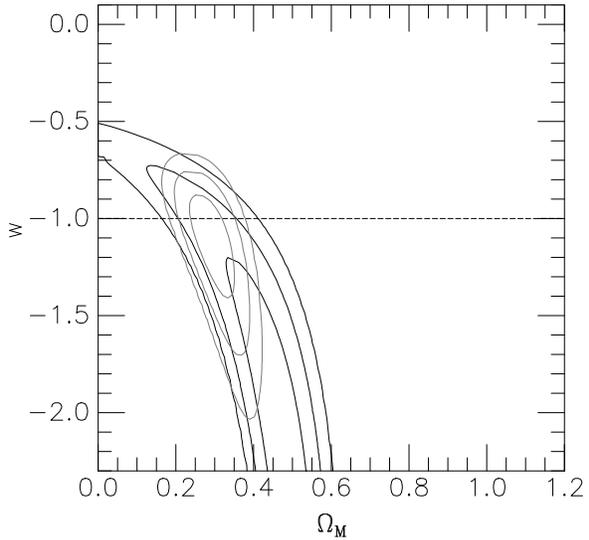}
\caption{Probability contours for dark energy parameter $w$ versus   
$\Omega_m$ are shown at 1$\sigma$, 2$\sigma$, and 3$\sigma$ when
$\Omega_{tot} = 1$. 
We also give 1$\sigma$, 2$\sigma$, and 3$\sigma$ contours when 
we adopt a prior of   
$\Omega_M h = 0.20\pm0.03$ from the 2dF survey (Percival et al. 2001).  
This sample includes all 172 \sna\ with $z>0.01$ and $A_V<0.5$ mag.
\label{omw}}
\end{figure}

We can also fit for the equation of state of the dark energy,
$p=w\rho$, where $w=1/3$ for radiation, $w=0$ for pressureless matter
(e.g., cold dark matter), $w=-1$ for a cosmological constant, etc.
Restricting ourselves to $\Omega_{tot}=1$, Figure~\ref{omw} shows the
contours for $\Omega_M$ and $w$, with and without the 2dF prior.

\begin{figure}[t]
\plotone{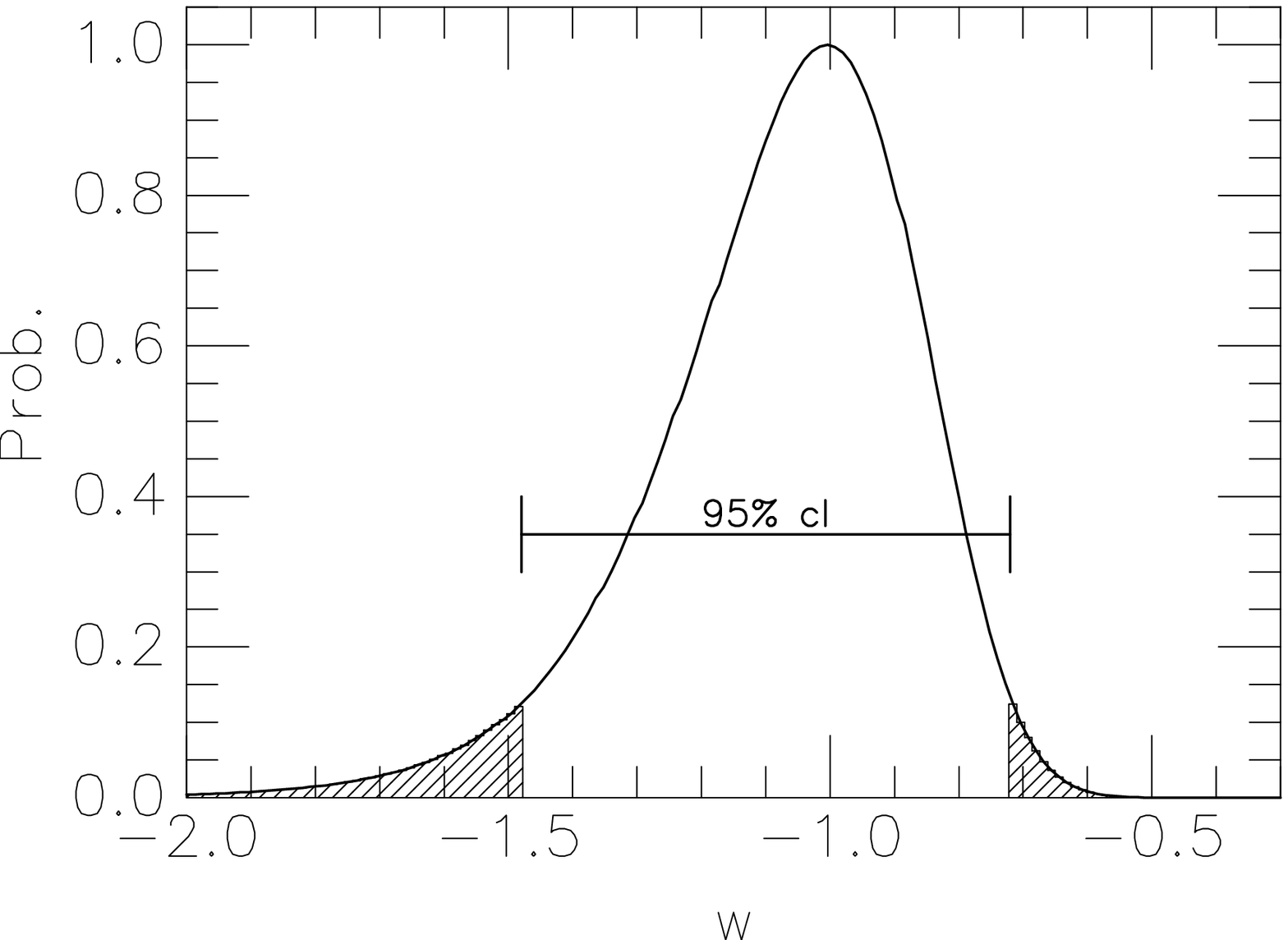}
\caption{The probability distribution for $w$ using the 2dF prior,
marginalized over $\Omega_M$.  The 95\% confidence limits are 
$-1.48<w<-0.72$.
\label{wprob}}
\end{figure}

Dark energy candidates with $w > -2/3$ are strongly disfavored by
these data.  Figure~\ref{wprob} shows the data from Figure~\ref{omw}
marginalized over $\Omega_M$.  The combination of \sna\ and
constraints on $\Omega_M$ from large scale structure puts a sharp
constraint on how positive $w$ can be.  The 95\% confidence limits on
$w$ lie in the range $-1.48<w<-0.72$.  If we additionally adopt a prior
that $w>-1$, we find that $w<-0.73$ at 95\% confidence.  These
constraints are again very similar to recent results reported using
the WMAP satellite: $w<-0.78$ at 95\% confidence for priors of $w>-1$
and $\Omega_M$ from the 2dF redshift survey (Spergel et al. 2003).

\begin{figure}[t]
\plotone{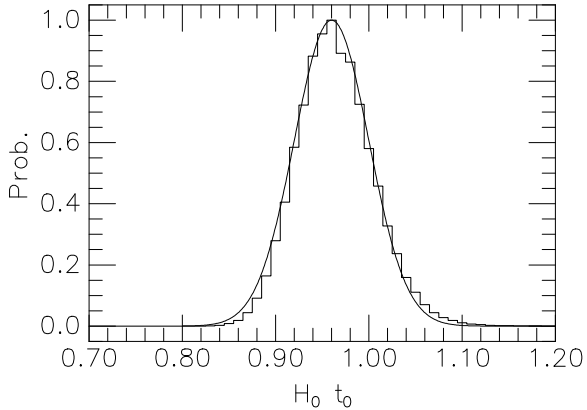}
\caption{The probability distribution for $H_0\,t_0$ given the \sna\
observations is tightly constrained to $0.96\pm0.04$, and an
approximating Gaussian curve.
\label{h0t}}
\end{figure}

Since the gradient of $H_0\,t_0$ is nearly perpendicular to the narrow
dimension of the $\Omega_\Lambda$--$\Omega_M$ contours, we obtain a
surprisingly good estimate of $H_0\,t_0$, illustrated in Figure~\ref{h0t}.
We find that $H_0\,t_0 = 0.96\pm0.04$.  Although independent
measurements of these parameters do not approach this precision yet,
for comparison we note that $h=0.72\pm0.08$ from Freedman \etal\ (2001)
corresponds to a Hubble time of $H_0^{-1} = 13.6\pm1.5$~Gyr.
Krauss \& Chaboyer (2002) have recently found a median age of 12.5 Gyr
for 17 metal-poor globular clusters.  If these globular clusters formed
0.6 Gyr after the Big Bang, then the universe has $t_0$ = 13.1 Gyr,
consistent with a world model with $h = 0.72$,
$\Omega_M$ = 0.3, and $\Omega_\Lambda$ = 0.7. The assumed globular
cluster incubation time of 0.6 Gyr corresponds to an epoch of star 
formation we would observe now at $z\approx8$.
Adopting an uncertainty of $\pm$ 2 Gyr for $t_0$, the product
$H_0\,t_0$ = 0.96 $\pm$ 0.19.  We also note that remarkable agreement
with the WMAP result for the age of the universe: $13.7\pm0.2$~Gyr
(Spergel et al. 2003).

\section{Discussion}

\subsection{Testing systematics}

Drell, Loredo, \& Wasserman (2000) (DLW) wrote a very thoughtful
critique of the then extant SN Ia data, questioning whether a
cosmological constant was required or a systematic error could account
for the observations.  We present here more than twice as many objects
with improved error estimates.  We believe that some of the points
noted by DLW are simply caused by selection effects, for example the
observed diminution of the range of $\Delta$ with redshift.  However,
their point that a large systematic shift in the luminosity of
supernovae with redshift, coupled with a large positive $\Omega_M$
could fit the data nearly as well as a universe with a cosmological
constant, is still valid."

Using our preferred sample of 172 \sna\ we reevaluate the Bayesian odds
factor between a model ``FRW'' with no systematic error and a cosmological
constant and a ``Model II'' with no cosmological constant but a systematic
error which goes as a power law of $(1+z)$ (i.e., the distance modulus
has a systematic error term of $\beta \ln(1+z)$ --- this is the DLW
``Model II'').
When we integrate both models over $H_0$, $\Omega_M$, and either
$\Omega_\Lambda$ or $\beta$ and divide by plausible prior ranges for
these parameters, we get an odds factor of 29 in favor of the FRW
model.  Most would interpret this as ``pretty good'' odds that the FRW
model is to be preferred over a systematic error, but not extremely
strong odds.

However, in this Bayesian framework, the ratio of the probabilities
for the two models also includes the ratio of the prior probabilities.
The systematic model can match the observations reasonably well, but
it does so only for $\Omega_M\approx 1$ and a $\beta$ which
corresponds to a systematic which rises to 0.7 mag at $z=1$.  Both of
these parameters seem out of step with other lines of investigation
and would be quite disfavored by reasonable priors.

We certainly do not expect that $\Omega_M$ could be anywhere near
unity.  The evidence against this comes from the observations that
the $H_0\,t_0$ product appears to be much larger than $2/3$, and there is
ample evidence on many scales that there is nowhere near as much mass
as $\Omega_M = 1$ in clustered form (e.g., Blakeslee et al. 1999;
Percival et al. 2001).

The argument against such a very large systematic error (0.7 mag at
$z=1$) is theoretical and empirical.  We work with \sna\ precisely
because they are dead stars which should have very little difference
between today and 7 Gyr ago, and we certainly do not have any
theoretical expectation that explosions have become a factor of 2
dimmer in the mean over the previous 7 Gyr.  

Empirically, we argue that the \sna\ we see locally should have at
least the range of properties seen at $z=1$ because there are low
metallicity systems today and there appear to be \sna\ which are
exploding quite promptly (e.g., NGC~5253 at 0.1~$L^*$ had a
starburst recently and has had 2 \sna\ in the past century).  Likewise,
if \sna\ were as much as a factor of 2 dimmer in the past, we would
expect there to be a significant dispersion --- surely {\it some} of
the \sna\ at $z=1$ would produce explosions comparable to those of
today.

We are always cognizant of the possibility of observational systematic
error or problems with our analysis, but 0.7 mag is considerably
beyond what we think might be possible.

The exact factor by which Model II would therefore be disfavored
according to these prior considerations is subjective since the
place where the systematic probability is becoming sizable is 
out on the tails of the priors we assign for $\Omega_M$ and
$\beta$.  We think it is not worthwhile to try to assign a
quantitative probability to these, other than to note that the factor
by which the net probability of Model II relative to Model FRW will
be disfavored will be significantly higher than the factor
of 29 from the integrated likelihoods.  

Therefore we believe that we have very strong evidence that a
systematic effect which goes as a power law of $(1+z)$ is not likely
to match the \sna\ data.  In particular, a simple gray or
intergalactic dust model does not appear to be viable, and
evolutionary effects such as metallicity, C/O ratio, or host dust
properties which change smoothly and monotonically with $z$ are also
likely not to fit the data as well as an FRW cosmology.

It is always possible to imagine a systematic error which is tuned to match
the FRW turnover near $z=1$, but it might be less a priori plausible
than a simple $(1+z)$ power.  However, a proper Bayesian analysis would
have to take into account the volume of the space from which such a
model was drawn and the model might be disfavored for that reason,
even though it might fit the \sna\ data very well.

\subsection{Selection and Other Effects}

The Fall 1999 survey went deeper and redder than any previous
campaign.  As a result we did start to see extinguished \sna\ at
$z\approx0.5$ and we expect that we are less subject to selection
bias.

Our model for rates seems to work remarkably well, suggesting
that there is not a rare population of luminous objects which we do
not know about from local samples.  The exceptions are SN~1999fo
(Gertie) and SN~1999fu (Alvin), which were not \sna, but were
definitely unusual, flux-variable objects.

At face value the average absorption of the distant supernovae is
significantly smaller than for the nearby sample. Out to $z\le0.3$ the
151 supernovae suffered an average absorption of $\langle A_V
\rangle$=0.35$\pm$0.50 mag, while the 37 distant supernovae with $z > 0.3$
average $\langle A_V \rangle$=0.20$\pm$0.23 mag.  There is a large scatter
in these averages, but the offset is 0.15 mag. The
distribution of absorption values for the nearby supernova sample has
a long tail to higher absorptions, which is missing for the distant
sample. However, the distribution of absorptions of the distant
supernovae is much narrower. Most probably this is a signature of the
magnitude-limited searches, where heavily obscured supernovae are not
detected. Selecting the 130 objects (172 less 42 SCP \sna\ without host
extinctions) which were used to construct the
likelihood functions ($z>0.01$ and $A_V <0.5$ mag), we find that the
averages decrease to $\langle A_V \rangle=0.17\pm0.14$ mag for the
nearby supernovae and $\langle A_V \rangle=0.14\pm0.11$ mag for the
distant ones. The difference has decreased to less than 0.03
mag and clearly the faintness of the distant supernovae cannot
be attributed to errors in the treatment of absorption.  In fact, more
absorption would make the distant supernovae more luminous and move
them to larger distances. The small average absorption argues that the
distant objects indeed are at larger distances than for an empty
universe model, in agreement with the results from Riess et
al. (2000), Leibundgut (2001), and Sullivan et al. (2003).

Of the objects at $z>0.8$ in our large data set, several objects
(principally those discovered before 1999) rely on rest-frame $U$-band data
to measure extinction. The rest-frame $U$-band data have a number of
uncertainties, including less certain K-corrections, greater susceptibility
to extinction, and less well studied behavior. Most importantly,
the \sna\ have less intrinsic uniformity in the $U$ bandpass. For
the data set presented here, the average measured distance to those
objects using the $U$ band as a primary source of information for distance
measurement is further than to those objects which do not rely on
$U$-band data, but the difference is not statistically significant. These
effects will be further discussed in Jha \etal\ (2003c) and Suntzeff
\etal\ (2003).

We have not corrected these data for gravitational lensing.  It
should be negligible at these redshifts, but will become much more
serious as we push to higher redshift.

Nevertheless, we do think we can start to perceive systematic errors
in the data at the 0.04 mag level.  The contours seen in Figures
\ref{omom_new}--\ref{omom_all} have their most probable
$\Omega_M,\Omega_\Lambda$ well offset from the $\Omega_M$ indicated by
the 2dF survey and a flat universe.  This is not a failure of our
probability analysis.  We have performed extensive Monte Carlo tests
on simulated data sets and we recover $\Omega_M,\Omega_\Lambda$ with
very little bias.  What drives our best fit solution to high values of
$\Omega_M,\Omega_\Lambda$ is the fact that the data make a very fast
transition from recent acceleration to earlier deceleration,
i.e. the transition from the faintness at $z\sim0.5$ to the brightness
at $z\sim1$ is very fast.  The contours in Figures
\ref{omom_new}--\ref{omom_all} are quite shallow in the
$\Omega_M+\Omega_\Lambda$ direction, and a systematic error of 0.04 mag
is quite sufficient to move our most probable values for
$\Omega_M,\Omega_\Lambda$ to these high values.
This may be a statistical fluke at the $1.5$-$\sigma$ level, although
the systematic offsets of the points in Figure~\ref{hubmed} from a
cosmological line may point to systematic errors.

We have described some of the possible culprits for such a systematic
above.  Selection biases because of UV selection or any of the other
practical difficulties in carring out a supernova search are always a
worry, particularly since they are so hard to quantify and because
\sna\ are so non-uniform in the UV.  Photometric and K-correction
errors are certainly possible, since absolute photometry of objects
with non-stellar SEDs at $m>20$ is {\it extremely} hard to achieve at
the 0.01 mag level.  It is conceivable that gravitational lensing is
more important than has been estimated; we expect that the \sna\ found
by the GOODS survey will offer a strong test of this.  Most likely
there is a variety of errors throughout this very heterogeneous set of
data, and the best way forward is acquisition of new data from well
designed observations which are taken and analyzed with more care.
Although the basic conclusion that the Universe is accelerating is not
in jeopardy since the $\Omega_M-\Omega_\Lambda$ direction is
relatively insensitive to systematic error, we remain wary of
over interpretation of the extant \sna\ data and urge others to do the
same.

\subsection{Comparison with the CMB}

As this paper was nearing completion, the WMAP team
reported new results from a high angular resolution all-sky map of
structure (Spergel et al. 2003). Their main conclusions, namely that
we live in a geometrically flat Universe with $\Omega_{total}$,
$\Omega_M$, and $\Omega_{\Lambda}$ of 1.00, 0.73, and 0.27, respectively,
are entirely consistent with the results presented here. This
convergence of results lends great strength to what is emerging as the
new ``Standard Model" of contemporary cosmology, in which the role of
matter is secondary to the accelerating expansion driven by dark
energy.

Ascertaining the physical mechanism that is responsible for this
accelerating expansion is one of the crucial next steps for
observational cosmology. Since supernovae are well suited to probing
precisely the redshift range over which $\Lambda$ begins to dominate,
new projects are being undertaken to exploit this.

\subsection{What Next?  The Future}

Future campaigns to observe \sna\ at high redshift will focus on three
themes: refining measures of luminosity distance, understanding and
controlling systematic errors, and learning about \sna\ physics.  The
payoff for measuring luminosity distances accurately is that we can
determine the equation of state of the dark energy and whether it has
changed as a function of time (redshift).  There is certainly a
systematic floor to our ability to determine distances from \sna, but
we do not yet know what it is.  It appears that gray dust is not an
issue, but gravitational lensing, for example, is certainly a
contaminant.

Opinions differ as to where the
incompletely understood \sna\ physics causes a cosmic scatter which
does not average out, but most would agree it lies in the range of
0.02--0.04 mag.  This implies that the luminosity distance at a given
redshift is not improved by observing more than 25--100 \sna\ at that
redshift.  There
are many outstanding questions about \sna\ physics which may best be
approached by observations of \sna\ at high redshift.
Measuring \sna\ rates and comparison with the star-formation rate may
elucidate the \sna\ explosion mechanism, \sna\ progenitors, and
gestation period.  The distribution of \sna\ parameters as a function
of redshift may also be useful for understanding \sna\ and host galaxy
physics.

The HZT, in collaboration with the Surf's-Up group at the IfA,
undertook a six-month survey of 2.5 deg$^2$ in $R$, $I$, and $Z$.
Many \sna\ as well as other transient objects were discovered and 
followed (Barris et al. 2003).  Members of the HZT have
begun the ESSENCE survey (Smith et al. 2002) to discover and
follow 200 \sna\ in the range $0.1<z<0.8$ with the goal of measuring the
equation of state parameter $w$ of the dark energy to $\pm$10\%.  
Riess et al. (2003) have an ongoing program with \hst\ and the 
GOODS survey, using the Advanced Camera for Surveys (ACS) to discover
and follow approximately 6 \sna\ at $z\approx 1.4$.  This will be
exceedingly powerful for measuring the effects of dark energy,
understanding systematics, and constraining any time evolution of $w$.
The CFHT Legacy Survey (http://www.cfht.hawaii.edu/Science/CFHLS/) has
the goal of finding and following $\sim2000$ \sna\ in multiple colors.
The Lick Observatory Supernova Search (LOSS; Filippenko et al. 2001)
is finding and monitoring record numbers of very nearby supernovae,
with 82 in 2002 alone (http://astro.berkeley.edu/$\sim$bait/kait.html). 
The Nearby Supernova Factory (Aldering et al. 2002) will obtain
spectrophotometry for $\sim300$ nearby \sna, suitable for
K-corrections of high-redshift supernovae and for exploring \sna\
physics and systematics.  In the near future, Pan-STARRS (Kaiser et
al. 2002) should discover and follow a vast number of \sna\ (${>}10^4$
per year).  Finally, in the more distant future, the proposed SNAP
satellite (Nugent 2001) would follow several thousand objects with an
experiment designed to minimize observational systematic errors,
limited only by the floor imposed by nature.  The future
indeed looks bright for cosmological and related studies of \sna.

\acknowledgments

We thank the staffs at many observatories for their assistance with
the observations.  Financial support for this work was provided by
NASA through programs GO-08177, GO-08641, and GO-09118 from the Space
Telescope Science Institute, which is operated by the Association of
Universities for Research in Astronomy, Inc., under NASA contract NAS
5-26555. Funding was also provided by National Science Foundation
grants AST-9987438 and AST-0206329.  A.C. acknowledges the support of
CONICYT (Chile) through FONDECYT grant Nos. 1000524 and 7000524.  This
paper uses data from the Sloan Digital Sky Survey (SDSS) early data
release.

\section{Appendix A}

\subsection{$N(N-1)/2$ differences}

We use the Alard \& Lupton (1998) code to convolve one image to match
the other.  We fit a suitable star in one of these images with the
Schechter ``Waussian'' (inverse of a truncated Taylor series for
$e^{x^2}$) to derive a standard PSF.  Since we know where the
supernova is located, we do a two-parameter fit (peak and sky) of this
standard PSF, and we scale the flux measured in the standard
star by the ratio of the fit peaks to derive a supernova flux
difference for these two images.  We also select roughly ten other
locations in the image and insert a copy of the standard star scaled
to match the supernova flux, and pick it up again using the same
procedure of two-parameter fits.  This tells us of any bias (we do not
see any) and gives us an estimate of the flux error for this pair of
images.

Applying this procedure to all $N(N-1)/2$ pairs creates an $N\times N$
antisymmetric matrix of flux differences, and a symmetric matrix of
errors.  We also choose approximately a dozen stars for each supernova
field which are bright enough to have good S/N in all images but faint
enough not to be saturated, and perform a PSF flux measurement for all
stars in all images.  Matching up these tables of stars, we can form a
mean flux scale between all pairs of images, and then assemble another
$N\times N$ antisymmetric matrix of magnitude differences.
The flux differences are all put on a common scale using the magnitude
differences.

Finally, we fit these antisymmetric matrices as the difference between
the $i^{th}$ and $j^{th}$ entries of an $N$ vector.  We derive two
sets of error estimates: the first is based on the pairwise flux error
estimates and the second comes from the residuals between the observed
difference matrix and the matrix assembled by differencing the terms
of the resultant vector.  These estimates are generally consistent
with one another, although the observed errors reflect the
uncertainties of the individual observations and the mismatch errors
give a broader view of the inconsistency of any observation with the
others.


\end{document}